\documentclass[12pt]{article}
\usepackage{amsmath, amssymb}
\usepackage{graphicx}
\usepackage{enumerate}
\usepackage{natbib}
\usepackage{comment}
\graphicspath{{figures}}

\usepackage{enumitem}
\usepackage{mathrsfs, booktabs, amsthm, amsfonts, bbold}
\usepackage{hyperref}
\usepackage{subfiles}

\usepackage{xcolor}
\protected\def\[#1\]{\begin{equation}\begin{aligned}#1\end{aligned}\end{equation}}
\protected\def\(#1\){\begin{equation*}\begin{aligned}#1\end{aligned}\end{equation*}}

\newtheorem{theorem}{Theorem}

\newtheorem{lemma}{Lemma}

\theoremstyle{remark}
\newtheorem{remark}{Remark}%

\theoremstyle{definition}

\usepackage{subcaption}
\usepackage{float}

\usepackage[plain,noend]{algorithm2e}
\usepackage[mathscr]{euscript}
\usepackage{overpic}
\usepackage{bm}

\begin{document}

\def\spacingset#1{\renewcommand{\baselinestretch}%
{#1}\small\normalsize} 
\spacingset{1}

  \title{\bf Graphical Model-based Inference\\on Persistent Homology}
  \author{
  	Zitian Wu\thanks{\href{email:zitianwu@ufl.edu}{zitianwu@ufl.edu}}\\
	Department of Biostatistics, University of Florida\\
    and \\
Arkaprava Roy\thanks{\href{email:arkaprava.roy@ufl.edu}{arkaprava.roy@ufl.edu}}\\
	Department of Biostatistics, University of Florida\\
		    and \\
	Leo Duan\thanks{\href{email:li.duan@ufl.edu}{li.duan@ufl.edu}}\\
	Department of Statistics, University of Florida
    }
  \maketitle
  \vspace{-0.5in}

\begin{abstract}
  Persistent homology is a cornerstone of topological data analysis, offering a multiscale summary of topology with robustness to nuisance transformations, such as rotations and small deformations. Persistent homology has seen broad use across domains such as computer vision and neuroscience. Most statistical treatments, however, use homology primarily as a feature extractor, relying on statistical distance-based tests or simple time-to-event models for inferential tasks. While these approaches can detect global differences, they rarely localize the source of those differences. We address this gap by taking a graphical model-based approach: we associate each vertex with a population latent position in a conic space and model each bar's key events (birth and death times) using an exponential distribution, whose rate is a transformation of the latent positions according to an event occurring on the graph. The low-dimensional bars have simple graph-event representations, such as the formation of a minimum spanning tree or the triangulation of a loop, and thus enjoy tractable likelihoods. Taking a Bayesian approach, we infer latent positions and enable model extensions such as hierarchical models that allow borrowing strength across groups. Applications to a neuroimaging study of Alzheimer's disease demonstrate that our method localizes sources of difference and provides interpretable, model-based analyses of topological structure in complex data.\\
  The code is provided and maintained at {\color{blue}\url{https://github.com/zitian-wu/graphPH}}.
\end{abstract}
\vfill

\addtolength{\textheight}{-0.8in}
\addtolength{\topmargin}{0.5in}

\newpage

\section{Introduction}

Persistent homology is a foundational tool in topological data analysis that captures topological features of data. Roughly speaking, homology can be thought as $k$-dimensional holes in a topological space, and persistent homology tracks the birth and death of these holes as a function of a scale parameter. Barcode is commonly used for representing the persistent homology \citep{ghrist2008barcodes,edelsbrunner2010computational}.

\begin{figure}[H]
  \centering
  \begin{subfigure}[t]{0.48\textwidth}
      \centering
      \includegraphics[width=\textwidth]{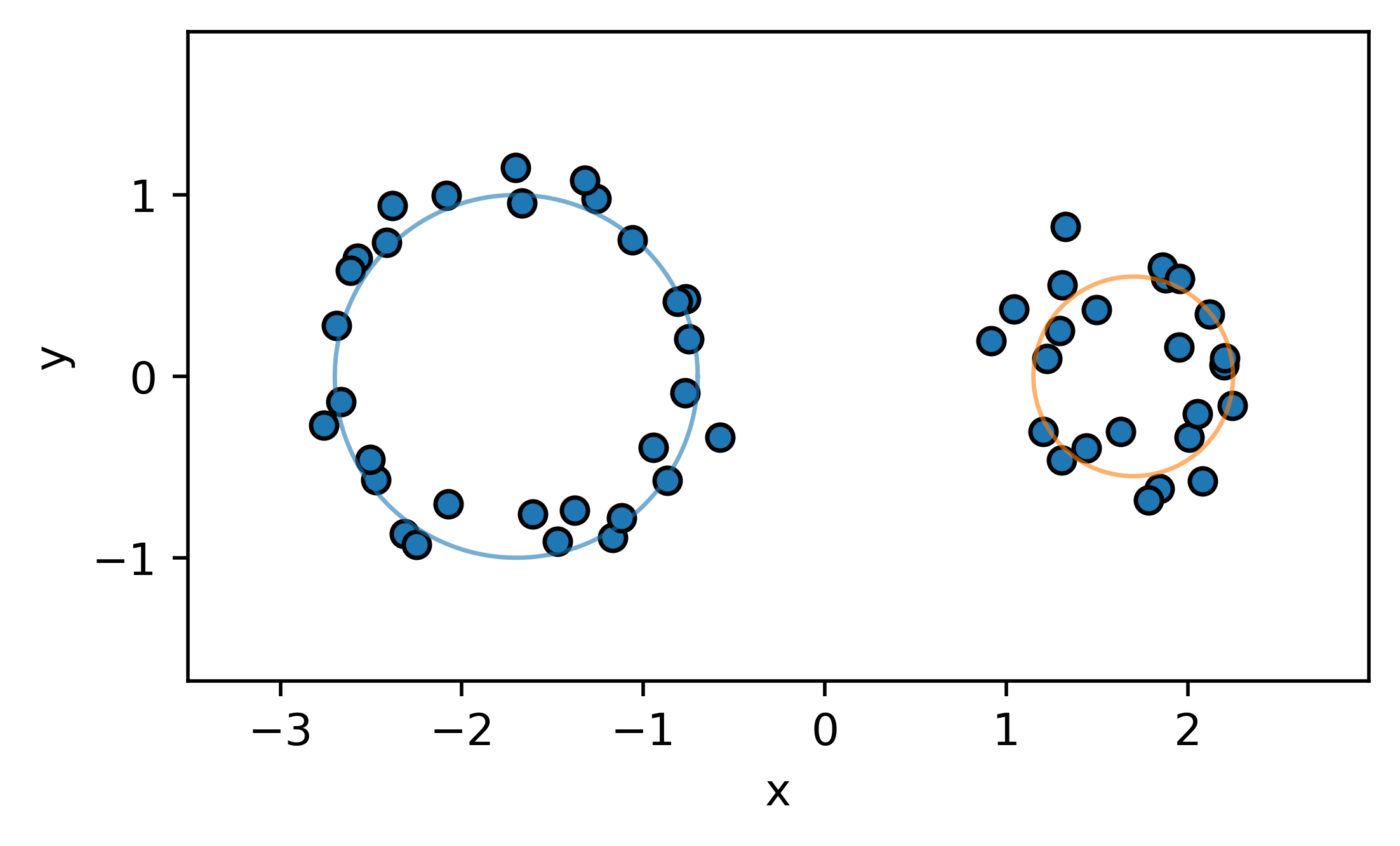}
      \caption{Points near two circles.}
  \end{subfigure}
  \hfill
  \begin{subfigure}[t]{0.51\textwidth}
      \centering
      \includegraphics[width=\textwidth]{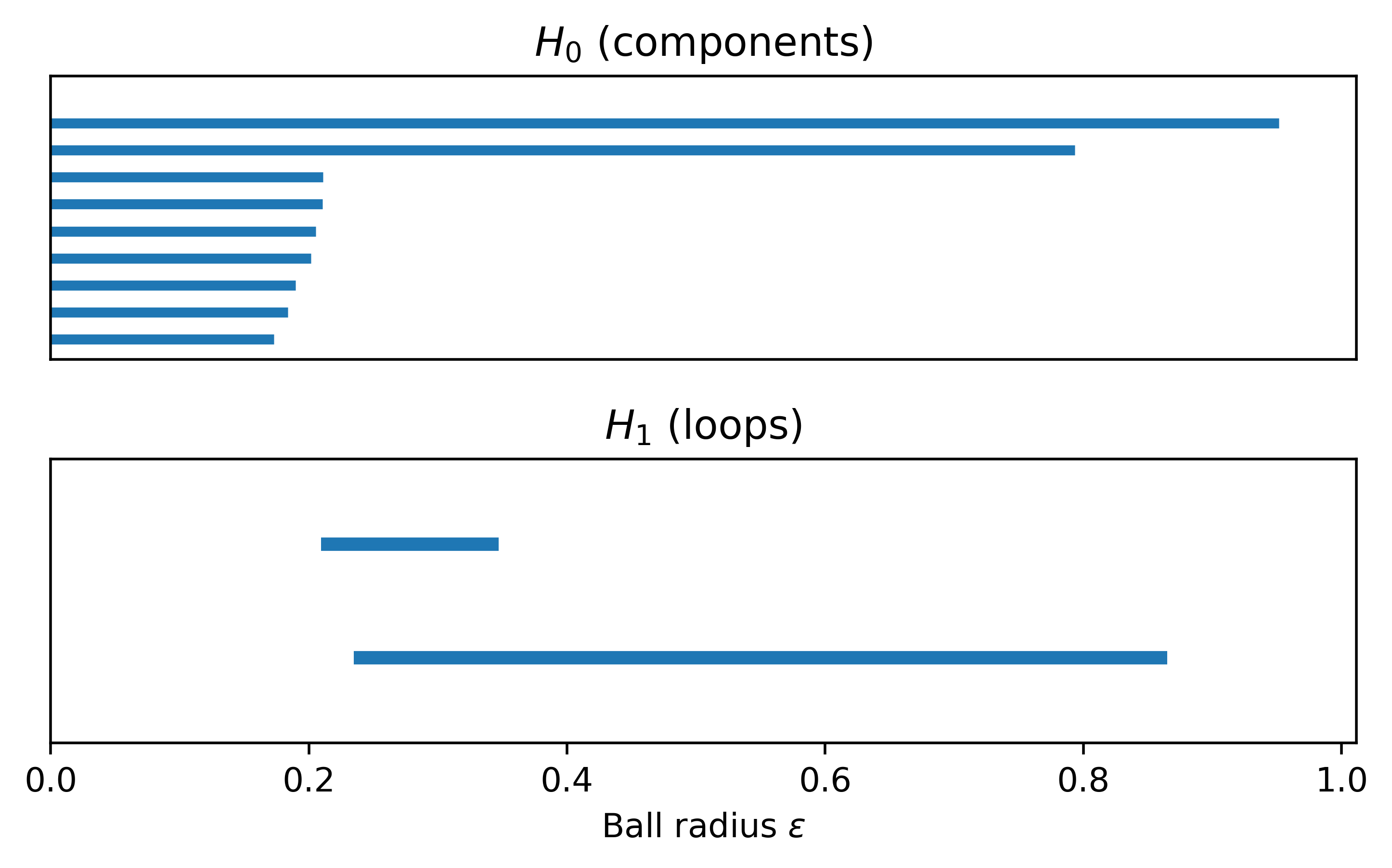}
      \caption{Persistence barcodes.}
  \end{subfigure}
  \caption{
Illustration of the persistent homology of two circles via the Vietoris-Rips filtration: a ball of radius $\epsilon$ is placed around each point and is gradually expanded. Zero-dimensional features tracks the formation of connected components, when two components merge, one component  persists and the other dies. In this example, although multiple components may form during the expansion, most are short-lived and only two components have a long persistence time (the bars with smaller or infinite death time are omitted). One-dimensional features tracks the formation of two loops, formed when connected components form cycles with holes.
\label{fig:barcode}
  }
\end{figure}

To illustrate, we now examine the formation of a barcode via the Vietoris-Rips filtration \citep{vietoris1927hoheren,gromov1987hyperbolic},  which places a ball of radius $\epsilon$ (the scale parameter mentioned above) around each point and gradually increases $\epsilon$. As the radius expands, these balls begin to overlap and merge, connecting nearby points and forming larger connected components, also known as the zero-dimensional ($H_0$) features. Further, during the growth of $\epsilon$, connected components combine, loops may emerge when multiple connections create cycles. The loop is a one-dimensional ($H_1$) feature, or intuitively, a one-dimensional hole. Similarly, higher dimensional features/holes could form. By observing how these topological features appear and disappear as $\epsilon$ changes, we identify which structures {\em persist} across multiple scales of $\epsilon$, characterizing the underlying shape and patterns within the data.  Figure \ref{fig:barcode} illustrates such a persistence barcode.

The persistent homology framework is versatile and enjoys a wide variety of filtrations to accommodate different needs for capturing topological features. For example, the Vietoris-–Rips captures the relative proximity of points and is invariant under rotation and translation of the data set; the radial sublevel filtration uses the distance from a designated central point, and characterizes radial symmetry or asymmetry.

With an origin in computational topology, persistent homology has gained a surge of interest in statistical analysis. Progress has been made on quantifying the distribution of features. To be clear on the terminology, one says there is a barcode (a collection of bars) for each subject (such as an image); and over multiple subjects sampled from a population, one obtains a distribution of barcodes.
\cite{mileyko2011probability} defined probability measures on persistence diagram, a transform of barcode;
\cite{cohen2005stability} showed the stability of diagram, as small perturbations of the sample lead to only small changes in the diagram, hence the diagram enjoys concentration of measure;
\cite{fasy2014confidence} studied bootstrap confidence intervals of persistence diagrams;
\cite{moon2023hypothesis} proposed hypothesis testing procedures on persistence images, a vector representation for diagram \citep{adams2017persistence};
\cite{bubenik2015statistical} established that the persistence landscapes (another barcode transform) enjoys appealing large sample properties, such as asymptotic normality of sample mean. Besides these methods, many appproaches have been developed for specific applications \citep{biscio2019accumulated,thomas2023feature, crawford2020predicting, moon2023using}.

Despite the rich literature, the aforementioned methods are largely based on treating the persistence barcode or its transforms as summary statistics, thereby characterizing the sampling distribution. In parallel, there is a nascent class of model-based approaches in which one imposes a statistical model on the persistence outputs. For example, \cite{maroulas2020bayesian} took a Bayesian approach and modeled the persistence diagrams as Poisson point processes; \cite{garside2021event} considered the barcode as equivalent to survival data, hence allowing for the estimation of the hazard function. Compared to the summary statistic-driven methods, the model-based ones are appealing because they provide model-based smoothing, allow straightforward extensions such as incorporating covariates, and are suitable for inference when the sample size is far from the asymptotic regime.

Our work follows the model-based approach, and is motivated by the goal of {\em localization}: to identify the source of difference between the barcodes in the original observation space. Clearly, the localization task requires a more fine-grained approach than simply treating the bars as independently generated from a simple distribution (for example, as one would do in a survival model). The key challenge is on how to represent the dependency between the bars in a statistical model.

To achieve the goal of modeling bar dependency, we first note that the  homology features are derived from a filtration (a monotonically changing set according to $\epsilon$) of simplicial complex, for which the one-dimensional complex is a graph consisting of vertices and edges, and the two-dimensional complex further involves triangles (binary relationship between three vertices). That means there is already rich information in the simplicial complex. However, a full-scale modeling of the simplicial complex carries a high modeling and computational burden, than focusing on features such as the barcode themself. For example, for $n$ vertices and $T$ discrete time points, a model that tracks all triangles has an $O(T n^3)$ computational complexity, which is prohibitive for large $n$ and $T$.

To address this issue, we propose a middle-ground approach that uses simple graph events to probabilistically represent the key time points of the filtration—the birth and death times of the bars. We show that the zero-dimensional bars can be represented by the formation of a minimum spanning tree, while the one-dimensional bars correspond to the triangulation of a loop. Importantly, we can obtain useful information, such as the edges of the tree and the vertices of the loop, by directly leveraging  high-performance computational tools developed by the computational topology community (such as GUDHI, \cite{gudhi:urm}), saving statisticians effort to deal with high-dimensional complexes. In this way, one can focus on the low-cost modeling of the remaining graph events. To localize events to specific vertices, we employ a Bayesian latent position model \citep{hoff2002latent} and propose a novel conic space representation that enjoys log-concavity of the posterior distribution, thus ensuring good statistical properties such as identifiability and consistency.

 Section~2 presents the proposed likelihood and prior specifications. In Section~3, we describe posterior computation, and in Section~4, we provide the asymptotic theoretical analysis. Section~5 reports simulation studies evaluating the method's ability to localize between-group differences and the benefits of incorporating higher-order topological features into the analysis. Finally, Section~6 applies the proposed method to structural connectome data from the Alzheimer's Disease Neuroimaging Initiative.

\section{Method}
To be informative to general statistical readers, we first give a brief introduction to complex filtration, and then describe our graphical model for persistent homology.

\subsection{Background on simplicial complex and filtration}\label{sec:filtration}
Let $V$ be a finite set of points, with each point a vertex.
A \emph{simplicial complex} (or complex for short) $\mathcal K$ on $V$ is a collection of subsets of $V$ such that:
(i) every vertex $v \in V$ appears as a singleton $\{v\} \in \mathcal K$;
(ii) the set is closed by subseting, if $\sigma \in \mathcal K$ and $\tau \subseteq \sigma$, then $\tau \in \mathcal K$.  Each member vertex set $\sigma \in \mathcal K$ is called a simplex.  For example, in a complex $\mathcal K=\{\{1\}, \{2\}, \{3\}, \{4\}, \{1,2\}, \{1,3\}, \{2,3\}, \{1,4\}, \{1,2,3\}\}$, each $\sigma\in \mathcal K$ is a simplex. We refer to a simplex of size $k$ as a $(k-1)$-simplex and treat it like a fully connected small graph. Thus, a $0$-simplex represents a vertex, a $1$-simplex does an edge, a $2$-simplex does a triangle, and so on. The nested sequence of complexes $\mathcal K_0\subseteq \mathcal K_1\subseteq \cdots \subseteq \mathcal K_{m}$ is called a \emph{filtration} of $\mathcal K_m$.

As discussed above, the standard Vietoris--Rips filtration is constructed from a set of points $W$ equipped with a metric in the space, $\text{dist}: \mathcal Y \times \mathcal Y \to [0,\infty)$. Given a parameter $\epsilon \geq 0$,  a set of $(k+1)$ points form a $k$-simplex in $\mathcal{K}_\epsilon$ if and only if $\text{dist}(\tilde y_j, \tilde y_{j'}) \leq \epsilon$ for all $j\neq j'$. We can see that  $\mathcal K_\infty=\lim_{\epsilon \to \infty}\mathcal{K}_\epsilon$ can be interpreted as all non-empty subgraphs of a complete graph formed among the data points. Therefore, one may refer to the standard Vietoris--Rips filtration as the Vietoris--Rips filtration of a complete graph.
The persistence barcode is a summary statistic of the filtration. We will discuss a few other common filtrations, including the Vietoris-- Rips filtration of a cubical complex, the alpha filtration, and the sublevel filtration in the supplementary materials. Since most of the filtrations enjoy almost the same graph model-based characterization (except for the sublevel filtration), we focus on the Vietoris-- Rips filtration in the main text.

\subsection{Persistent homology likelihood}\label{sec:method-lik}
We now take a likelihood-based approach to model the persistent homology outputs. Consider the observed data as $\tilde y^{s}$ with superscript $s$ as the subject index, in some space $\mathcal{Y}$. We assume the data to be generated from the following likelihood:
\[\label{eq:generative}
\tilde{\mathcal L} \{ \tilde y^{s} \mid (\theta, \rho) \} =  \mathcal{G}( \tilde y^{s} \mid y^{s};\rho) \mathcal{L}( y^{s} ; \theta),
\]
where $y^{s}=T(\tilde y^{s})$ is the persistent homology features (such as the barcode and the complexes at birth and death times) generated via a filtration $T$. We consider $\theta$ as the parameter of interest, governing the generative likelihood $\mathcal{L}$ of the features. The change from the features to the observed data is captured by $\mathcal{G}$, a conditional distribution given $y^{s}$, such as those presenting random rotations or small deformations.

In modern statistical and machine learning literature, it has become increasingly easy to model the generative distribution $\mathcal{G}$, for example by using generative adversarial networks \citep{goodfellow2020generative} with barcode-based distances.
Since our primary focus is not on data prediction but on characterizing the between-group differences, we assume the parameters of interest reside only in $\mathcal{L}$, thus allowing us to ignore $\mathcal{G}$ in our inferential task. We now refer to $\mathcal L$ as the persistent homology likelihood.

To be clear, we treat $\mathcal{L}$ as a \emph{partial likelihood}, as inspired by \citet{cox1975partial}. The filtration transforms $\tilde y^{s}$
into an ordered sequence of graph events that characterize the topology unique to subject/group. By giving up the attempt to model the remaining factors,
we gain robustness against misspecification of the parts of the data
generating mechanism that are not of interest \citep{doksum1990consistent}.
The robustness obtained is dictated by the filtration, since the features
depend on the data only through what the filtration reads. For the
Vietoris--Rips and alpha filtrations this is the matrix of pairwise distances,
so the likelihood is invariant to rotation, translation, reflection and any
isometric deformation of the data; for a sublevel filtration it is the values
of the filtering function at the vertices, so the likelihood is invariant to
any transformation leaving those values unchanged, such as a rotation about
the reference point for a radial filtration. Each filtration thus contributes
an invariance of its own, as we discuss in the supplementary materials.
Using a partial likelihood is a long-standing practice. For the
Cox proportional hazards model, substantial literature justifies its use
as the marginal of a full sampling model
\citep{raftery1996accounting, sinha2003bayesian, kim2009bayesian,
ventura2016pseudo}. Rank likelihoods are another example, modeling
only the ranks of the observations, so that inference on the
parameter does not depend on the tail or asymmetry of the distribution
\citep{pettitt1983likelihood, hoff2007extending}. Closely related are quantile \citep{lavine1995approximate, dunson2005approximate} and copula
likelihoods \citep{genest1995semiparametric}. Distance and dissimilarity likelihoods \citep{duan2021bayesian,oh2001bayesian} model the pairwise information in place of
the point coordinates, discarding information that
is not needed for the clustering or embedding task. More broadly, our approach is related to the recent work of conditioning on insufficient statistics for robust consideration \cite{lewis2014bayesian,luciano2024insufficient}, although we assume $y^{s}$ is sufficient for $\theta$.

There are other approaches similar in spirit to partial likelihoods. Composite likelihoods multiply simple, low-dimensional marginal or conditional components \citep{varin2011overview, smith2009extended, pauli2011bayesian, ribatet2012bayesian, friel2012bayesian}.
Moment-based likelihoods specify only the mean and variance relationships based on estimating equations \citep{yin2009bayesian,ventura2010default}. Compared to the above alternatives, partial likelihood is distinct in the assumption that $\mathcal{L}(y^s;\theta)$ contains all the terms involving $\theta$. We adopt this assumption here.

For ease of notation, we temporarily suppress the index $s$. Each barcode is a collection of paired time points $(b^k_i, d^k_i)$, where $b^{k}_i\ge 0$ is a scalar denoting the birth time, and $d^k_i$ is the death time, $d^k_i\ge b^k_i$ and could take infinite values. For each bar, we augment with a list of edges/edge sets $\mathcal E^k_i$ that are directly relevant to the birth/death event. As we show later, $\mathcal E^k_i$ can be obtained as some transform of two complexes at the birth and death times. Therefore, our persistent homology features take the following form:
\[
y = \{ (b^k_i, d^k_i, \mathcal E^k_i): i=1,\ldots,n_k, k=0,\ldots,K\},
\]
and the bar is indexed by $i\in [n_k]=(1,\ldots,n_k)$ in each dimension. Following the convention of persistent homology literature, we start counting $k$ from zero.
In this work, we primarily focus on zero- and one-dimensional topological features, i.e., $k=0,1$, as higher-dimensional features are rarely used in practice. Nevertheless, the proposed likelihood naturally extends to any dimension $k \ge 2$, as illustrated by a simulation in the supplementary materials.

Suppose there are $n$ vertices for each subject. We assume the likelihood of all subjects in the same group is dependent on a shared matrix parameter $\Lambda\in (0,\infty)^{n\times n}$ with each element $\lambda_{j,k}=\lambda_{k,j}>0$. That is, we now consider $\Lambda$ (or its variants) as the parameter of interest $\theta$ from now on. We can now specify the likelihood for the zero-dimensional bars and the one-dimensional bars.

\noindent\textbf{Zero-dimensional bars and minimum spanning tree formation}

In the Vietoris-Rips filtration, the zero-dimensional bars represent connected components. We know the number of zero-dimensional bars has $n_0=n$, and $y$ is a transformation of the pairwise distance matrix $D=\{\Delta_{j,k}:(j,k)\in[n]^2\}$ with $\Delta_{j,k}= \text{dist}(\tilde y_j, \tilde y_k)$ such as the Euclidean distance. There is an interesting relationship between the zero-dimensional bars and the minimum spanning tree:
\(
\hat T = {\arg\min}_{T\in \mathbb{T}} \sum_{(j,k)\in E_T} \Delta_{j,k}
\)
where $\mathbb{T}$ is the set of all spanning trees of the complete graph for $n$ vertices, $T=([n], E_T)$, with $E_T$ containing $n-1$ edges that connect all $n$ vertices. The connection holds regardless of whether there are ties in the non-zero distances in $D$. Nevertheless, for ease of understanding, we will focus on no ties for now.

\begin{figure}[H]
    \centering
    \begin{subfigure}[t]{0.45\textwidth}
      \centering
      \includegraphics[width=\textwidth,height=5cm]{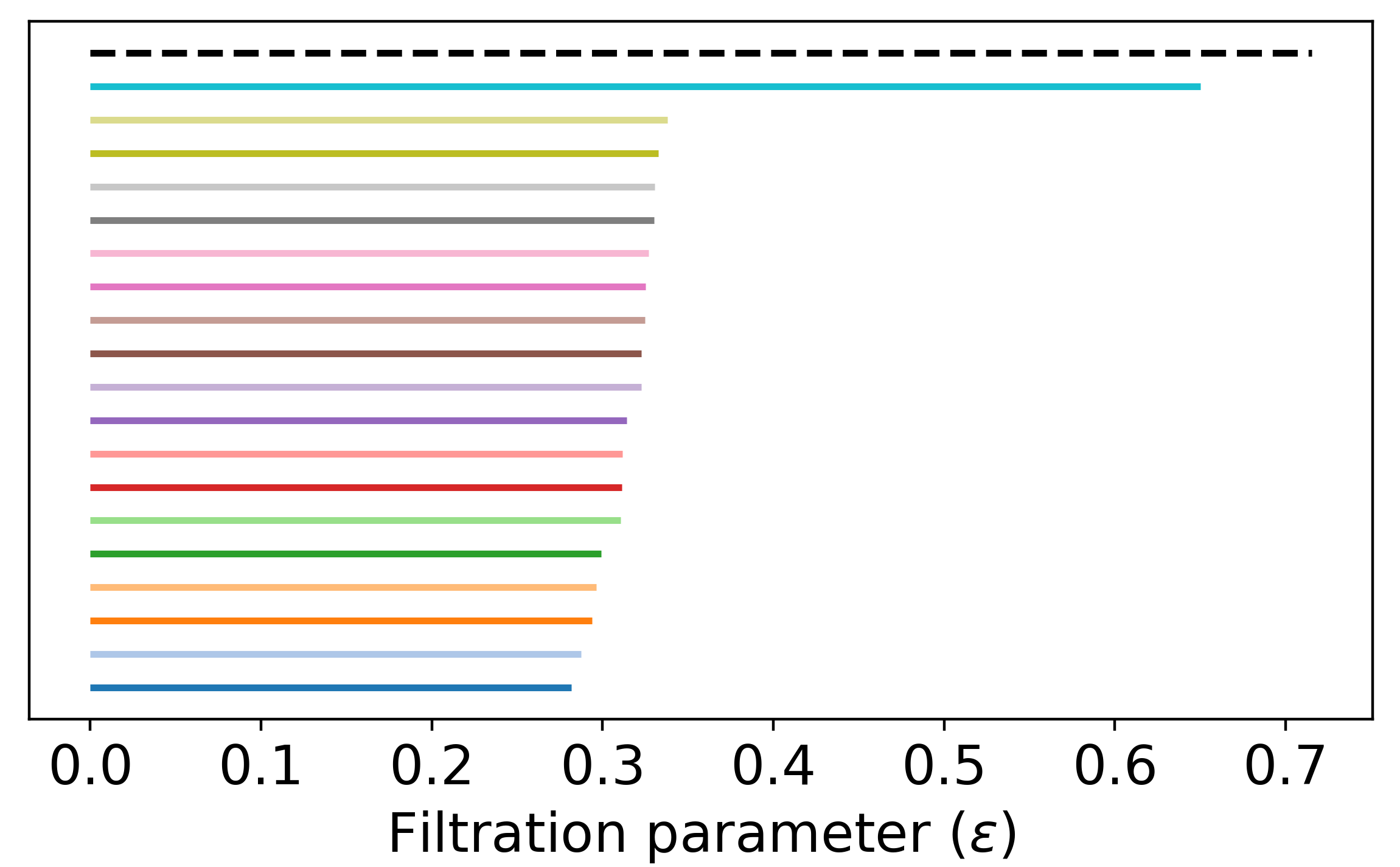}
      \caption{Zero-dimensional persistence barcode.}
  \end{subfigure}
    \begin{subfigure}[t]{0.45\textwidth}
        \centering
        \includegraphics[width=\textwidth,height=5cm]{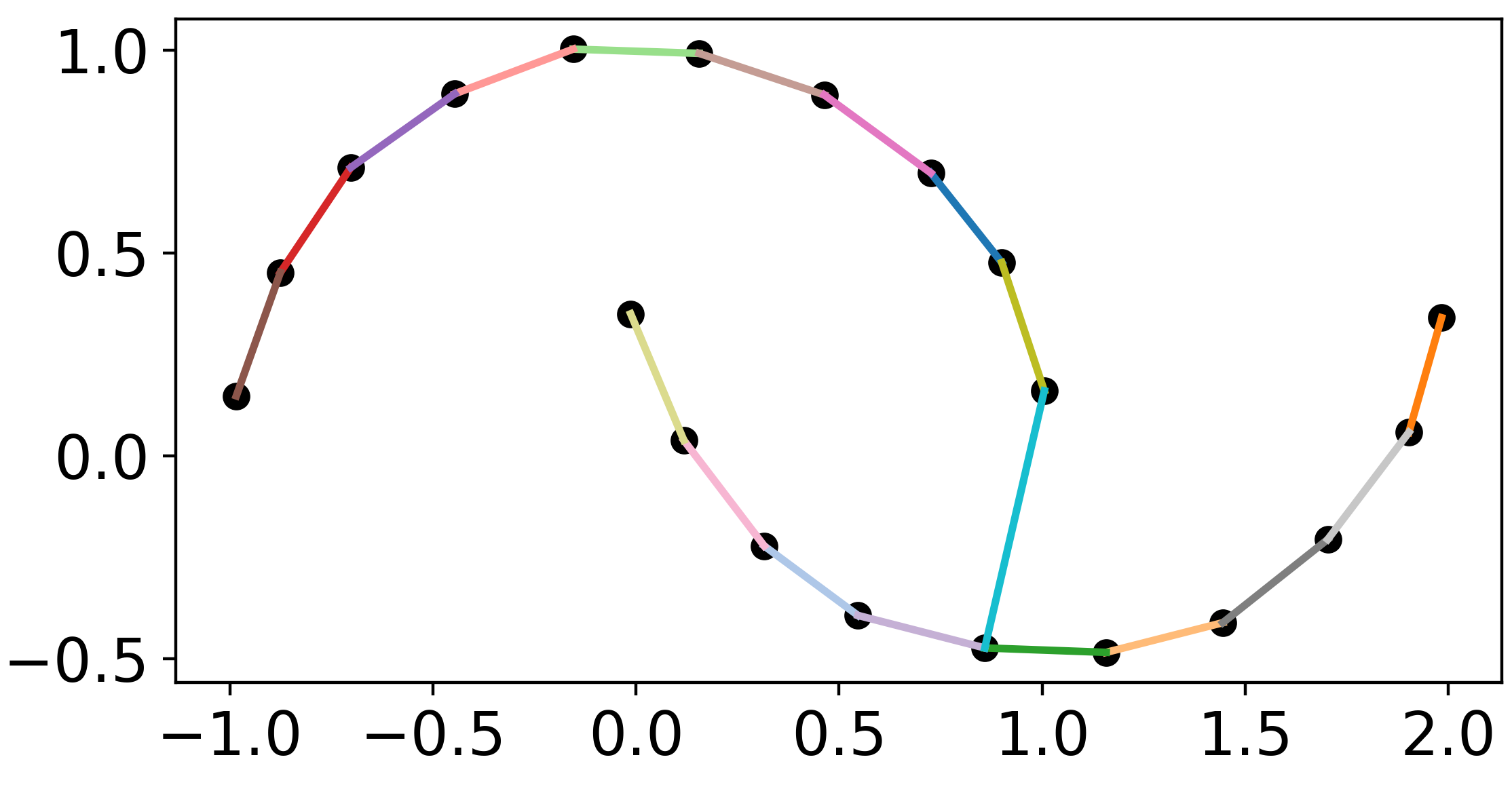}
        \caption{Minimum spanning tree (MST) overlaid on the two moons dataset.}
        \label{fig:two_moons_mst}
    \end{subfigure}
    \hfill
    \caption{Illustration of using the minimum spanning tree to represent the zero-dimensional bars. Each bar (except for the dashed line representing the infinite bar) has a matching edge with the same color on the minimum spanning tree.}
\end{figure}

The minimum spanning tree is uniquely formed (algorithmically determined given $D$) by the following $(n-1)$-step process \citep{kruskal1956shortest}. Denoting a forest by $\mathcal F$, as a graph containing disjoint component trees,  we index a forest by step $t$. We start with an $\mathcal{F}_1$ containing $n$ disjoint trees, each tree having one vertex. From $i=1$ to $i=n-1$, we perform the following steps:
\begin{itemize}
  \item Obtain the admissible edge set $$A_i \;=\; \{\,e=(j,k):\ j\ \text{  and}\ k\ \text{are in different components of } \mathcal{F}_{i}\,\}.$$
  \item Add the shortest edge
  $e_i \;=\; \arg\min_{f\in A_i} \Delta_f,$
  into $\mathcal{F}_{i}$, and change notation to $\mathcal{F}_{i+1}$.
\end{itemize}
The iterative process terminates at $\mathcal{F}_{n}$, which corresponds to a single spanning tree.

It is not hard to see that $A_i$ is a monotonically decreasing set; hence the shortest edge $e_i$ is monotonically increasing. Each time an edge is added, a component tree dies due to being merged into another tree --- corresponding to the death time of a zero-dimensional bar in persistent homology. Therefore, we have
\(
b^0_i = 0, \qquad d^0_i = \Delta^{\hat T}_{(i)}/2,
\)
for $i=1,\ldots,n-1$, where $\Delta^{\hat T}_{(i)}$ is the $i$-th smallest edge length of the minimum spanning tree $\hat T$.
 Lastly, for $i=n$, $b^{0}_{n} = 0$ and $d^{0}_{n} = \infty$ hence do not need modeling.

In order to represent the above partial ordering relationship, we assign  a {\em competing exponentials model} for the death times of zero-dimensional bar:
\[\label{eq:competing_exponentials}
\mathcal{L}[ (b^0_i, d^0_i, A_i,e_i)_{i=1}^{n_0}; \Lambda]
  & =  \prod_{i=1}^{n-1}  \bigg [   ({\sum_{f\in A_i}\lambda_f})\exp\!\Big(-\,d^0_i {\sum_{f\in A_i}\lambda_f}\Big) \frac{\lambda_{e_i}}{\sum_{f\in A_i}\lambda_f}\,\bigg]
  \\ & = \prod_{i=1}^{n-1}{\lambda_{e_i}}\exp\!\Big(-\,d^0_i {\sum_{f\in A_i}\lambda_f}\Big)
\]
where $e_i$ is the $i$-th edge (with length $d^0_i$) added to the spanning tree. We use shorthand notation $\lambda_f \equiv \lambda_{j,k}$, as an edge is a pair-index $f=(j,k)$.

In the first line of \eqref{eq:competing_exponentials}, the exponential density corresponds to length $d^0_i$ via the first-order statistic transformation of $|A_i|$-many conditionally independent exponential random variables; the next term corresponds to the marginal probability of the edge $e_i$ winning over the other competing edges in $A_i$. The detailed derivation can be found in \cite{balog2017lost}.

\begin{remark}\label{rmk:ordering}
Our likelihood targets the relative ordering of edge lengths in the filtration \citep{luce2005individual,plackett1975analysis,hunter2004mm,guiver2009bayesian,caron2012bayesian}. This ordering determines the sequence of graph events that form the barcode. The exponential model is a parametric form with all its order statistics having closed-form density.

Besides convenience, there are two further justifications. First, for data arising from a sub-exponential distribution, the distance is guaranteed to be sub-exponential as well; hence the exponential density under a certain non-trivial $\lambda_f>0$ is guaranteed to be not lighter-tailed compared to the truth. Second, note that each edge $f$ has its own exponential rate $\lambda_{f}$; hence, the form \eqref{eq:competing_exponentials} is very flexible. Restriction does arise when we let different subjects within the same group share a common $\Lambda$ parameter, potentially under-characterizing the dispersion within subjects; hence, a multi-parameter density may be pursued as an extension. Nevertheless, in simulation studies, we show that the competing exponentials model still preserves the between-group discriminability, even if the data are generated from a multi-parameter distribution.
\end{remark}

\noindent\textbf{One-dimensional bars and loop triangulation}

\begin{figure}[H]
  \centering
  \includegraphics[width=0.5\textwidth]{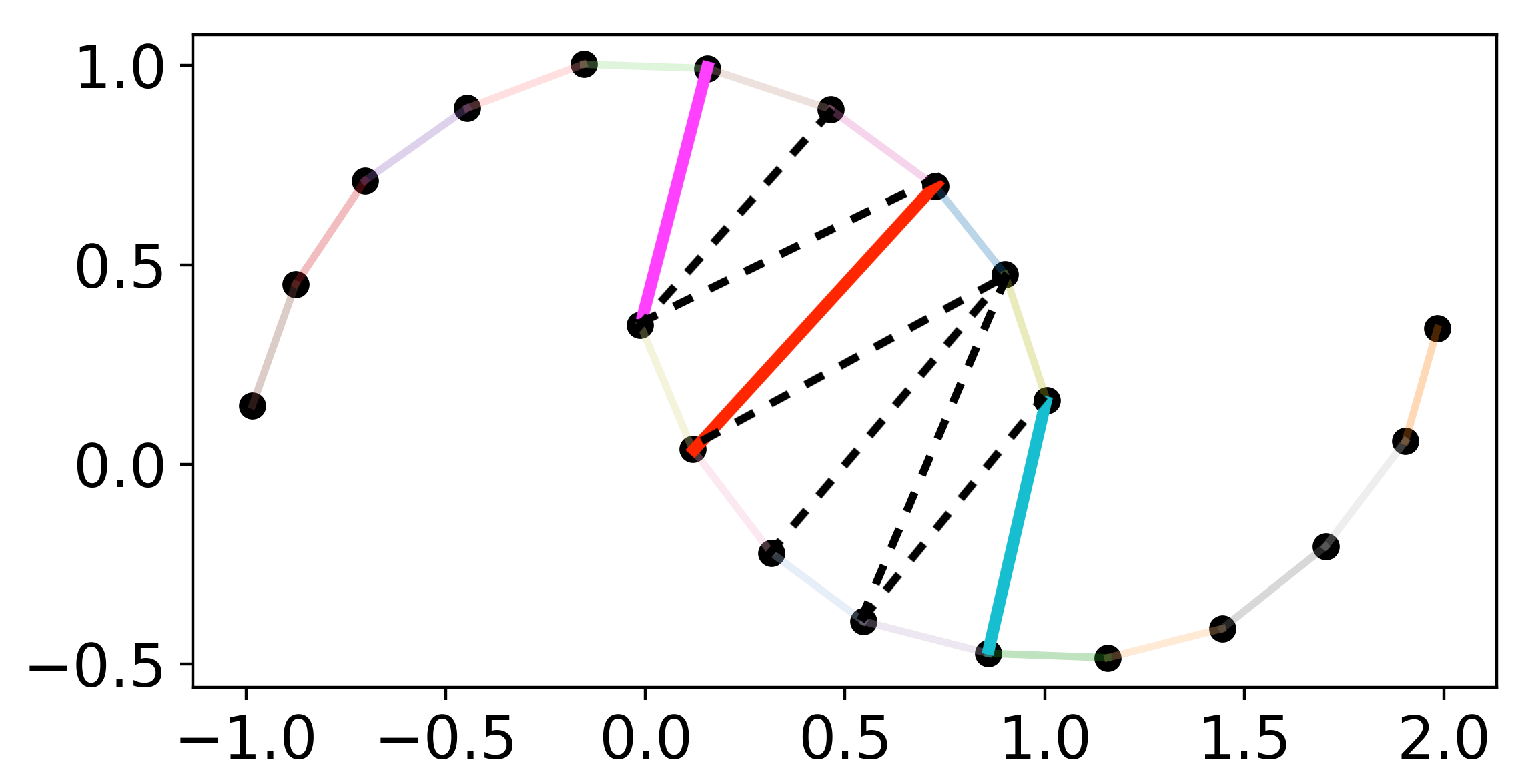}
  \caption{Illustration of the loop triangulation that corresponds to a one-dimensional bar. The magenta edge is associated with the birth of the loop (magenta and the existing edges on the minimum spanning tree, with blue the longest edge on the minimum spanning tree), and the red edge is associated with the loop death. The dashed lines are the triangulation edges with length larger than the birth time and smaller than the death time.}
  \label{fig:two_moons_mst_h1}
\end{figure}

The one-dimensional bars correspond to loops. The birth of a loop is when (i) an edge $e_i=(j,k)$ with length $b^1_i$ forms and is not on the minimum spanning tree $\hat T$, and (ii) there is a path of edges on $\hat T$ starting from $j$ and ending at $k$ formed by time $b^1_i$. The death happens when the loop is triangulated by the time the last edge $f_i=(j',k')$ with edge length $d^1_{i}$. By triangulation of a loop of $m$ vertices ($m>3$, as $m=3$ means immediate triangulation), we mean that at least $(m-3)$ edges are formed, via which the loop can be divided into non-intersecting triangles.
 Figure \ref{fig:two_moons_mst_h1} illustrates the loop triangulation event.

There are a few potential complications associated with the loop triangulation: (i) the loop could be triangulated with additonal vertex/vertices inside the loop, hence the number of triangulating edges could be more than $(m-3)$; (ii) one loop could be split into two or more loops at a certain time, hence it is possible that a new loop is born while an existing loop is still alive, leading to edge-dependency between two one-dimensional bars; (iii) by the time a loop dies, there could be more than one set of edges that can triangulate the loop. Although finding the solution of triangulating edges (and the alternative solutions, if not unique) is computationally possible, as to be described in Section 3.2, we propose a lower-bound likelihood modeling all the triangulation-contributing edges for each loop.

To be specific, we compare the two complexes at the birth and death times of a loop, and use the newly added edges connected to the loop vertices. By connected edges, we mean those edges that are directly incident to the loop vertices, or indirectly connected via other vertices. At birth time $b^1_i$ of a loop, we denote the loop vertices as $L_i$, and record all the edges of the component graph containing $L_i$ as $E(L_i,b^1_i)$; at the death time $d^1_i$, we denote the edges of the component graph containing $L_i$ as $E(L_i,d^1_i)$. We can then see that the edges that contribute to the triangulation event, except the death edge and those already in the minimum spanning tree, are in $\tilde B_i=E(L_i,d^1_i) \setminus E(L_i,b^1_i)\setminus (\{f_i\} \cup E_{\hat T})$.

It is not hard to see that forming $\tilde B_i$ leads to triangulation; therefore, the probability of forming $\tilde B_i$ is smaller than or equal to the probability of triangulation. As an edge may contribute to the death of more than one loop, to avoid double counting, we let $B_i = \tilde B_i \setminus  \bigcup_{j: d^1_j < d^1_i} \tilde B_j$, so it excludes the edges that have already contributed to the death of other loops. Further, note that $B_i$ could include edges that are formed before the loop birth time; we let $B_i= B^{1}_i \cup B^{2}_i$, where $B^{1}_i$ corresponds to the edges that are formed during $(0,b^1_i)$, and $B^{2}_i$ corresponds to the edges that are formed during $(b^1_i, d^1_i)$. We assign the following likelihood:
\[\label{eq:triangulation_likelihood}
  & \mathcal{L}[ (b^1_i, d^1_i, B_i)_{i=1}^{n_1}; \Lambda]  \\ & =    \prod_{i=1}^{n_1}
  \bigg\{
    {\lambda_{e_i}} {\lambda_{f_i}} \exp(-\lambda_{e_i} b_{i}^1-\lambda_{f_i} d_{i}^1)
    \prod_{g \in B^{1}_i} ( 1- e^{-\lambda_{g} b_{i}^1})
    \prod_{h \in B^{2}_i} ( e^{-\lambda_{h} b_{i}^1} - e^{-\lambda_{h} d_{i}^1})
  \bigg\},
\]
where the first term in the factor is the exponential density for the birth and death times, and the remaining term represents the probabilities for the set containing the other triangulating edges.
In the above equation \eqref{eq:triangulation_likelihood}, $(1-e^{-\lambda_g b_i^1})$ is the probability that edge $g\in B_i^1$ appears by the birth time $b_i^1$, whereas $e^{-\lambda_h b_i^1}-e^{-\lambda_h d_i^1}$ is the probability that edge $h$ appears between the birth and death times, $b_i^1$ and $d_i^1$.

\noindent\textbf{Higher-dimensional bars}

The form of the likelihood in \eqref{eq:triangulation_likelihood} extends to bars of any dimension greater or equal to 2, with the birth and death times $(b^1_i,d^1_i)$ replaced by $(b^k_i,d^k_i)$, and with $B^1_i$ and $B^2_i$ the edges arriving before $b^k_i$ and between $b^k_i$ and $d^k_i$. This is simply due to the birth and death of a bar being sufficiently guaranteed by edge arrivals. We provide an illustration of modeling two-dimensional bar in the supplementary materials.

\subsection{Bayesian modeling with conic latent coordinates}
Since $\Lambda$ is the matrix that governs the rate of the exponential distribution, we now parameterize it using the latent position model, where each vertex is associated with a latent coordinate $z_j$ in some latent space.

We consider the latent space as a strict acute cone $\mathcal C\subset \mathbb R^m$ such that
\(
   z_j ^\top z_k \;>\; 0 \quad \text{for all } z_j,z_k\in \mathcal C .
\)
That is, any two vectors in $\mathcal C$ have an angle less than $90^\circ$. For example, any cone generated by vectors lying in a closed spherical cap of radius at most $45^\circ$  around some unit vector $u$,
$\mathrm{cone}\{v:\|v\|=1,\ v^\top u > \cos(45^\circ)\}$, is an acute cone; the interior of the first quadrant is another acute cone.
Although in our model, we do not restrict the orientation of the cone.

We take a Bayesian approach and assign the following prior for $Z=(z_1,\ldots,z_n)$:
\[\label{eq:prior_z}
\pi_0( Z \mid \kappa, \alpha) \propto \kappa^{\{n m+\alpha n(n+1)\}/2} \exp\left(-\frac{\kappa}{2} \sum_{j=1}^n \|z_j\|^2_2\right) \prod_{j\le k}
(z_j^\top z_k)^{\alpha}\, \mathbb{1}( z_j \in \mathcal C \text{ for all } j)
\]
with $\kappa>0$, $\alpha>0$; the power term of $\kappa$ is based on integration over a cone. The indicator confines the support to the cone: outside $\mathcal C$ some inner product is negative, where $(z_j^\top z_k)^{\alpha}$ is not defined for non-integer $\alpha$, so the prior is supported on $\mathcal{C}^n$ and the indicator is what rules the exterior out. The exponential kernel regularizes the scale of the latent coordinates, and the product of inner products makes the boundary of the cone repulsive, rather than merely excluded. For the latter, its negative logarithmic transform yields $- \alpha \log(z_j^\top z_k)$, which is commonly referred to as the log-barrier function and acts as a strong penalty if the inner product gets too close to $0$. We choose a small $\alpha$ to ensure its effect is not pronounced when the inner product is far away from 0. Since the density vanishes as any inner product tends to $0$, the boundary is unattainable, and the indicator is inactive except on a set the prior already gives zero mass; this is why we do not carry it in the algebra that follows.
The form of \eqref{eq:prior_z} resembles non-local priors \citep{fuquene2019choosing}, in that the density vanishes on a degenerate set with $z_j^\top z_k=0$. In addition, we include $j=k$ for the benefit of identifiability as described soon.

Since both the likelihood and the prior of $Z$ are invariant to the orthonormal transformation on the rows of $Z$ ($\tilde Z = Z P$ for some $P^\top P = I_m$), $Z$ is not directly identifiable in the model. Nevertheless, one may wonder if $Z$ can be identified as an equivariant member in the set $\{Z \in \mathbb{R}^{n\times m}: ZZ^\top = \hat\Lambda, m\ge m^0\}$ for some optimal value of $\hat\Lambda\succeq 0$ with $\text{rank}(\hat\Lambda)=m^0$? We answer affirmatively --- in fact, we show that one equivariant posterior mode of $Z$ can be easily found as a second-order stationary point.

\begin{theorem}\label{thm:identifiability}
  Under the likelihood \eqref{eq:competing_exponentials}, \eqref{eq:triangulation_likelihood}, and prior \eqref{eq:prior_z}, consider the conditional posterior of $Z$ given $\kappa$ and $\alpha$, reparametrized by $\Lambda=ZZ^\top$, $q_{\kappa, \alpha}(\Lambda)= \Pi(Z\mid \kappa, \alpha, y)$, then the following properties hold:
  \begin{enumerate}
    \item $q_{\kappa, \alpha}(\Lambda)$ is a strictly log-concave function in $\Lambda\succeq 0$ (ignoring the rank constraint), for any finite $\alpha>0$;
    \item The unique maximizer $\hat\Lambda= \arg\max q_{\kappa, \alpha}(\Lambda)$ is of rank $m^0\le n$ under suitable value of $\kappa$.
    \item Let $r_{\kappa,\alpha}(Z)= \log\Pi(Z\mid \kappa, \alpha, y)$ with $Z\in\mathbb{R}^{n\times m}$, and each $z_i \in \mathcal C$. Suppose given $(\kappa,\alpha)$, the rank of $\hat\Lambda$ is $m^0$, then as long as $Z$ has column dimension $m \ge m^0+1$, any $Z^*$ that satisfies the second-order stationarity condition:
    \(
    \nabla r_{\kappa,\alpha}(Z) = 0, \quad \nabla^2 r_{\kappa,\alpha}(Z) \succeq 0
    \)
    has $(Z^*)  (Z^*) ^\top = \hat\Lambda$ and $\text{rank}(Z^*)=m^0$.
  \end{enumerate}
  \end{theorem}

  \begin{remark}We would like to clarify that our focused $q_{\kappa, \alpha}(\Lambda)$ is still the density of $Z$, except viewed as a function of $\Lambda$. The transformed density of $\Lambda$ would be $q_{\kappa, \alpha}(\Lambda) \big(\det^{\!*}\Lambda\big)^{\{\text{rank}(Z)-n-1\}/{2}}$ (with $\det^{\!*}$ pseudo-determinant), which is rank-dependent, hence not log-concave.
  \end{remark}

The above equivariant identifiability for $Z$ is not only an encouraging message, but also a principled guide for choosing the dimension of $Z$ in practice: we want to choose $m$ to be sufficiently large, so that it can be {\em strictly} larger than the rank of the maximizers of $\hat\Lambda$ under a reasonable range in the posterior. The theoretical justification for strict inequality is provided in the proof. This is in line with the common practice in Bayesian modeling of using an overfitted model \citep{bhattacharya2011sparse,van2015overfitting,legramanti2020bayesian}, where one specifies a dimension larger than the expected one and relies on the Bayesian model to shrink the extra dimensions.

In our model, the shrinkage on the rank of the posterior mode is due to $\sum_{j=1}^n \|z_j\|^2_2= \|\Lambda\|_*$, the nuclear norm commonly used in optimization to induce a low-rank optimal solution. Since we use a continuous prior for $Z$, the posterior $\Lambda$ has rank $m$ almost everywhere, but is concentrated near $\hat \Lambda$ with rank $m^0$. We use prior $\kappa\sim \text{Ga}(2,3)$ so it has prior mean $6$, while the prior support $\kappa$ vanishes at $0$. We set $m=5$ for the column dimension of $Z$, and $\alpha=0.1$ in the positive barrier term. Empirically, these values lead to good mixing performance for the Markov chains.

\subsection{Hierarchical model for multi-group data analysis}
\label{sec:multigrp}
We next extend our model to accommodate multiple subjects drawn from several groups, indexed by $p=1,\ldots,P$. For the $S_p$ subjects belonging to the $p$-th group, each subject $s\in \mathcal I_p$ is associated with its own barcode and edge set, denoted $y^s$. Assuming that the data $\{y^s: s\in \mathcal I_p\}$ are conditionally independent, the likelihood can be expressed as
\(
\mathcal{L}( \{y^s: s\in \mathcal I_p\}; \Lambda^p) = \prod_{s \in \mathcal I_p} \mathcal{L}[ (b^{[s]0}_t, d^{[s]0}_t, A^{[s]0}_t)_{t=1}^{n^{[s]}_0}; \Lambda^p] \;\; \mathcal{L}[ (b^{[s]1}_t, d^{[s]1}_t, B^{[s]1}_t)_{t=1}^{n^{[s]}_1}; \Lambda^p].
\)
For different groups, we assign each an individual $\Lambda^{p}$.

For statistical modeling, it is very useful to borrow strength across groups, so that we can shrink the   $\Lambda^{p}$ toward a common $\bar\Lambda$, while having the significant differences that remain show the source of the across-group variability.
For this goal, we develop a Bayesian hierarchical prior for modeling a collection $\{\Lambda^{1},\ldots,\Lambda^{P}\}$ based on the Procrustes-type distance that minimizes the negative impact of free orthonormal transformation on the difference between $\Lambda^{p}$ and $\bar\Lambda$.

Consider a latent matrix $\bar Z\in \mathbb{R}^{n\times m}\cap \mathcal C^n$ with $m$ sufficiently large, we define the distance squared between $Z^p: Z^p Z^{p\top}= \Lambda^p$ and $\bar Z:\bar Z \bar Z^\top= \bar\Lambda$ as
\(
\text{dist}^2(Z^p, \bar Z) = \min_{R^\top R = I, s\in \mathbb{R}}\| Z^p - s \bar Z R\|_F^2
= \|Z^p\|_F^2 - \|Z^{p\top} \bar Z\|^2_*/\|\bar Z\|_F^2
\)
 where $R$ represents any orthonormal transform, and $s$ does any scaling. The second equality is due to the fact that the closed-form solution for the minimizer is $\hat R= UV^{\top}$, with $UDV^\top=\bar Z^{\top} Z^p $ the singular value decomposition, and $\hat s={\operatorname{tr}(Z^{p\top} \bar Z \hat R)}/{\|\bar Z\|_F^2} = \|Z^{p\top} \bar Z\|_*/\|\bar Z\|_F^2$, with $\| \cdot \|_*$ the nuclear norm.

 Since $\|Z^{p\top} \bar Z\|_* \;=\; \operatorname{tr}\!\big((\bar Z^\top \Lambda^p \bar Z)^{1/2}\big)$, and $\|Z^p\|_F^2 = \text{tr}(\Lambda^p)$, we have the following hierarchical prior that extends \eqref{eq:prior_z}:
\(\label{eq:hierarchical_prior}
\Pi_0(Z^p \mid \bar Z) & \propto (\kappa)^{\{n m+\alpha n(n+1)\}/2} \exp\left(-\frac{\kappa}{2} \bigg [\|\bar Z\|_F^2 \text{tr}(\Lambda^p)
- [\operatorname{tr}\!\big((\bar Z^\top \Lambda^p \bar Z)^{1/2}\big)]^2 \bigg ]
\right) \prod_{j\le k}
(z^{p\top}_j z^{p}_k)^{\alpha}\\
& \text{ for } p =1,\ldots,P,\\
 \Pi_0(\bar Z) & \propto
 \exp( - \frac{1}{2} \kappa_0  \|\bar Z\|_F^2)
 \prod_{j\le k} (\bar z_j^{\top} \bar z_k)^{\alpha},
\)
in which $\Pi_0(Z^p \mid \bar Z)$ is again a strictly log-concave function in $ \Lambda_p\succeq 0$. Since the Procrustes-type distance is invariant to the scaling of $\bar Z$, we fix $\kappa_0=6$ in the prior for $\bar Z$.

\section{Posterior Computation}

We now discuss the posterior computation for our proposed model. One may notice that in our likelihood, the edge sets are not directly known from the barcode, hence need to be calculated from the data. Fortunately, these sets only need to be pre-computed once for each subject $s$, and are invariant to the value of the parameters; given the edge sets, the posterior computation can be carried out efficiently. We now first discuss the posterior sampling algorithm, then discuss the algorithm for finding the edges.

\subsection{Markov chain Monte Carlo with warm-start}
 We use the No-U-Turn Sampler (NUTS, \cite{hoffman2014no}) to estimate the posterior distribution. For the implementation, we use the numpyro package \citep{bingham2019pyro} for its low implementation cost of the proposed model using the automatic differentiation capabilities. Due to the popularity of No-U-Turn Sampler algorithms in the Bayesian literature and the high level of automation provided by software packages, we refer the reader to \cite{hoffman2014no} for the details. Here we provide a brief summary of the algorithm, and use the remaining space to discuss a few key aspects specific to our approach.

The No-U-Turn Sampler algorithm is a variant of the Hamiltonian Monte Carlo algorithm, which uses a discretized approximation of Hamiltonian dynamics to propose new samples. The Hamiltonian dynamics describe the evolution of a system in terms of position ${\tilde q}$ (equivalent to our parameter $\theta$) and momentum ${\tilde p}$ (an equal-dimensional vector of auxiliary variables) by
\(
\frac{d{\tilde q}}{dt} = \nabla_{\tilde p} H, \qquad
\frac{d{\tilde p}}{dt} = -\nabla_{\tilde q} H,
\)
where $H({\tilde q}, {\tilde p})$ is the Hamiltonian energy function of the system, as the sum of the negative log-posterior (except for the constant term) and augmented kinetic energy for ${\tilde q}$, typically of form ${\tilde q}^\top \tilde M {\tilde q}$ with $\tilde M$ a positive definite matrix. The No-U-Turn Sampler algorithm extends Hamiltonian Monte Carlo by adaptively determining the trajectory length, automatically stopping the simulation when the path begins to double back, which avoids the need to manually tune path lengths and improves exploration of the posterior distribution.

There are two points specific to our proposed model.  First, on handling the acute cone constraint on the matrix $Z$, thanks to the $F(Z)=- \alpha \sum_{i\le j} \log(z_i^\top z_j)$ term in the negative log-prior, we see that the gradient of the log-prior with respect to $Z$
\(
\nabla_Z F(Z)\;=\;-\alpha\,C\,Z,\qquad
C_{jj}=\frac{2}{\lambda_{jj}},\quad C_{jk}=\frac{1}{\lambda_{jk}}\ (j\neq k).
\)
Therefore, the momentum change direction in $-\nabla_{\tilde q} H$ contains
\(
  -\{\nabla_{Z} F(Z)\}_j
  = \alpha\Big(\frac{2}{\lambda_{jj}}\,z_j+\sum_{k\neq j}\frac{1}{\lambda_{jk}}\,z_k\Big).
\)
If $\lambda_{jj}=\left\|z_j\right\|^2 \downarrow 0$, the term $\left(2 / \lambda_{jj}\right) z_j$ will explode along $z_j$, making $z_j^\top z_j$ larger and away from zero. Similarly, if an off-diagonal $\lambda_{jk}=z_j^{\top} z_k \downarrow 0$, the term $\left(1 / \lambda_{jk}\right) z_k$ will explode along $z_k$ (and symmetrically for $z_k$ along $\left.z_j\right)$, aligning the two rows hence making $z_j^\top z_k$ more positive and further away from zero. Although the discretization error of the Hamiltonian dynamics may create $z_j^\top z_k<0$, the Metropolis-Hastings step rejects the proposal. Therefore, provided the initial value of $Z$ is in the acute cone, the algorithm can efficiently maintain the constraint in the subsequent iterations.

Second, we can exploit the conditional log-concavity of the posterior distribution to warm-start the No-U-Turn Sampler. The warm-start strategy has two major benefits: first, it reduces the time for burn-in, and; second, it improves the stability of the algorithm as the automatic tuning of the algorithm (such as choosing the step size) is carried out in the high posterior density region. Specifically, fixing $\kappa=6$ (prior mean), we first run gradient ascent (with $\Lambda= Z Z^\top$ parameterized) to find a conditional mode $\hat Z$ of the posterior distribution, then use  $\hat Z$ and $\kappa=6$ as the initial value for the Markov chain. We find the strategy empirically effective in inducing fast convergence and good mixing performance.

\subsection{Finding simplices associated with birth and death events}\label{sec:edges}
We now describe useful algorithms for finding the simplices (such as edges and triangles) important for the birth and death of bars. To be general, we introduce the boundary matrix-based algorithm that is nearly universally applicable to all persistent homology features, including two- or higher-dimensional ones.

For ease of understanding, we focus on the simplex where the edge orientation does not matter. At time $t$, the boundary matrix $\partial^t_k$ has the rows corresponding to all the $(k-1)$-simplices,  and the columns corresponding to all the $k$-simplices:
\(
(\partial^t_k)_{\sigma^{k-1},\,\tau^{k}}
=
\begin{cases}
1 & \text{if }\sigma^{k-1}\text{ is a face of }\tau^{k},\\
0 & \text{otherwise.}
\end{cases}
\)
in which, face means a nonempty vertex set. For example, vertex $\sigma^0$ is a face of edge $\tau^1$ if the edge is incident to the vertex, and $\partial_1$ is the vertex-edge incidence matrix of the graph at time $t$; edge $\tau^1$ is a face of triangle $\tau^2$ if the triangle is incident to the edge, and $\partial_2$ is the edge-triangle incidence matrix at time $t$.
The boundary matrix is useful as it allows us to identify the simplices using linear algebra modulo 2.

To illustrate, we now focus on the birth and death events of a loop. First, to identify the edges (among $|E_t|$ many edges) that form a new loop at a birth time $t$ due to the newly added edge $\tau_t$, we use the matrix $\partial^t_1$, and the edges correspond to the ones in the binary vector $x\in\{0,1\}^{|E_t|}$:
\(
  \partial^t_1 x \equiv 0 \pmod 2 \quad \text{and} \quad x_{\tau_t}=1,
\)
where $\equiv 0 \pmod 2$ means equal $0$ after the modulo 2 operation. Solving the system can be done using Gaussian elimination modulo 2, which replaces subtraction with the XOR operation.

Second, to identify the triangles that fill the loop (formed earlier at a birth time $t'$, with the edges denoted by $x^{t'}$) due the newly added triangle $\tau_t$ at a death time $t$, we can use matrix $\partial^t_2$ and we know the triangles corresponding to ones in the binary vector $y$:
\(
  \partial^t_2 y \equiv x^{t'} \pmod 2 \quad \text{and} \quad y_{\tau_t}=1.
\)
To clarify, the solution may not be unique, and some solutions may use more triangles than a minimal filling, which we refer to as triangulation in a previous section and can be found by simply taking the solution(s) with the least number of triangles.

The above two computational strategies are easy to implement, and importantly, directly extensible for higher-dimensional simplices. For example, at the birth of a void (two-dimensional feature) at time $t$, the list of triangles can be found via the solution of $
  \partial^t_2 x \equiv 0 \pmod 2 \quad \text{and} \quad x_{\tau_t}=1
$. For the death of a void (filled by tetrahedra) when $\tau_t$ is added, the list of tetrahedra can be found via the solution of $  \partial^t_3 y \equiv x^{t'} \pmod 2 \quad \text{and} \quad y_{\tau_t}=1.$

With the great generality of the linear algebra modulo 2 solution, the Gaussian elimination associated with a $k_1\times k_2$ matrix has a cost of $\mathcal O(\min(k_1 k^2_2, k_2 k^2_1))$. Therefore, application of the above algebraic solutions may have an expensive cubic cost. On the other hand, the cost can be tolerable, since the calculation is only needed once for each subject, before running Markov chain Monte Carlo. In the supplementary materials, we also discuss some alternative solutions that are specific to our proposed model, hence less general but with much lower cost.

 \section{Asymptotic theory for persistent homology posterior}\label{sec:theory}
In this section, we study the asymptotic behavior of the posterior distribution under the persistent homology model. We establish the asymptotic result under two related but different settings. First, we focus on the ideal case where the model is correctly specified, and potentially, a diverging number of zero-dimensional features are collected. This setting allows us to obtain not only the consistency of the model parameters, but also the convergence rate of the posterior distribution. Second, we then generalize and consider the case where the persistent homology may contain higher-dimensional features, but the model could be misspecified to some extent. We establish a set of sufficient conditions for the consistent recovery of feature membership, an important characteristic of the topology, and discuss possible extensions on how to ensure the robust consistency result holds.
\subsection{Consistency and convergence rate}
In this section, we establish posterior consistency and a convergence rate
for the strict lower-triangular entries of $\Lambda$. We allow $n$ to grow
with the number of independent subjects $S$. For a filtration that eventually
connects all $n$ vertices, there are $n-1$ finite zero-dimensional features.
For one subject, define $E=\{(j,k):1\le k<j\le n\}$,
$N_f=\sum_{i=1}^{n-1}\mathbf{1}\{e_i=f\}$, and
$T_f=\sum_{i:f\in A_i^0}d_i^0$.
The joint likelihood in \eqref{eq:competing_exponentials} can be written as
\[
 \prod_{s=1}^{S}\left[
   \left(\prod_{i=1}^{n-1}\lambda_{e_i^{[s]}}\right)
   \exp\!\left\{-\sum_{f\in E}\lambda_f
                  \sum_{i:f\in A_i^{[s]0}}d_i^{[s]0}\right\}
 \right].
\]
We use the prior \eqref{eq:prior_z} on the unbounded set
$\mathcal C_n=\{Z\in\mathbb R^{n\times m}:z_j^\top z_k>0,
1\le j\le k\le n\}$, with $1\le m\le n$, fixed $\kappa>0$, and
$\alpha>0$, which may depend on $n,S$. We assume the following conditions
on the common true law $P_0$:
\begin{itemize}
 \item[(C1)] There exists $Z_0\in\mathcal C_n$, with
 $\Lambda_0=Z_0Z_0^\top$, such that
 $\mu_f:=\mathbb E_0N_f=\lambda_{0,f}\mathbb E_0T_f>0$
 for every $f\in E$.
 \item[(C2)] $T_f$ has finite variance under $P_0$ for every $f\in E$.
\end{itemize}
Condition (C1) calibrates the likelihood to the truth. Condition (C2)
requires only finite exposure variances, whose magnitude enters the rate
below; no uniform variance bound or exponential moment is assumed.

\begin{lemma}[Large support for the prior of $Z$]
For the prior \eqref{eq:prior_z}, every $Z_*\in\mathcal C_n$ and every
$\epsilon>0$ satisfy $\Pi(\|Z-Z_*\|_F<\epsilon)>0$.
Moreover, $\kappa\|Z\|_F^2\sim\chi^2_\nu$, where
$\nu=nm+\alpha n(n+1)$. Consequently, for $0<\varepsilon<1$,
$\Pi(\|Z\|_F>R_\varepsilon)\le\varepsilon$, where
$R_\varepsilon^2=\kappa^{-1}\{\nu+2\sqrt{\nu\log(1/\varepsilon)}
+2\log(1/\varepsilon)\}$.
\label{lem:supportZ}
\end{lemma}
Although we assign a prior to $\kappa$, we treat it as fixed to simplify the calculations.

Let $\textup{LTri}(B)=(B_{jk}:1\le k<j\le n)$, and denote the posterior
based on $Y_S=(Y_1,\ldots,Y_S)$ by $\Pi_S(\cdot\mid Y_S)$.
Write $p=nm$, $q=n(n-1)/2$ and $b=\alpha n(n+1)/2$, so that
$\nu=p+2b$. For the truth, define $M_0=\max_j\|z_{0,j}\|_2$,
$\tau=\min_{j\le k}z_{0,j}^\top z_{0,k}$,
$\mu_{\min}=\min_{f\in E}\mu_f$, and
$\bar\lambda_0=\max_{f\in E}\lambda_{0,f}$.
Set $v_n=\sum_{f\in E}\operatorname{Var}_0(\lambda_{0,f}T_f)/\mu_f$.
Let
\[
 \begin{split}
 D_{n,S}={}&(p+q+v_n)\log(e+nS)+p\log(e+M_0)
       +(p+b)\log(e+\tau^{-1})\\
       &+\kappa\|Z_0\|_F^2
       +b\log\!\left(e+\frac{\nu}{\kappa n}\right).
 \end{split}
\]
Set $\gamma_{n,S}=D_{n,S}/(S\mu_{\min})$ and
$r_{n,S}^2=\bar\lambda_0^2\gamma_{n,S}$.

\begin{theorem}[Posterior consistency for $\Lambda$]
Under (C1)--(C2), if $\gamma_{n,S}\to0$, then, for a sufficiently large
constant $M$ independent of $n,S$,
\[
 \mathbb E_0\Pi_S\!\left(
 \|\textup{LTri}(\Lambda-\Lambda_0)\|_2^2>Mr_{n,S}^2
 \,\middle|\,Y_S\right)\longrightarrow0.
\]
If also $r_{n,S}^2\to0$, the posterior is consistent for the strict lower
triangle of $\Lambda_0$.
\label{thm:consistency_wellspecfied}
\end{theorem}

For fixed $n,m,\alpha$ and truth, $r_{n,S}^2=O(\log S/S)$.
For growing $n$, if $\mu_{\min}\gtrsim n^{-1}$ and
$D_{n,S}\lesssim n^2\log(e+nS)$, then
$r_{n,S}^2\lesssim\bar\lambda_0^2n^3\log(e+nS)/S$.
Thus, in this case, $(1+\bar\lambda_0^2)n^3\log(e+nS)=o(S)$ suffices
for consistency. Proofs are provided in the Supplementary Material.
\subsection{Consistent recovery of the topological feature-membership under some model misspecification}
Next, we consider a general setting where the features can have dimension greater or equal to one, and the true generating law of the homology may fall outside our specified class of models. Due to the possibility of model misspecification, we do not hope to recover the true generating law of the homology exactly, but we still want to correctly quantify some aspects of the topology. In particular, we focus on feature membership, defined as whether a vertex belongs to a certain topological feature, such as in the formation of a loop.

To formalize the quantity of interest under the ground truth, for each subject $s$, let $D^{(s)}=\{\Delta^{(s)}_{j,k}\}_{j<k}$ be the subject’s distance matrix, and we assume the data-generating distribution $D^{(s)}\stackrel{iid}{\sim}F^0$, where $F^0$ is some unknown distribution. We define the subject-level indicator vector
$$\iota^{(s)} \;=\; g(D^{(s)})\in\{0,1\}^{|V|}$$
where $g$ returns which vertices form a certain topological feature. Across $S$ subjects, we know that $\hat p_S(v)=\tfrac1S\sum_{s=1}^S \iota^{(s)}_v \to p_{F^0}(v)=\mathbb{P}_{F^0}(\iota_v=1)$ as $S\to\infty$ by the law of large numbers.

On the other hand, under our specified model parameterized by $\Lambda$. Let $Q_{\Lambda}$ be the model-implied law of $D$ from which we can simulate new distances. We have a model-induced distribution $\mathbb{P}_{Q_{\Lambda}}(\iota_v=1)$ for the feature-membership. A natural question is when can we have $\mathbb{P}_{Q_{\Lambda}}(\iota_v=1)\to  p_{F^0}(v)$ as the posterior distribution of $\Lambda$ concentrates? The following theorem provides a set of sufficient conditions.

\begin{theorem}\label{thm:consistency_misspecfied}
Let $E$ be a finite index set of distances and let $\mathcal C$ be a finite collection of rank relations on $E$.
Let $\Omega_{\mathcal C}$ denote the finite set of partial orderings induced by $\mathcal C$, and let
$R:\mathrm{Perm}(E)\to\Omega_{\mathcal C}$ be the restriction map. There exists a subset $\mathcal A_v\subseteq \Omega_{\mathcal C}$ such that $\iota_v=\mathbf 1\{R\in \mathcal A_v\}$ almost surely, hence under $F^0$, $\iota_v$ is a random event determined by $R\in\Omega_{\mathcal C}$. If the following conditions hold:
\begin{enumerate}
\item[(C3)]
For any $D$, there exists a neighborhood $C_D=\{D': g(D')=g(D)\}\ni D$ with $\mathrm{radius}(C_D)\ge \delta_0>0$,
and under the true law $F^0$ ties occur with probability $0$.
\item[(C4)]
The feature $\iota_v=g(D)$ is determined solely by the relative ordering among $\{\Delta_{i,j}\}_{i<j}$.
\item[(C5)]
There exist weights $\eta\in(0,\infty)^{|E|}$ such that the full ranking of $\{\Delta_e\}_{e\in E}$ under $F^0$
follows the Plackett--Luce law $\mathrm{PL}(\eta)$; equivalently, for any permutation $\pi$ of $E$,
\(
\mathbb{P}_{F^0}\!\big(\Delta_{\pi(1)}<\cdots<\Delta_{\pi(|E|)}\big)
= \prod_{k=1}^{|E|}\frac{\eta_{\pi(k)}}{\sum_{\ell\ge k}\eta_{\pi(\ell)}}.
\)
\item[(C6)]
There is a continuous map $\Lambda\mapsto \theta(\Lambda)\in(0,\infty)^E$ such that the model-implied law of $R$ is
\(
p^{\mathcal C}_{\bar\theta(\Lambda)} \;=\; R\!\left[\mathrm{PL}\!\left(\bar\theta(\Lambda)\right)\right]
\quad \text{on } \Omega_{\mathcal C},\qquad
\bar\theta(\Lambda)=\theta(\Lambda)\Big/\sum_{e\in E}\theta_e(\Lambda).
\)
The restricted family $\bar\theta\mapsto p^{\mathcal C}_{\bar\theta}$ is identifiable on the simplex
$\Delta^{|E|-1}$.
\item [(C7)] The prior for $\Lambda$ on the symmetric matrix space $\mathbb{S}^n$ is absolutely continuous w.r.t.\ Lebesgue measure and has
a strictly positive density on every nonempty open set.
\item [(C8)] The following Kullback-Leibler divergence is minimized at some $\Lambda:\theta(\Lambda)=c\eta$ for some constant $c>0$:
\(
\mathrm{KL}\!\big(F^0 \,\|\, Q_\Lambda\big)
=\mathrm{KL}\!\big(F^0_R \,\|\, Q_R(\cdot;\bar\theta(\Lambda))\big)
+\; \mathbb{E}_{F^0_R}\!\Big[\mathrm{KL}\!\big(F^0_{M|R}\,\|\,Q_{M|R}(\cdot\mid R;\Lambda)\big)\Big],
\)
where $M$ denotes the lengths of bars in the barcode.
\end{enumerate}
Let $\widehat\Lambda_S$ be a draw from the posterior based on $S$ i.i.d.\ observations of $R$.
Then, for every $v$,
\(
\mathbb{P}_{Q_{\widehat\Lambda_S}}(\iota_v=1)\;\xrightarrow[S\to\infty]{P}\;
p_{F^0}(v):=\mathbb{P}_{F^0}(\iota_v=1).
\)
\end{theorem}

Our consistency of partial information recovery is inspired by the results established for the rank likelihood framework \citep{hoff2014information}.
	Most of the above conditions are considered standard in the literature, except for (C3) and (C6). Condition (C3) constrains the true law of the partial orderings to a family. Despite being a restriction, the Plackett--Luce family is routinely used in the literature \citep{guiver2009bayesian,caron2012bayesian}. Condition (C6) assumes that the misspecified model on the barcode lengths does not impact the recovery of the ranking law. To make (C6) more generally applicable, one could extend our model by first treating $b_i^k$ and $d^k_i$ as noisy realizations near latent $\beta_i^k$ and $\delta_i^k$, and using the competing exponential-based likelihood of $\beta_i^k$ and $\delta_i^k$ using the observed partial ordering.

\section{Simulation studies}\label{sec:sim}
We conduct simulation studies to evaluate the performance of our method. We first generate three group oracles (sets of points in $\mathbb{R}^2$) that are related to each other: in Group 1, we draw $\mu^{(1)}_i\sim \text{N}(0,0.25^2 I)$ for $i=1,\ldots,150$; in Group 2, we have $\mu^{(2)}_i=\mu^{(1)}_i$ for $i=1,\ldots,75$, while $\mu^{(2)}_i \sim \text{N}(\Delta 1, 0.25^2 I)$ for $i=76,\ldots,150$, with $\Delta>0$ a scalar; in Group 3, we have $\mu^{(3)}_i = \mu^{(1)}_i $ for $i=1,\ldots,45$, $\mu^{(3)}_i \sim \text{N}(\Delta 1, 0.25^2 I)$ for $i=46,\ldots,75$, and $\mu^{(3)}_i = \mu^{(2)}_i $ for $i=76,\ldots,150$. Effectively, at the oracle, Group 1 and Group 2 are 50\% different, Group 2 and Group 3 are 40\% different, and Group 1 and Group 3 are 70\% different. Then, in each group, we add independent Gaussian noise from $\text{N}(0, 0.07^2 I)$ to each point in the oracle, and obtain the data points $\tilde y^{s}_i$.
We simulate 10 subjects in each group and conduct repeated experiments under three different values of $\Delta\in \{0.5, 1, 2\}$.

\begin{figure}[H]
  \centering
  \begin{subfigure}[b]{0.32\textwidth}
      \centering
      \includegraphics[width=\textwidth]{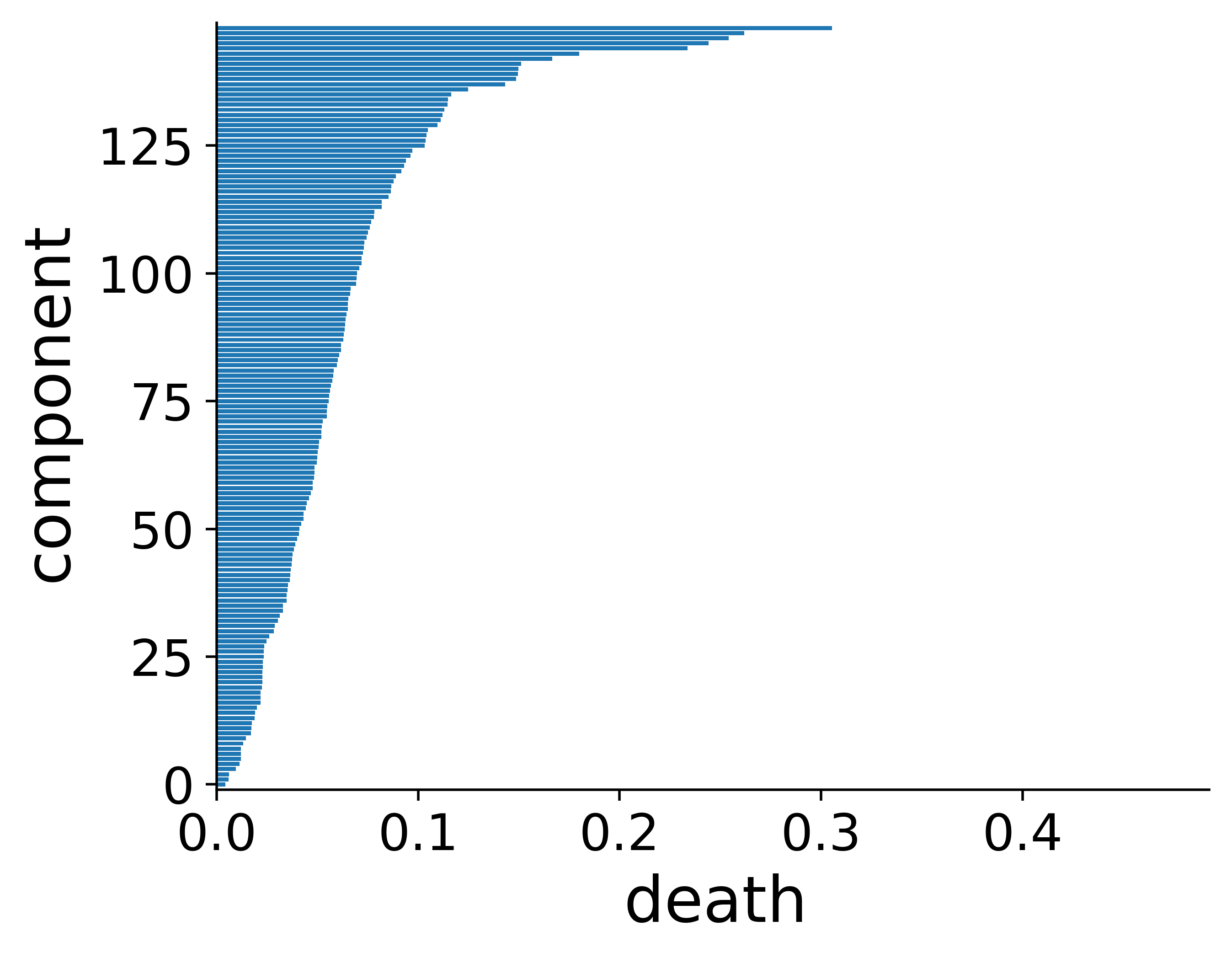}
      \caption{Group 1}
      \label{fig:barcode1}
  \end{subfigure}
  \hfill
  \begin{subfigure}[b]{0.32\textwidth}
      \centering
      \includegraphics[width=\textwidth]{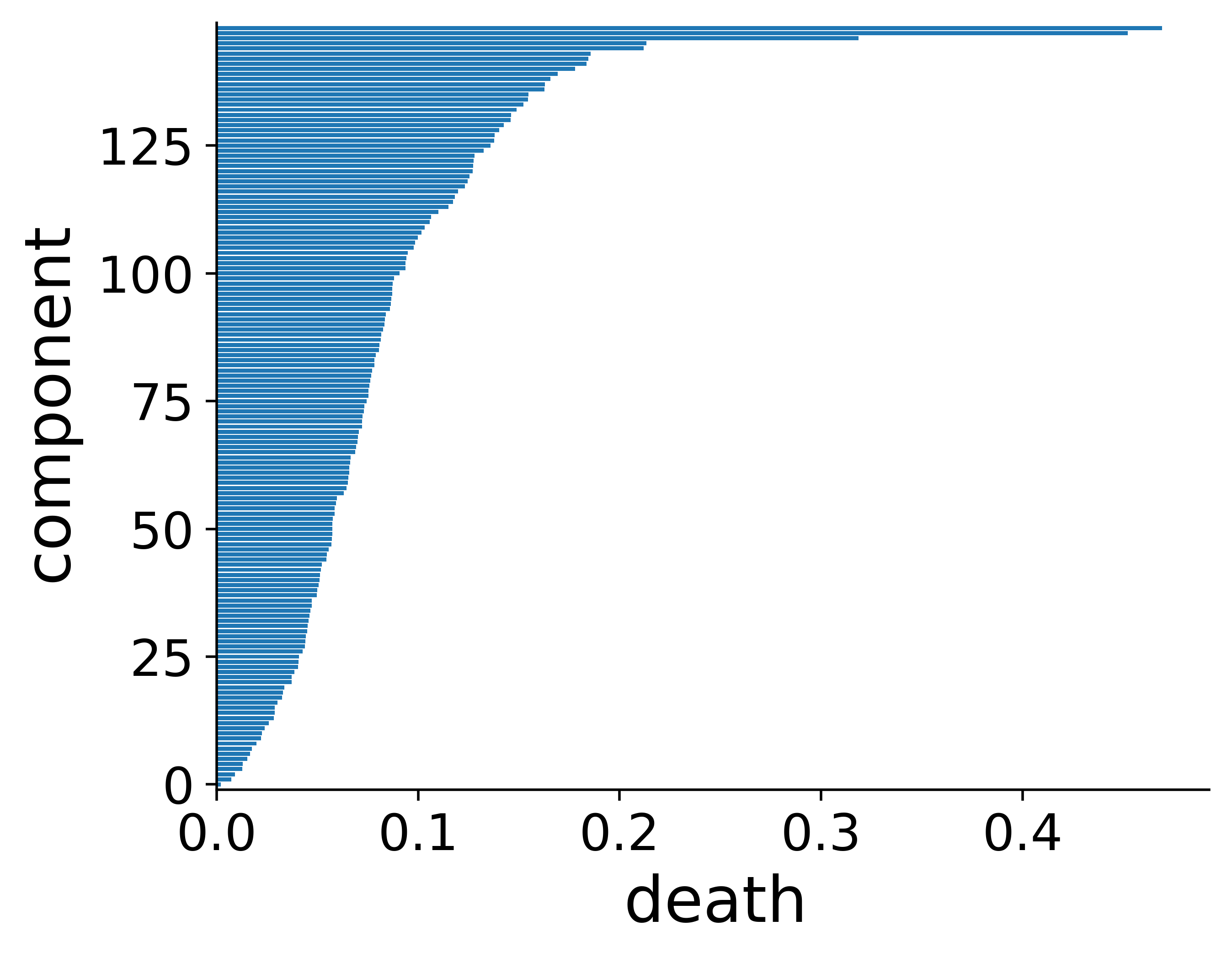}
      \caption{Group 2}
      \label{fig:barcode2}
  \end{subfigure}
  \hfill
  \begin{subfigure}[b]{0.32\textwidth}
      \centering
      \includegraphics[width=\textwidth]{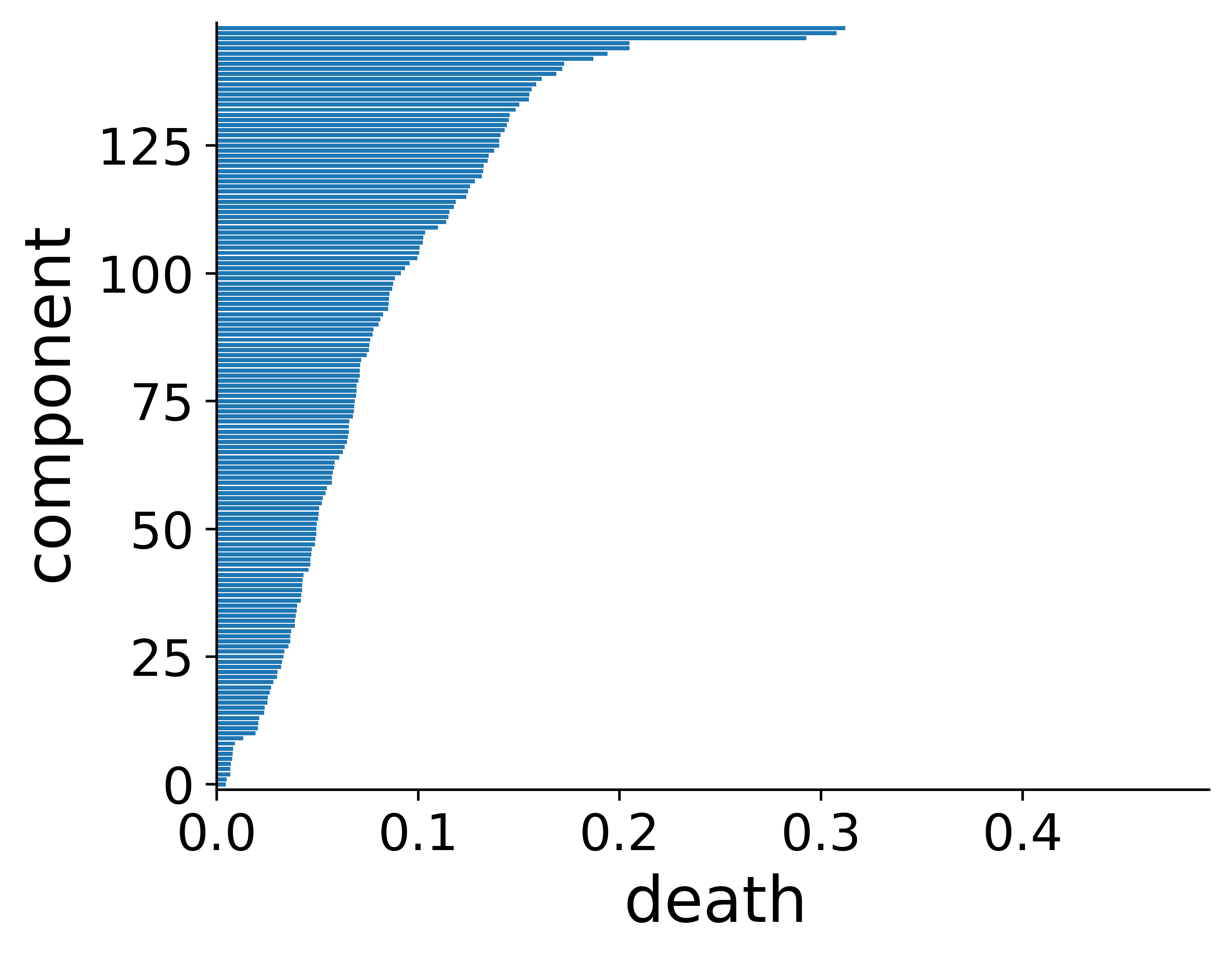}
      \caption{Group 3}
      \label{fig:barcode3}
  \end{subfigure}
  \caption{Zero-dimensional persistence barcodes for vertices generated near (a, Group 1) one point in $\mathbb{R}^2$, (b, Group 2 and c, Group 3) near two points separated by vector$(\Delta 1)$ in $\mathbb{R}^2$. Group 1 has bar death times much smaller than those in Group 2 and Group 3. On the other hand, the difference between Groups 2 and 3 is not obvious in the barcode, since the two groups differ only in the proportion of points near the two points.
  }
  \label{fig:barcodes}
\end{figure}

Figure \ref{fig:barcodes} shows the persistence barcode for one subject in each group under $\Delta=1$.
We can see clear differences between Group 1 and the other two, but the differences between Groups 2 and 3 are not obvious since they differ only in the proportion of points near locations $0$ and $\Delta 1$.

 We fit our model to the simulated data and estimate the posterior. In each setting, we run 1000 iterations of the sampler, with the first 500 iterations discarded as burn-in. For visualizing the latent coordinates, we first calculate the posterior mean of $\hat\Lambda^{p}$ for $p=1,2,3$ (under $m=5$), then obtain a low-rank approximate decomposition of $\hat\Lambda^p \approx (\tilde Z^p \tilde Z^p)^T$ using truncated SVD at rank $2$. We apply $O(2)$ orthonormal transformations, so that the distances $\|\tilde Z^1- \tilde Z^2\|_F$ and $\|\tilde Z^2 - \tilde Z^3\|_F$ are minimized.

\begin{figure}[H]
  \centering
  \begin{subfigure}[b]{0.32\textwidth}
      \centering
      \includegraphics[width=\textwidth]{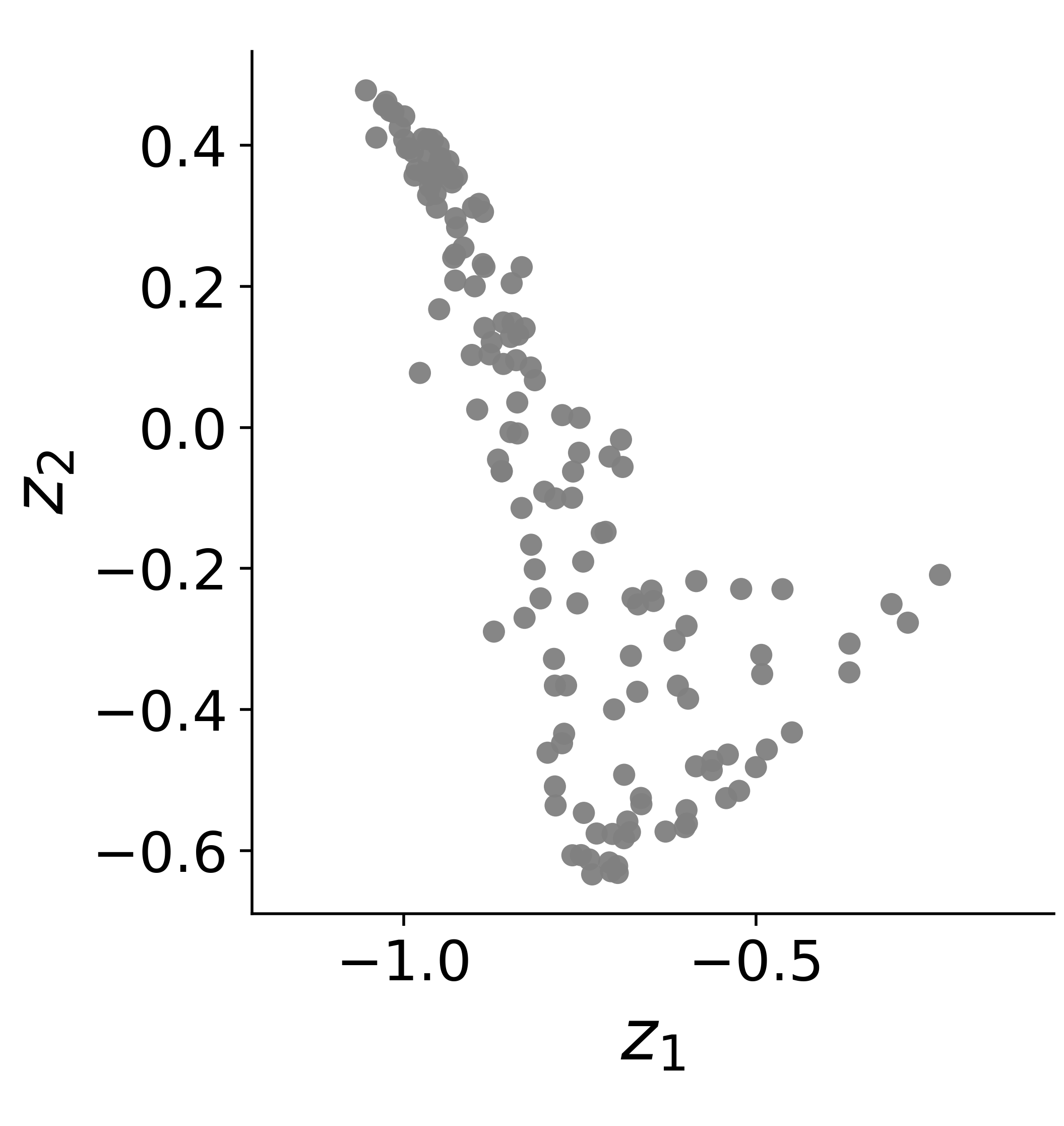}
      \caption{Group 1}
  \end{subfigure}
  \hfill
  \begin{subfigure}[b]{0.32\textwidth}
      \centering
      \includegraphics[width=\textwidth]{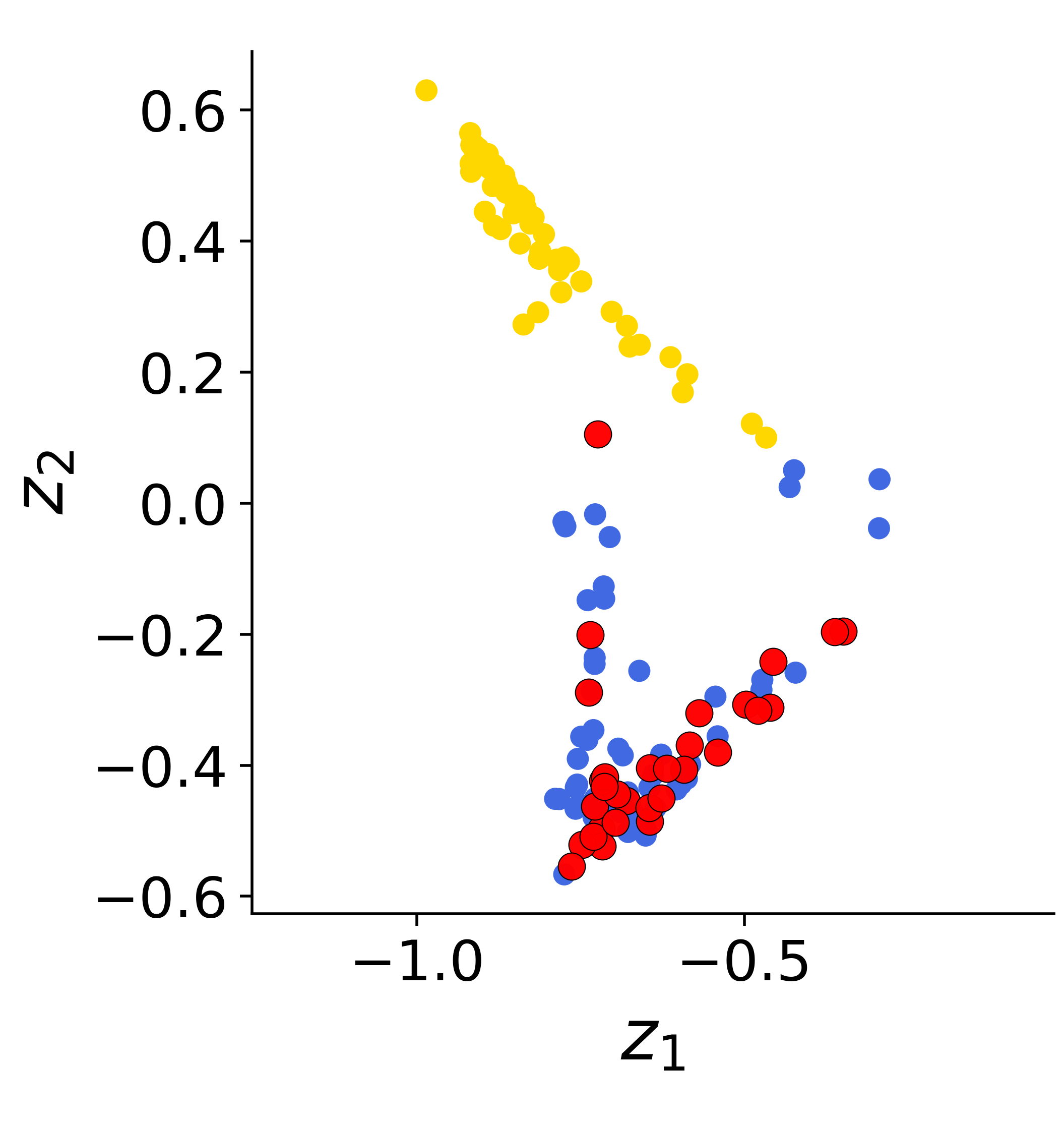}
      \caption{Group 2}
  \end{subfigure}
  \hfill
  \begin{subfigure}[b]{0.32\textwidth}
      \centering
      \includegraphics[width=\textwidth]{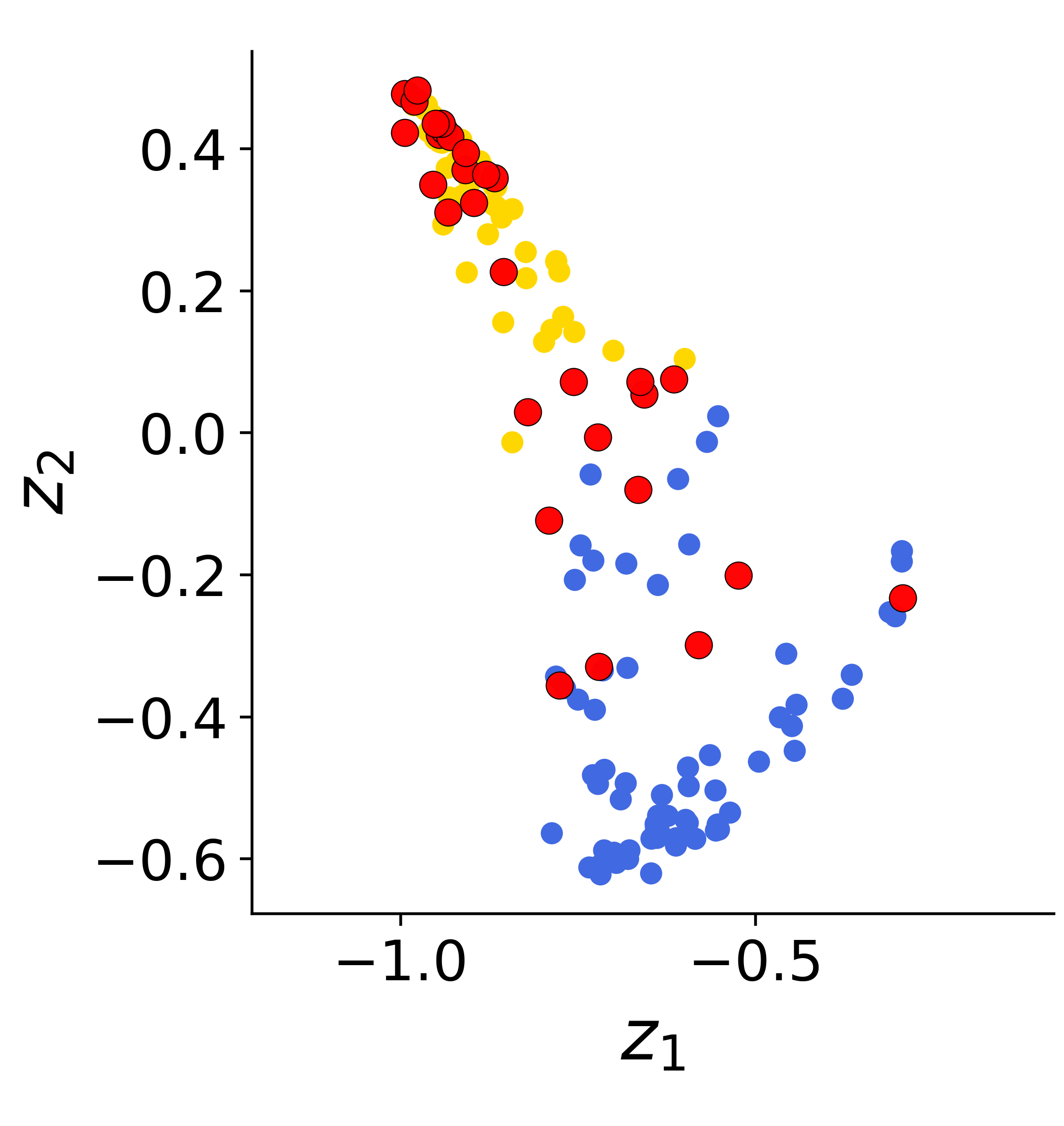}
      \caption{Group 3}
  \end{subfigure}
  \caption{Visualization of estimated latent coordinates in two dimensions. The difference between Groups 2 and 3 can now be localized to the 30 vertices (colored in red) that are mixed in cluster 1 (blue) under Group 2, whereas the vertices of the same indices are mixed in cluster 2 (yellow) under Group 3.
  }
  \label{fig:sim_procrutes}
\end{figure}

 Figure \ref{fig:sim_procrutes} plots the coordinates $\tilde Z^p_i$ in the three groups. Clearly, the latent coordinates in Group 1 are close to each other, while those in Group 2 and Group 3 have a two-cluster structure (colored in yellow and blue).
Comparing the latent coordinates in Group 2 and Group 3, we can see that there are 30 vertices (colored in red) that are mixed with the blue cluster under Group 2, whereas the vertices of the same indices are mixed with the yellow cluster under Group 3. The result is the effect of localization on which vertices contribute to the difference between the two groups.

In addition, Table~\ref{tab:localization} quantifies the localization aspect over $R=100$ replicates for different choices of $\Delta$.
Each replicate repeats the same data generation procedure as above with $10$ subjects per group, and refits the model to each group setting $m=5$.
For a pair of groups $(p,p')$ and each posterior draw $t$, we obtain the latent coordinates $z^{(p),t}_i$ from $\Lambda^{(p),t}$, align $z^{(p'),t}$ to $z^{(p),t}$ by the transformation minimizing the Frobenius distance as above, and compute the per-vertex latent difference $\delta^{t}_i=\bigl\lVert z^{(p),t}_i-z^{(p'),t}_i\bigr\rVert_2, i=1,\ldots, 150$. The same quantity is also used later to localize brain regions in Section~\ref{sec:data}.
Vertices are ranked by the posterior median $\hat\delta_i$ based on $\delta^{t}_i$'s.
The true set of the vertices whose oracle positions differ for the $(p,p')$ group pair is $G=\{i:\mu^{(p)}_i\neq\mu^{(p')}_i\}$, thus for group pair (2,3) $G=\{46,\ldots,75\}$ and that for (1,2) is $G=\{76,\ldots,150\}$.
Here we evaluate the performance of the proposed method in identifying the true set of vertices in $G$ for different group pairs. We measure how well $\hat\delta_i$ recovers the set $G$ by the area under the receiver operating characteristic curve,
$\mathrm{AUC}=\{|G|\,(150-|G|)\}^{-1}\sum_{i\in G}\sum_{j\notin G}\mathbb{1}\bigl(\hat\delta_i>\hat\delta_j\bigr),
$
which equals the probability that a randomly chosen vertex in $G$ has a larger $\hat\delta_i$ than a randomly chosen vertex outside $G$.
In addition to reporting the AUC, we also report precision, which is the fraction of the $|G|$ highest-ranked vertices that belong to $G$, whose chance level is $|G|/150$.
Both are computed in each replicate and averaged over the $R$ replicates.
As a null control, we generate an additional group from the oracle of Group 1 with independent subject noise, so that no vertex differs, and score it against an arbitrary set of the same $30$ indices as for Groups 2 and 3.

\vspace{-1cm}

\begin{table}[H]
\centering
\small
\begin{tabular}{cccc}
\hline
$\Delta$ & $|G|$ & AUC & precision \\
\hline
\multicolumn{4}{l}{\emph{Groups 2 vs 3} --- shared two-cluster structure; 30 vertices swap cluster} \\
$0.5$ & $30$ & $0.968\,(0.003)$ & $0.870\,(0.006)$ \\
$1$   & $30$ & $0.994\,(0.001)$ & $0.945\,(0.004)$ \\
$2$   & $30$ & $0.998\,(0.000)$ & $0.977\,(0.003)$ \\[0.3em]
\multicolumn{4}{l}{\emph{Groups 1 vs 2} --- one cluster becomes two, so the global structure itself changes} \\
$0.5$ & $75$ & $0.889\,(0.003)$ & $0.813\,(0.004)$ \\
$1$   & $75$ & $0.785\,(0.004)$ & $0.714\,(0.005)$ \\
$2$   & $75$ & $0.758\,(0.004)$ & $0.694\,(0.005)$ \\[0.3em]
\multicolumn{4}{l}{\emph{Null control} --- both groups from one configuration; no vertex differs} \\
---   & $30$ & $0.503\,(0.007)$ & $0.201\,(0.007)$ \\
\hline
\end{tabular}
\caption{Localization accuracy over $R=100$ replicates, each with $n=150$ vertices. Vertices are ranked by the posterior median latent difference $\hat\delta_i$. $G$ is the set of vertices whose oracle positions differ between the two groups. AUC is the area under the ROC curve of this ranking against $G$. Precision is the fraction of the $|G|$ top-ranked vertices that lie in $G$. Standard errors in parentheses.}
\label{tab:localization}
\end{table}

Groups 2 and 3 share the same two-cluster structure and differ only in which $30$ vertices belong to which cluster, so the difference between them is local. Localization is near exact in this case: the AUC is $0.968$, $0.994$ and $0.998$ at $\Delta=0.5$, $1$ and $2$, and the precision rises from $0.870$ to $0.977$ over the same range, against a chance level of $0.2$. The ranking therefore recovers almost all of $G$, and sharpens as the two groups separate. In the null control, where no vertex differs, the AUC is $0.503$ and the precision is $0.201$, both at chance; the separation reported above thus comes from the group difference.

\begin{figure}[H]
    \centering
    \begin{subfigure}[t]{0.48\textwidth}
        \centering
        \includegraphics[width=\textwidth]{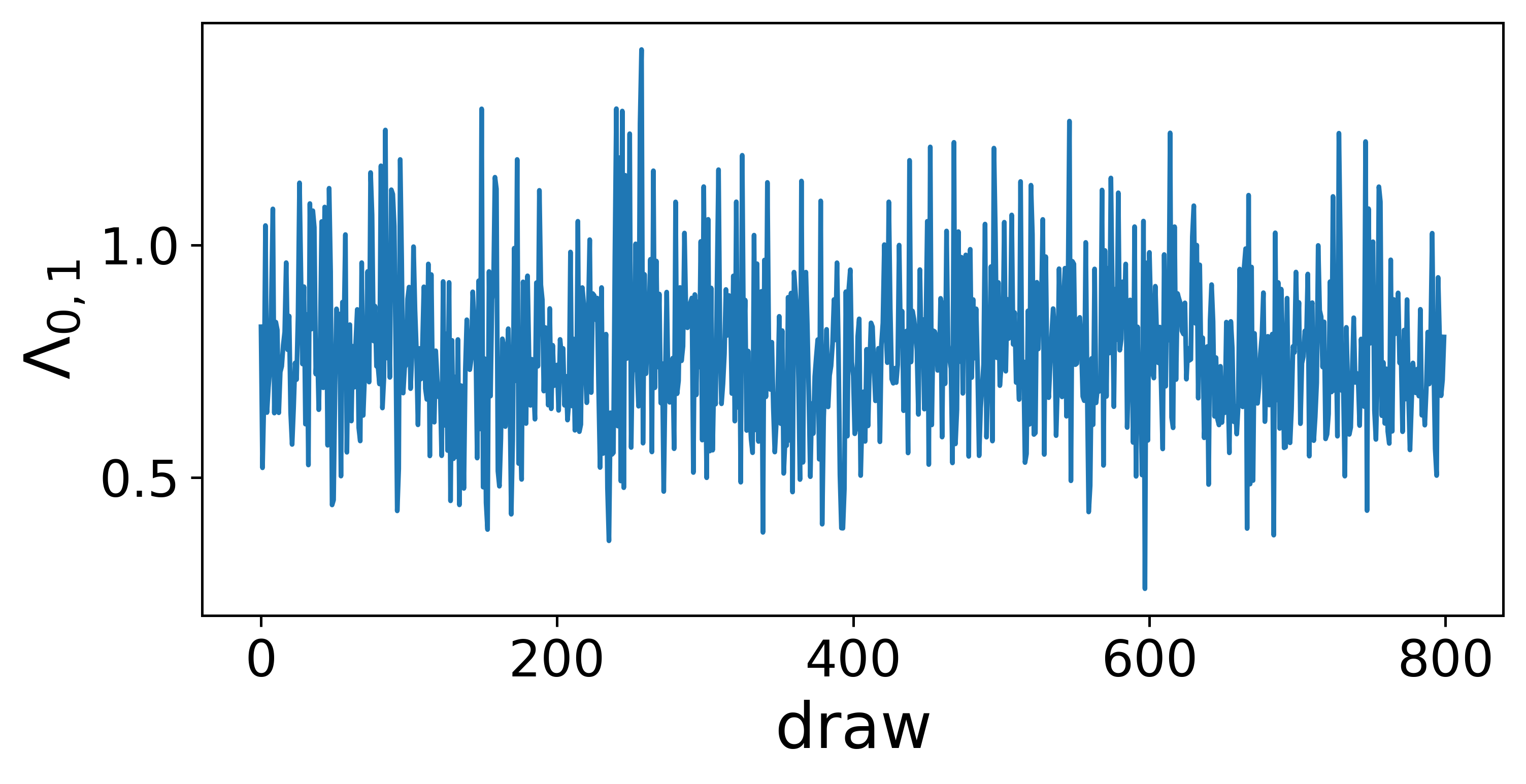}
        \caption{Trace plot of $\Lambda_{0,1}$}
        \label{fig:trace_plot}
    \end{subfigure}
    \hfill
    \begin{subfigure}[t]{0.48\textwidth}
        \centering
        \includegraphics[width=\textwidth]{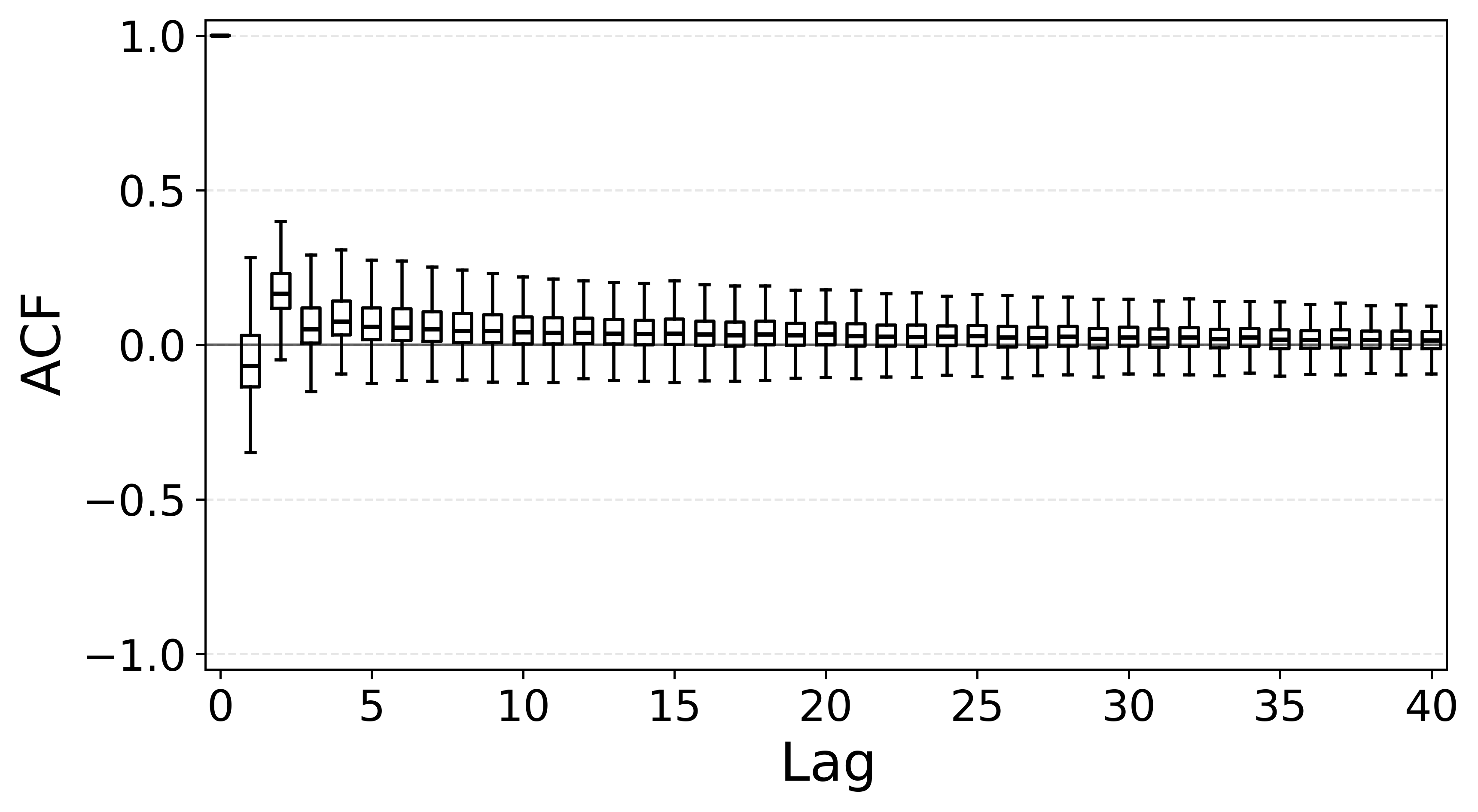}
        \caption{Boxplots of the autocorrelation function for all elements of $\Lambda$.}
        \label{fig:autocorrelation_plot}
    \end{subfigure}
    \caption{Markov chain diagnostics shows that the posterior distribution can be efficiently sampled by the No-U-Turn Sampler.
    }
    \label{fig:sim_plots}
\end{figure}

The No-U-Turn Sampler shows excellent mixing of Markov chains and the autocorrelation function decays rapidly to near zero. Figure \ref{fig:sim_plots} shows the trace plot of one element of $\Lambda$, and the boxplot of the autocorrelation function for all elements of $\Lambda$, under $\Delta=1$. We find similar results under $\Delta=0.5$ and $\Delta=2$. {In addition, we conduct experiments (i) where the data are generated near two circles of different radii,   and around two tori, hence producing one-dimensional and two-dimensional features;
(ii) against a more flexible competing Weibull model, which produces nearly identical results on the quantities of interest; (iii) when points go through transformations such as rotation and reflection, leading to worsening performance of the complete likelihood approach from a generative model, but no impact on the persistent homology likelihood.
We provide the details in the supplementary materials.}

\section{Data application}\label{sec:data}
We now apply our method to analyze the structural connectivity data related to Alzheimer's disease. Alzheimer’s disease (AD) is a major neurodegenerative disorder, with early detection of brain changes considered essential for effective treatment. A key research focus is how white matter changes as the disease progresses \citep{phillips2015graph}. Studies show that structural connectivity alterations spread with disease progression \citep{tucholka2018structural} and tend to impact reduced connectivity brain regions \citep{daianu2015}. White matter changes are also linked to amyloid plaque buildup, a hallmark of Alzheimer's disease that emerges before clinical symptoms \citep{prescott2014alzheimer}. Most prior work relies on scalar connectivity metrics derived from brain imaging, but these can be sensitive to how the metric is defined \citep{phillips2015graph}. In this study, we use data from the Alzheimer’s Disease Neuroimaging Initiative (ADNI), which provides connectome information for $n = 83$ brain regions. For each region pair, both the count and mean length of connecting white matter fibers are available. There are $S=104$ subjects in the dataset, consisting of three groups: cognitively normal (CN) with 21 subjects, mildly cognitively impaired (MCI) with 46 subjects, and Alzheimer's disease (AD) with 37 subjects.

\begin{figure}[H]
  \centering
  \begin{subfigure}[t]{0.32\textwidth}
      \centering
      \includegraphics[width=\textwidth]{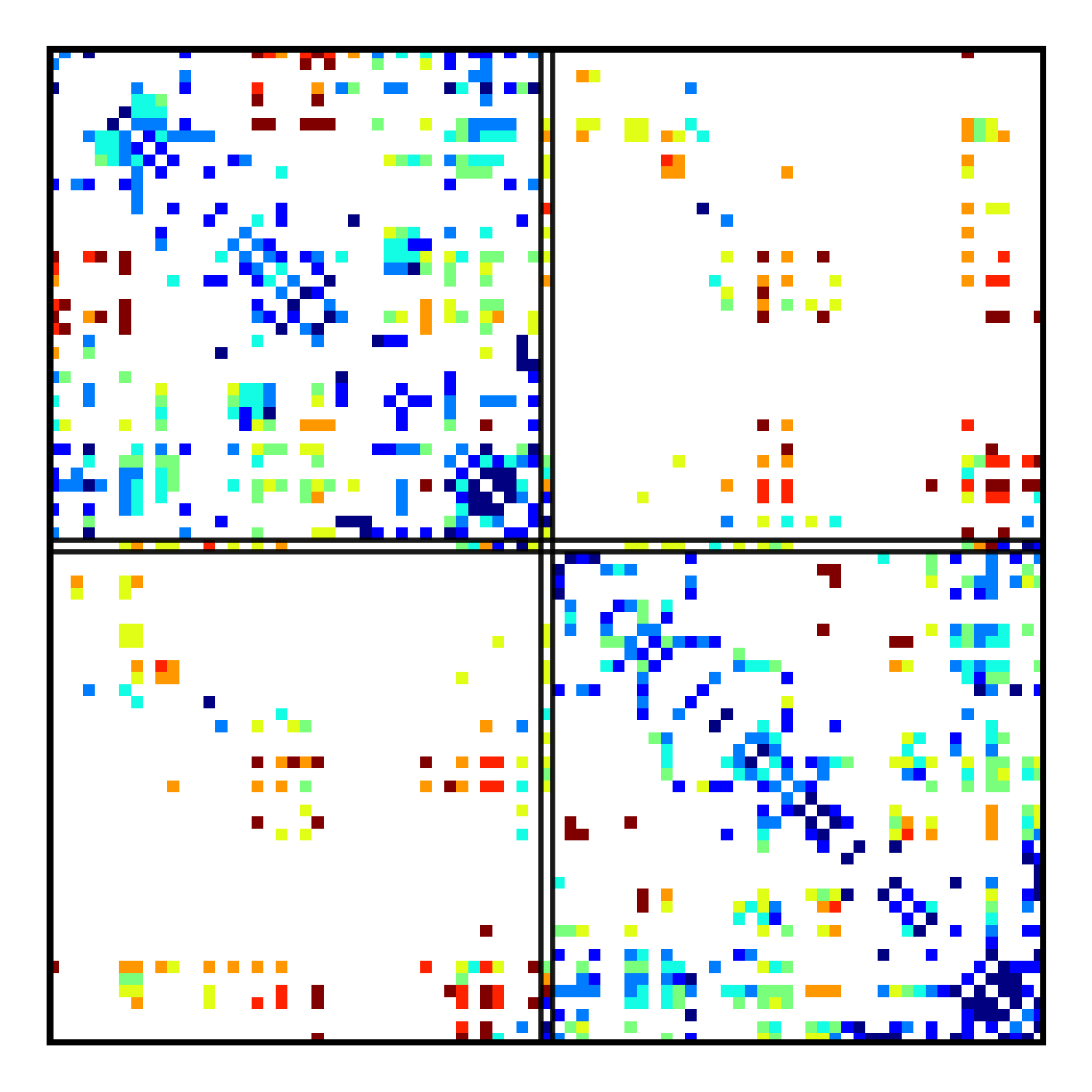}
      \caption{Structural connectivity matrix for an Alzheimer's disease subject}
  \end{subfigure}
  \hfill
  \begin{subfigure}[t]{0.32\textwidth}
      \centering
      \includegraphics[width=\textwidth]{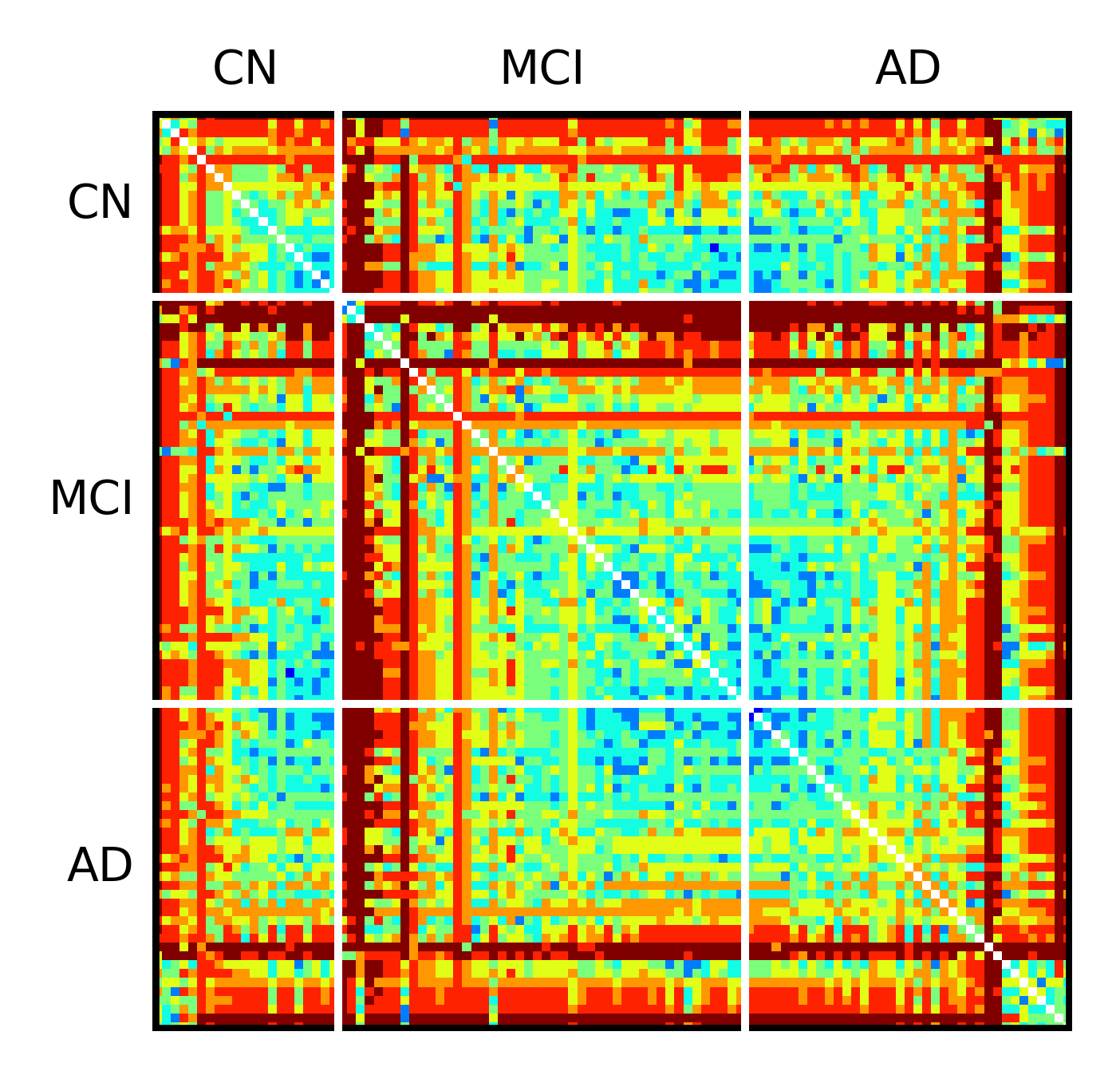}
      \caption{Bottleneck distance matrix between all subjects' persistence diagrams}
  \end{subfigure}\hfill
  \begin{subfigure}[t]{0.32\textwidth}
    \centering
      \includegraphics[width=\textwidth]{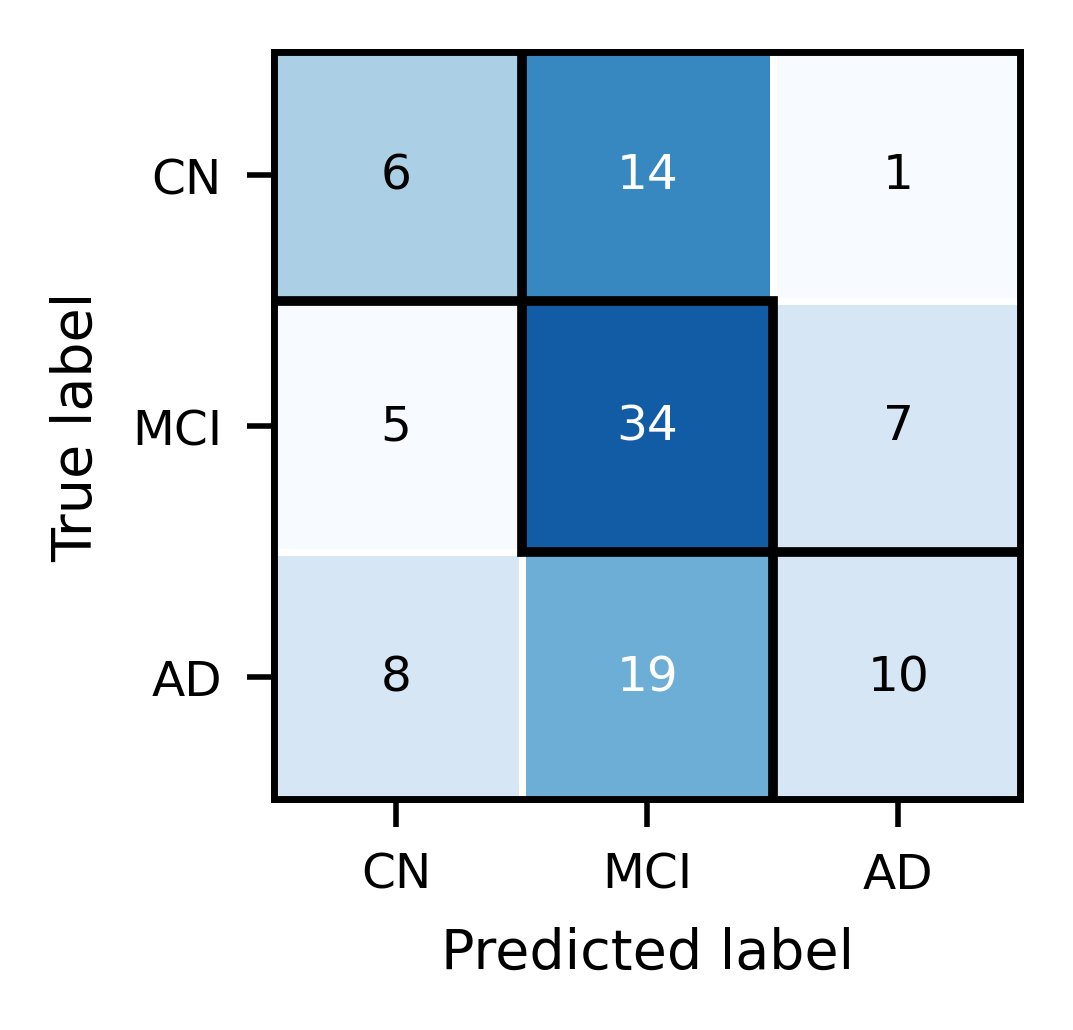}
      \caption{Confusion matrix of using $K$-nearest neighbor classifier on the bottleneck distance matrix}
  \end{subfigure}
  \caption{Exploratory analysis using the bottleneck distance.
  \label{fig:adni_exploratory_analysis}}
\end{figure}

We follow the idea of \citet{lee2011discriminative} and convert the pairwise information in brain
networks into a distance representation, then analyze the resulting barcode-based summaries.
The structural connectomes are pre-processed according to the pipeline in \cite{roy2019bayesian}, each subject connectivity score matrix is symmetric and consists of $n(n-1)/2$ entries, with each score greater than or equal to 0 and representing the strength of the connection between the two regions. We apply a spectral embedding via Laplacian eigenmaps to convert the connectivity scores to distances, then apply the Vietoris--Rips filtration to obtain the persistent homology. The embedding uses the normalized Laplacian with $k = 5$ coordinates,
fixed using the normalized Laplacian spectrum that distance matrices are comparable across
subjects (see the supplementary materials for details). The embedded clouds satisfy the
condition under which Remark~\ref{rmk:ordering} justifies the exponential
form. The embedding retains the eigenvectors of the normalized Laplacian
belonging to its $k$ smallest eigenvalues, and these eigenvectors have unit
norm, so every coordinate of every embedded region lies in $[-1,1]$ and all
pairwise distances are at most $2\sqrt k$. A bounded random variable is
sub-Gaussian, hence sub-exponential \citep{vershynin2018high}, so the distances
produced by this pipeline are sub-exponential however the connectivity scores
themselves are distributed. The embedding step also leaves the barcode stable:
the stability result of \citet{cohen2005stability} bounds the change in the
persistence diagram by the change in the distance matrix, and the Laplacian
eigenmap coordinates vary continuously in the connectivity weights, so a small
perturbation of a connectome produces only a small change in the barcode, and
hence in the homology features.

In the exploratory analysis, we first compute the bottleneck distance between the persistence diagrams of all the subjects, and plot the distance matrix in Figure \ref{fig:adni_exploratory_analysis}(b). The subjects are arranged according to their group label; however, we can hardly see any group pattern in the bottleneck distance matrix. In order to quantitatively assess the difference between the groups, we fit a $K$-nearest neighbor classifier to the data and report the prediction accuracy in confusion matrix \ref{fig:adni_exploratory_analysis}(c), corresponding to the prediction accuracy of a 3-class classifier only at 48.1\%, and the one between normal (CN) and diseased (MCI and AD combined) at 73.1\%.

With the existing scientific finding that established across-group differences in the structural connectivity, the lack of group-discriminative patterns in the bottleneck distance matrix is somewhat surprising. This clearly suggests that useful signals are likely distorted in the calculation of the bottleneck distance (as a global metric comparing two homologies) hence motivating localizing approach.

\begin{figure}[H]
  \centering
  \begin{subfigure}[t]{0.32\textwidth}
      \centering
      \includegraphics[width=\textwidth]{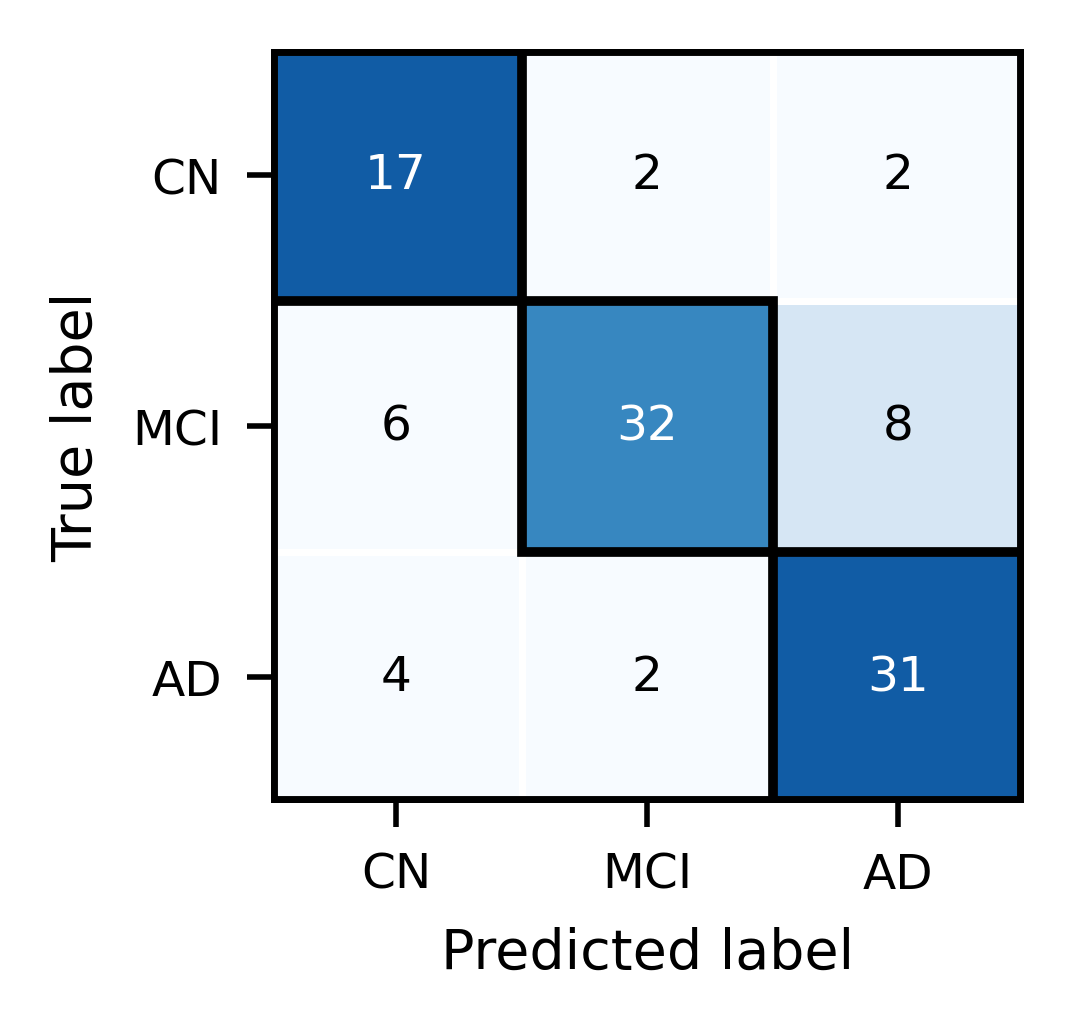}
      \caption{Persistence homology likelihood model show much lower misclassification rate in the confusion matrix, compared to the classifier based on the raw connectivity scores.}
    \end{subfigure}
  \hfill
\begin{subfigure}[t]{0.32\textwidth}
  \centering
  \includegraphics[width=\textwidth]{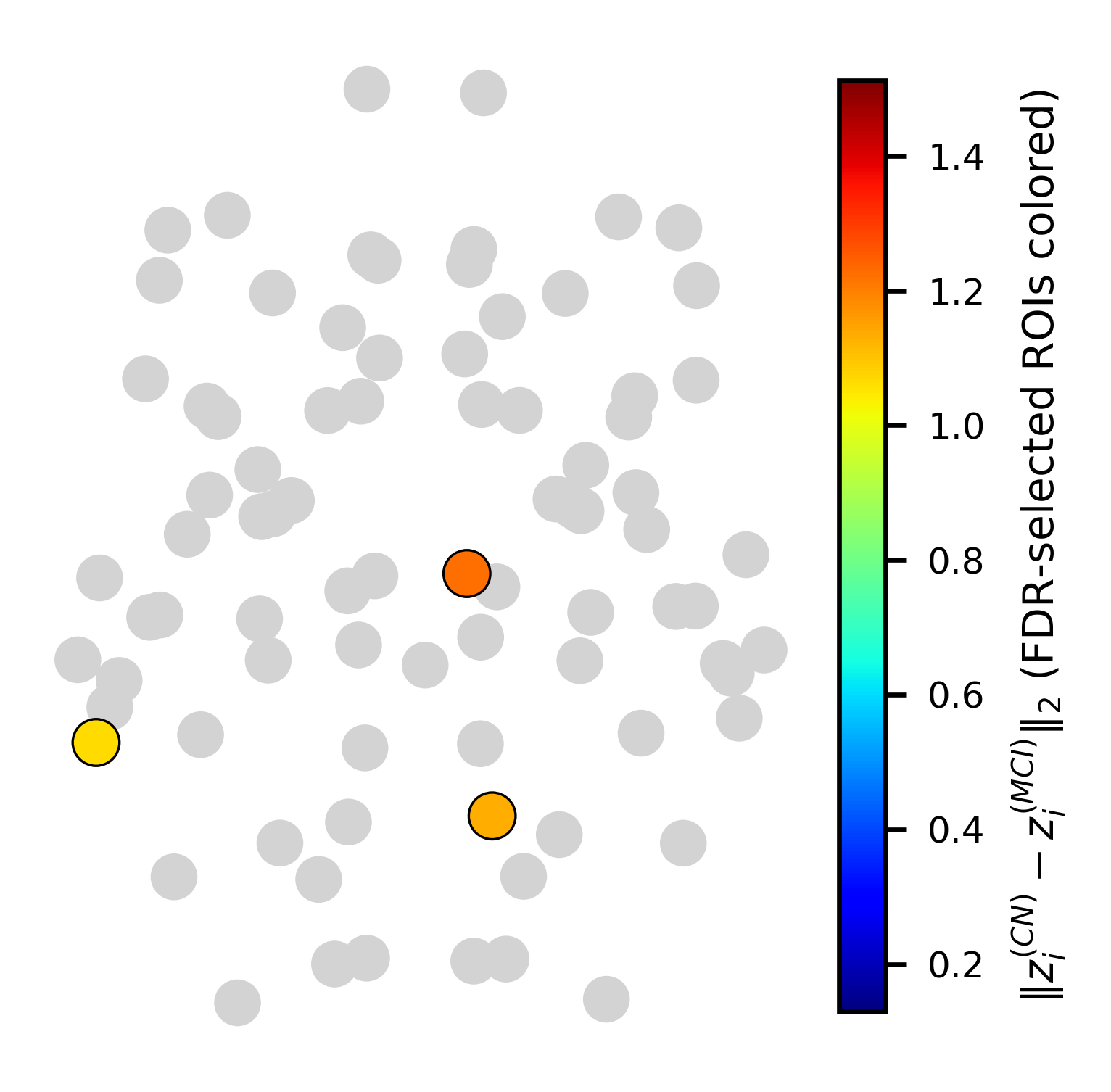}
  \caption{Dorsal view of the brain regions of interest with significant differences between the cognitive normal and mildly cognitively impaired groups.}
\end{subfigure}\hfill
\begin{subfigure}[t]{0.32\textwidth}
  \centering
  \includegraphics[width=\textwidth]{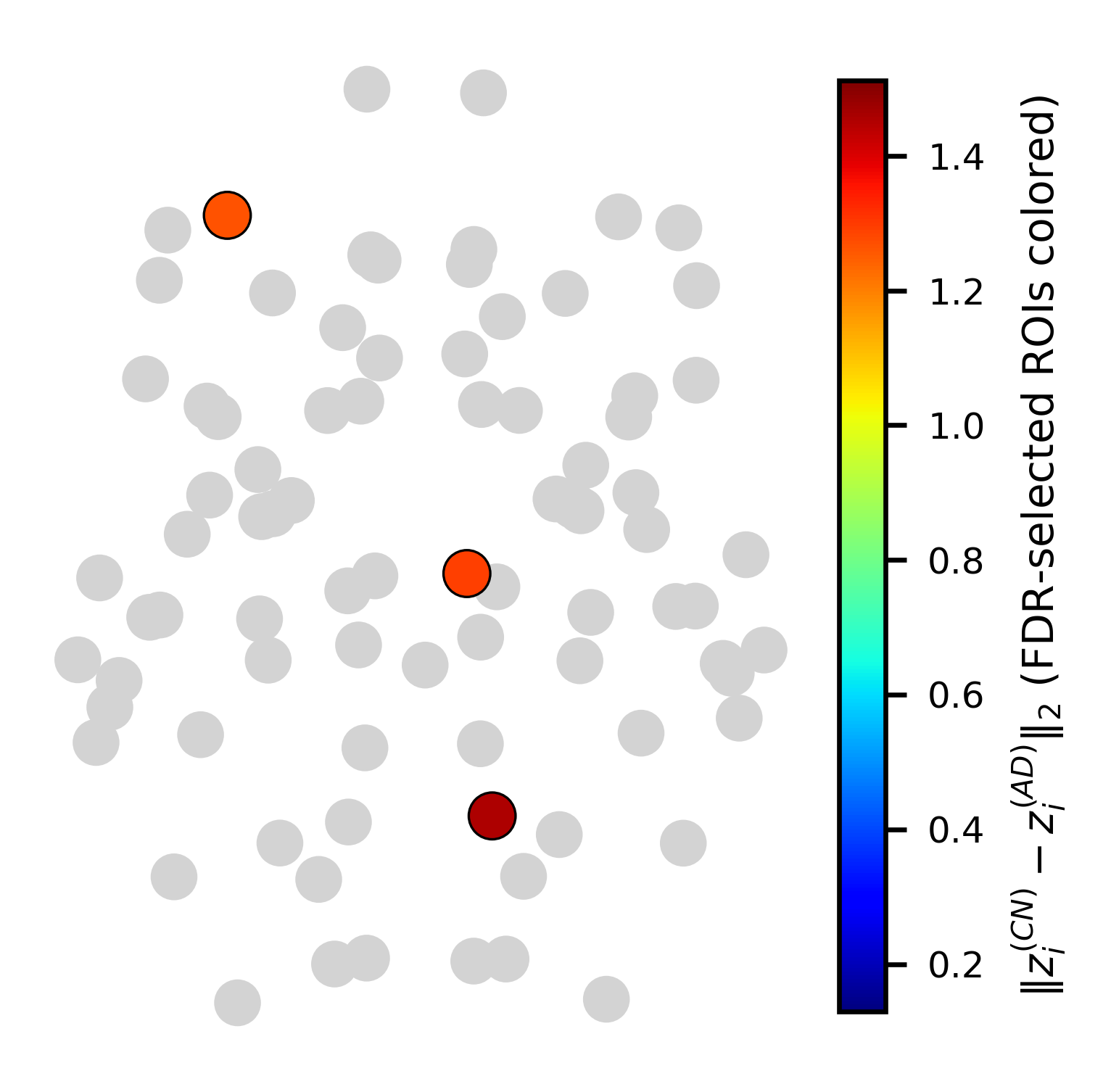}
  \caption{Dorsal view of the brain regions of interest with significant differences between the cognitive normal and Alzheimer's disease groups.}
\end{subfigure}\hfill
  \begin{subfigure}[t]{0.45\textwidth}
      \centering
      \includegraphics[height=4cm, width=\textwidth]{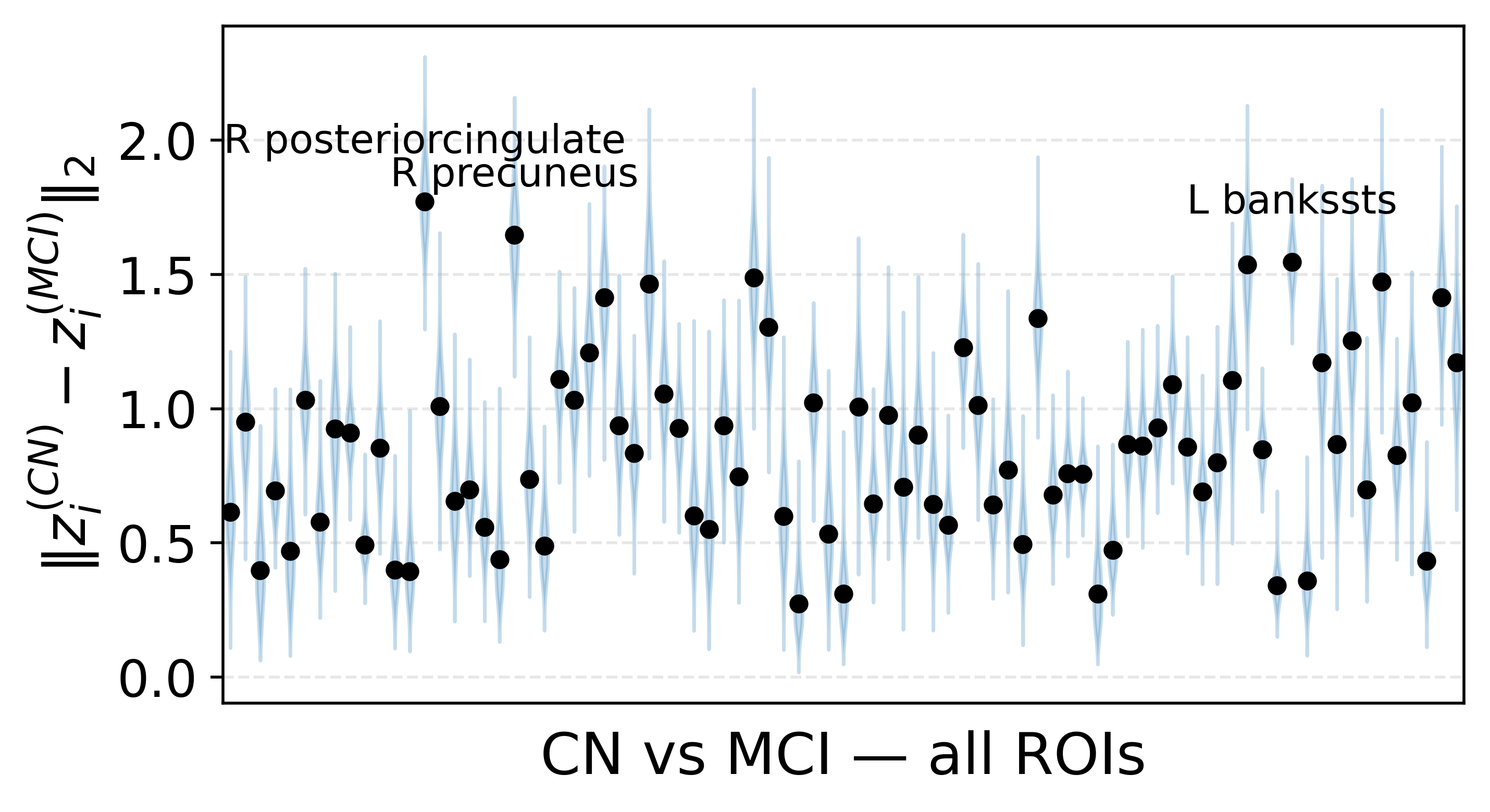}
      \caption{Violin plot showing the posterior distribution of the distance between the latent coordinates of the regions, between the cognitively normal and mildly cognitively impaired groups.}
  \end{subfigure}\hfill
  \begin{subfigure}[t]{0.45\textwidth}
    \centering
    \includegraphics[height=4cm, width=\textwidth]{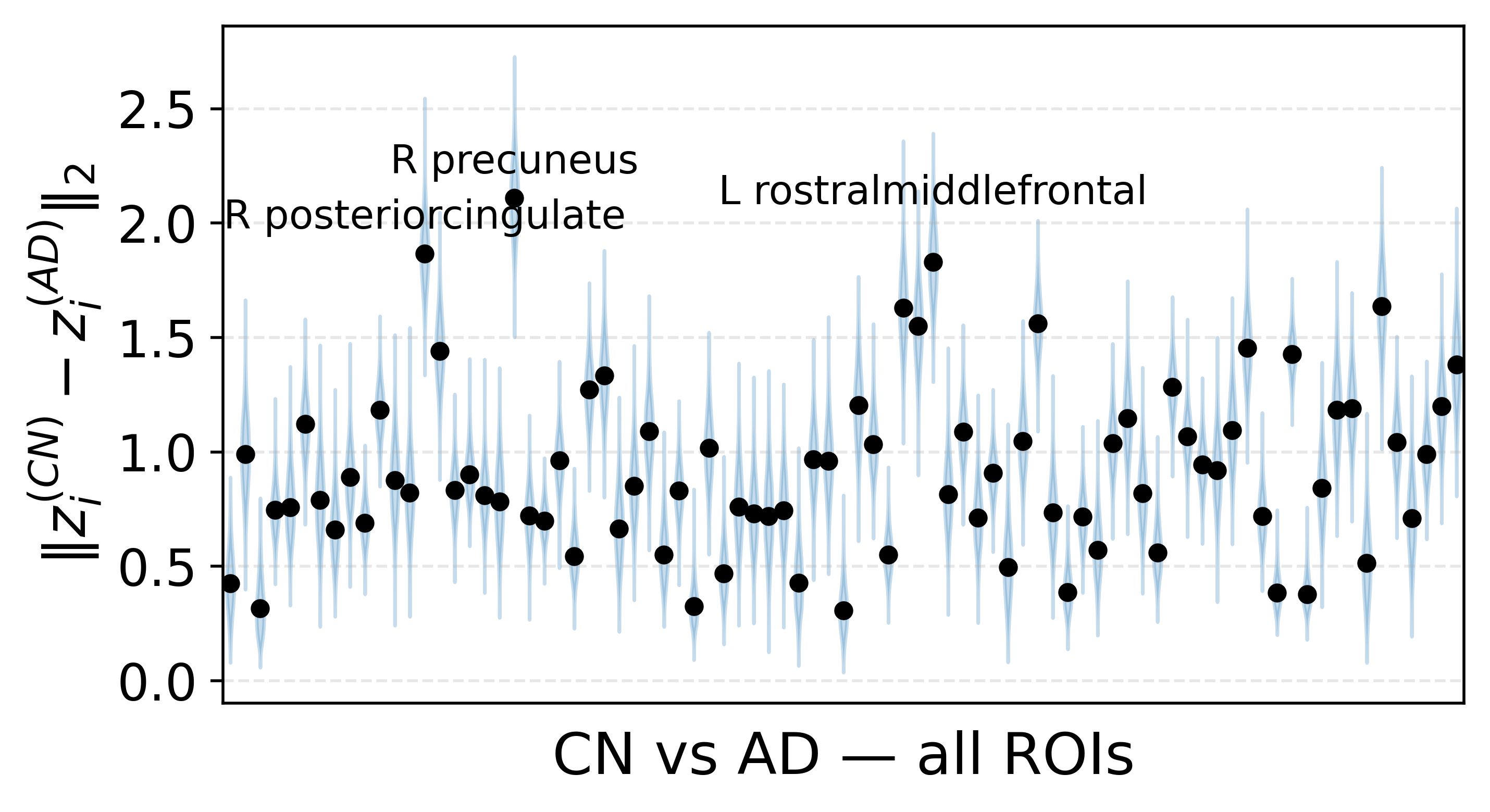}
    \caption{Violin plot showing the posterior distribution of the distance between the latent coordinates of the regions, between the cognitive normal and Alzheimer's disease groups.}
\end{subfigure}

\caption{The persistent homology model using a hierarchical prior shows good discriminability between the three groups and allows us to localize the source of the difference down to a few regions.
}
  \label{fig:adni_model_fit}
\end{figure}

We fit our model to the data and estimate the posterior. We use an optimization-based warm start to find the conditional mode of the posterior distribution, then run 10000 iterations of the sampler, with the first 500 iterations discarded as burn-in. Using the posterior mean of the parameter $\tilde\Lambda^p$, we use the maximum likelihood classifier $\hat c_i= \arg\max_{c\in \{1,2,3\}} \hat p(y^{s} \mid \tilde\Lambda^p)$ to predict the group assignment for each subject. We see a clear improvement in the prediction accuracy compared to the $K$-nearest neighbor classifier, with the prediction accuracy of the 3-class classifier reaching 76.9\%, and the one between normal and diseased reaching 86.5\%.

For comparison, we also fit a 3-class logistic regression model with lasso regularization on the raw connectivity scores, and a 3-class ordinal regression model with 5-fold cross-validation. The prediction accuracy of the logistic regression model is 43.3\% (78.8\% for normal and diseased), and that of the ordinal regression model is 48.1\% (78.8\% for normal and diseased). We can see that our proposed method achieves much better performance than the two baseline models.
We further compare our method with low-rank matrix regression on the symmetric connectivity matrix \citep{zhou2014regularized} and with kernel ridge regression on persistence landscapes \citep{bubenik2015statistical}. Under 5-fold cross-validation, their 3-class accuracies are $50.0\%$ and $44.2\%$ ($77.9\%$ for both when normal and diseased subjects are compared); details are given in the supplementary materials.

We now localize the source of the difference down to a few regions. We first calculate the pairwise distance between the latent coordinates of the regions, $\|z_i^{(p)} - z_i^{(p')}\|_2$ in each posterior sample, then we plot the posterior distribution of the coordinate-wise distance in Figure \ref{fig:adni_model_fit}(d,e). Then we follow the Bayesian false discovery rate (FDR) control approach in \citep{muller2007fdr, kiran2021bayesian} to identify the regions with significant differences between the groups. At a false discovery rate level less or equal to 0.1, we identify 3 regions with significant differences between the cognitive normal and mildly cognitively impaired groups, and 3 regions with significant differences between the cognitive normal and Alzheimer's disease groups. Figure \ref{fig:adni_model_fit}(b)(c) highlights the regions corresponding to major differences.

\section{Discussion}

In this article, we propose a likelihood-based approach that models the birth and death events of persistent homology features in terms of corresponding graph events. The introduction of latent coordinates for the vertices in the simplicial complex allows us to localize the between-group differences, while quantifying the uncertainty for statistical inference. There are a few interesting directions that can be pursued in the future. First, since one could simultaneously obtain more than one type of filtration (such as the sublevel filtration from different radial centers) from the data, there is a need for filtration selection based on the level of informativeness. Extending our hierarchical model-based approach, one could consider a group shrinkage prior that jointly shrinks group-specific latent coordinates toward some commonality while allowing heterogeneity. There is a large literature in hierarchical, multiscale, or structural shrinkage models \citep{guhaniyogi2020large,kundu2024flexible,boss2024group} that can be adapted to our setting. Second, one could extend our approach to multi-parameter persistent homology \citep{botnan2022introduction}, under which the complex may evolve under several simultaneous scales. Although the birth and death events must be defined differently, which requires some modifications to the order-based likelihood representations discussed above, the core idea can be applied: we can use relatively simple graph events to capture the key changes occurring in the potentially high-dimensional simplicial complex, enabling feasible statistical estimation and inference.

\section*{Declaration of the use of generative AI and AI-assisted technologies}
During the preparation of this work, the authors used ChatGPT and Claude Code for assisted writing of Python code. After using this tool/service, the authors reviewed and edited the content as necessary and take full responsibility for the content of the publication.

\section*{Acknowledgement}
The authors would like to thank Peter Bubenik for helpful discussions.

\bibliographystyle{chicago}
\bibliography{ref_fixed.bib}

\newpage
\appendix
\section*{Supplementary Materials}

\section{Extension to Other Filtrations \label{sec:background}}

\noindent\textbf{Generalized Vietoris--Rips Filtration}

Naturally, we do not have to force $\mathcal K_\infty$ to correspond to a complete graph, potentially leading to more interpretable filtrations in practice. For example, in image data analysis, it is common to have $\mathcal K_\infty$ as a cubical complex, where the largest simplex has each vertex only connected to its neighboring vertices in the spatial grid. In this case, we can define a Vietoris--Rips filtration of a cubical complex, $\mathcal{K}_\epsilon=\mathrm{VRC}(W, \epsilon)$, such that each $\sigma$ has $\text{dist}(w_i, w_j) \leq \epsilon$ and $(i,j)$ are neighbors in the spatial grid, for all pairs $(i, j)\in \sigma$.

With the potentially different choice of $\mathcal K_\infty$, we can see the  graph-event representations developed above are still valid, except that there would be fewer edges in the admissible sets for forming spanning tree or triangulation, compared to the complete graph case.

\noindent\textbf{Alpha Filtration}

We now extend to other type of filtrations. First, we consider the alpha filtration on Delaunay complex:
$
\text{Delaunay}(W)=\mathcal{K}_\infty = \{\,\sigma \subseteq V  : \; \cap_{i\in\sigma} \mathcal{V}(w_i) \neq \varnothing \,\},
$
where $\mathcal{V}(w_i) = \{w \in \mathcal Y : \text{dist}(w, w_i) \leq \text{dist}(w, w_j) \text{ for all } w_j \in V\}$ is commonly called the Voronoi cell of $w_i$. The alpha filtration is then formed with a ball $B(w_i,\epsilon)=\{w \in \mathcal Y : \text{dist}(w, w_i) \leq \epsilon\}$.
\(
\text{Alpha}(W,\epsilon)=\mathcal{K}_\epsilon = \{\,\sigma \subseteq V  : \; \cap_{i\in\sigma} [\mathcal{V}(w_i) \cap B(w_i,\epsilon)] \neq \varnothing \,\}.
\)
Further, one may consider a weighted version of the alpha filtration \cite{edelsbrunner2023simple}, where the distance is replaced by a power distance function (which is not a metric): $\text{power-dist}(w,w_i) = \|w-w_i\|^2 - \nu_i$, with $\nu_i\ge 0$ an assigned weight associated with the point $w_i$. We can immediately see a connection to the Vietoris--Rips filtration of the Delaunay complex: it is the exactly the same as the unweighted alpha filtration, and equivalent to the weighted alpha filtration after replacing the metric by the power distance function. Hence again, the graph-event representations developed above are directly applicable to the alpha filtrations.

\noindent\textbf{Sublevel Filtration}

Next, we consider the sublevel filtration  (or lower-star filtration, as in some literature). Given a function $\phi:\mathcal Y \to \mathbb{R}$, the canonical definition of sublevel filtration for the set $W$ is
\(
\text{Sublevel}(W,\epsilon)=\mathcal K_{\epsilon} = \{\sigma  \subseteq V : \phi(w_i) \leq \epsilon \text{ for all } i \in \sigma\}.
\)
Hence $\mathcal K_\infty$ is the power set of $V$ minus the empty set. For $\phi$, one could consider some score function such as a probability density estimate (kernel density filtration), or the distance to a reference point (radial filtration).

Unlike the previously discussed filtrations, this definition of sublevel filtration lacks a graph interpretation. Fortunately, when it comes to computing persistent homology features (such as connected components and loops), one does consider an auxilliray graph $H_\epsilon=(V,E_{H}(\epsilon))$, with $(i,j)\in E_H(\epsilon)$ only if (i) there is an edge $(i,j)\in E_G$ on another underlying graph $G$ that is predefined and invariant to $\epsilon$ (such as the cubical grid graph), and (ii) $i,j \in \sigma$ for some $\sigma \in \mathcal K_{\epsilon}$.

We can see it is more convenient to a vertex-centric distribution to model the bars in the sublevel filtration. We again use an exponential distribution with vertex-wise rate $\lambda_i$.

Clearly, some modifications are needed for the sublevel filtration. For simplicity, we assume $\phi(w_i)>0$ for all $w_i\in W$ and no ties in $\phi(w_i)$. First of all, each vertex $i$ could become a zero-dimensional bar with a positive birth time $\phi(w_i)$, with a positive length if $\phi(w_i) < \phi(w_j)$ for all the adjacent vertices $j:(i,j)\in E_G$. That is $w_i$ corresponds to a strict local minimum of $\phi$ on the graph $G$. To find out the death time, we again consider a minimum spanning tree $\hat T$, formed via the Kruskal procedure except with $\Delta_{j,k}=\max[\phi(w_j),\phi(w_k)]$. Then the death time for the zero-dimensional bar is associated with the addition of a vertex $j$ that simultaneously connects to the tree component containing $i$ and to an earlier component. That is, $\phi(w_j) > \phi(w_k)$ for all $k$ in the ealier component and the component(s) that just died --- equivalently, as we observe $d^i_0=\phi(w_j)$ for some $j$, $k$ is simply all the other vertices in the same component as $j$ at time $d^i_0$.

On the other hand, the one-dimension bars for the sublevel filtration are similar to those of the Vietoris--Rips filtration. The birth time is associated with the formation of a loop, and the death time is associated with the fill-in of a loop, except using  vertices instead of edges.

\section{Proof of the Theorems}

\subsection{Proof of Theorem 1}
\begin{proof}
  For (i), we first note the log-likelihood of $\Lambda$ as
  \(
  \log \mathcal{L}(y;\Lambda) = & \text{const} + \sum_{k=0}^K
  \bigg\{  \sum_{e \in \mathcal{E}_{1,k}}\log \lambda_{e} - \sum_{e \in \mathcal{E}_{2,k}} \alpha_{e,k}\lambda_{e}
  \\
  & +
  \sum_{e \in \mathcal{E}_{3,k}}\log( 1 - e^{-\lambda_{e} \beta_{e,k} })
  +
  \sum_{e \in \mathcal{E}_{4,k}}\log( e^{-\lambda_{e} \beta_{e,k}} - e^{-\lambda_{e} \gamma_{e,k}})
  \bigg\},
  \)
  with $\mathcal{E}_{i,k}$, $i=1,2,3,4$, the appropriate sets of edges, and $\alpha_{e,k}$, $\beta_{e,k}$, $\gamma_{e,k}$ the birth and death times, and $c$ a constant. It is not hard to see that the log-likelihood is a concave function in $\lambda_e$.

  The log-prior is equivalent to
  \(
    \log \pi_0(Z \mid \kappa, \alpha) = \text{const} -\frac{\kappa}{2} \sum_{i=1}^n \Lambda_{i,i} + \alpha \sum_{i\le j}
    \log(\Lambda_{i,j})
  \)
  in which, the sum of logarithmic function is strictly concave when $\alpha>0$, and covers all elements of $\Lambda_{i,j}$. Therefore, the log-posterior is a strictly log-concave function in $\Lambda\succeq 0$.

  For (ii), note that when $\Lambda\succeq 0$, $\sum_{i=1}^n \Lambda_{i,i} =\|\Lambda\|_*$, the nuclear norm of $\Lambda$. Therefore, $(\kappa/2)$ can be viewed the Lagrange multiplier for the nuclear norm constraint $\|\Lambda\|_{*} \le r_{m^0}$, hence by strong duality of a concave function, there exists a suitable value of $\kappa$, for the maximizer $\hat\Lambda$ to be of rank $m^0$.

  For (iii), we note that $-r_{\kappa,\alpha}(Z)$ is twice differentiable in $Z\in \mathbb{R}^{n\times m}\cap \mathcal C^n$. \cite{burer2005local} shows that for any convex and twice differentiable problem of positive semi-definite matrix $\Lambda$ with $\text{rank}(\Lambda)=m^0$, the solution of the second-order stationary point under reparametrization by $\Lambda= ZZ^\top$, with $Z\in\mathbb{R}^{n\times m}$ has $(Z^*)  (Z^*) ^\top = \hat\Lambda$, provided $m>m^0$ strictly.
  \end{proof}

 \subsection{Proof of Lemma 1}
\begin{proof}
Write $Z=(z_1^\top,\ldots,z_n^\top)^\top\in\mathbb R^{n\times m}$,
$\mathcal C_n=\{Z:z_j^\top z_k>0,\ j\le k\}$, and
$h(Z)=\prod_{j\le k}(z_j^\top z_k)^\alpha$.
Set $p=nm$, $b=\alpha n(n+1)/2$ and $\nu=p+2b$.
The prior density on $\mathcal C_n$ is
$\pi(Z)=Z_\pi^{-1}e^{-\kappa\|Z\|_F^2/2}h(Z)$.
By Cauchy--Schwarz and AM--GM, with $R=\|Z\|_F$,
$h(Z)\le\prod_{j=1}^n\|z_j\|^{\alpha(n+1)}\le(R^2/n)^b$.
Thus $0<Z_\pi<\infty$; positivity follows because $\mathcal C_n$
contains a neighborhood of any configuration with all $z_j$ equal and
nonzero. Gaussian radial integration also gives
\begin{equation}
 Z_\pi\le
 \left(\frac{2\pi}{\kappa}\right)^{p/2}
 \left(\frac{2}{\kappa n}\right)^b
 \frac{\Gamma(p/2+b)}{\Gamma(p/2)}
 \le\left(\frac{2\pi}{\kappa}\right)^{p/2}
       \left(\frac{\nu}{\kappa n}\right)^b,
 \label{eq:prior_normalizer_bound}
\end{equation}
using $\Gamma(a+b)/\Gamma(a)\le(a+b)^b$ for $a,b>0$.

Since $\mathcal C_n$ is a cone and $h(RU)=R^{2b}h(U)$,
polar coordinates give the exact radial density
$f_R(r)\propto r^{\nu-1}e^{-\kappa r^2/2}$, $r>0$.
Consequently $\kappa R^2\sim\chi^2_\nu$ and
$Z_\pi(\kappa)=\kappa^{-\nu/2}Z_\pi(1)$.
The moment generating function $(1-2t)^{-\nu/2}$ and Chernoff's
inequality imply $\Pi\{\kappa R^2>\nu+2\sqrt{\nu x}+2x\}\le e^{-x}$
for $x>0$.
Taking $x=\log(1/\varepsilon)$ proves the stated tail bound.

For $Z_*\in\mathcal C_n$, put
$\tau_*=\min_{j\le k}z_{*,j}^\top z_{*,k}>0$.
Continuity gives $\varepsilon_0>0$ such that every inner product is
at least $\tau_*/2$ on $B_F(Z_*,\varepsilon_0)$.
With $v_p=\pi^{p/2}/\Gamma(p/2+1)$, for $0<\varepsilon\le\varepsilon_0$,
\begin{equation}
 \Pi\{\|Z-Z_*\|_F<\varepsilon\}
 \ge Z_\pi^{-1}
 e^{-\kappa(\|Z_*\|_F+\varepsilon)^2/2}
 (\tau_*/2)^b v_p\varepsilon^p>0.
 \label{eq:prior_small_ball}
\end{equation}
Larger balls contain such a ball, proving the support assertion.
\end{proof}
\subsection{Proof of Theorem 2}
\begin{proof}
Use the notation of Theorem~\ref{thm:consistency_wellspecfied}, and write
$J=n-1$, and $D=D_{n,S}$ for brevity.
Bars denote averages over the $S$ independent subjects.
The specified product likelihood gives
$\ell_S(\lambda)=S\sum_{f\in E}
\{\bar N_f\log\lambda_f-\bar T_f\lambda_f\}$, with
$\sum_f\bar N_f=\sum_f\mu_f=J$.
The posterior is proper because
$\mathcal L_S(Z)\le\|Z\|_F^{2SJ}$ has finite prior expectation.
For $\psi(u)=u-1-\log u$, condition (C1) gives
$\mathbb E_0\{\ell_S(\lambda)-\ell_S(\lambda_0)\}
=-S\sum_f\mu_f\psi(\lambda_f/\lambda_{0,f})$,
so the population criterion uniquely targets $\lambda_0$.

Set $Q_S=S\sum_f\{(\bar N_f-\mu_f)^2+
(\lambda_{0,f}\bar T_f-\mu_f)^2\}/\mu_f$.
Each edge enters the spanning tree at most once, so
$N_f\in\{0,1\}$ and $\operatorname{Var}_0(N_f)\le\mu_f$.
Independence across subjects and (C1)--(C2) give
$\mathbb E_0Q_S\le q+v_n$.
By Markov's inequality, $\mathcal E_S=\{Q_S\le D\}$ satisfies
$P_0(\mathcal E_S^c)\le(q+v_n)/D\le1/\log(e+nS)\to0$.
On $\mathcal E_S$, the relative errors
$|\bar N_f-\mu_f|/\mu_f$ and
$|\lambda_{0,f}\bar T_f-\mu_f|/\mu_f$ are at most
$\sqrt{D/(S\mu_{\min})}=\sqrt{\gamma_{n,S}}\to0$ uniformly in $f$.
Hence, eventually $\bar N_f\ge\mu_f/2$,
$\widehat\lambda_f:=\bar N_f/\bar T_f
\in[\lambda_{0,f}/2,2\lambda_{0,f}]$, and
$\sum_f\lambda_{0,f}\bar T_f\le2J$.
The unrestricted maximizer $\widehat\lambda$ gives the exact identity
\begin{equation}
 \ell_S(\lambda)-\ell_S(\widehat\lambda)=-SW_S(\lambda),\qquad
 W_S(\lambda)=\sum_f\bar N_f\psi(\lambda_f/\widehat\lambda_f).
 \label{eq:likelihood_contrast}
\end{equation}
Since $\psi(u)\lesssim(u-1)^2$ near $1$, there is a fixed
$C_0>0$ such that, on $\mathcal E_S$ and for sufficiently large $S$,
$SW_S(\lambda_0)\le C_0Q_S\le C_0D$.

Take $\delta=c(\tau\wedge1)/\{SJ(1+M_0)\}$ and
$B=\{Z:\|Z-Z_0\|_F<\delta\}$, with a sufficiently small fixed $c>0$.
The bound
$|z_j^\top z_k-z_{0,j}^\top z_{0,k}|
 \le2M_0\delta+\delta^2$
shows that all inner products on $B$ are at least $\tau/2$ and
$\|\log\lambda-\log\lambda_0\|_\infty\le(10SJ)^{-1}$.
Combining
\eqref{eq:prior_normalizer_bound}--\eqref{eq:prior_small_ball}
with the chosen $\delta$ gives $\Pi(B)\ge e^{-C_\kappa D}$.
Indeed, $p\le n^2$ implies
$-\log v_p\lesssim p\log(e+nS)$, and the remaining radius and
prior-normalizer terms are controlled by the corresponding terms in $D$.
Here $C_\kappa$ is independent of $n,S$ for fixed $\kappa$.
On $\mathcal E_S$, the log-rate bound on $B$ gives
$\ell_S(\lambda)-\ell_S(\lambda_0)\ge-1$ throughout $B$.
Consequently, on $\mathcal E_S$,
\[
 \int e^{\ell_S(\lambda)-\ell_S(\widehat\lambda)}\,d\Pi
 \ge e^{-SW_S(\lambda_0)-1}\Pi(B)\ge e^{-C_1D}.
\]
For a fixed $B_1>C_1+2$, equation~\eqref{eq:likelihood_contrast}
therefore gives
$\Pi_S\{W_S(\lambda)>B_1D/S\mid Y_S\}
\le e^{-(B_1-C_1)D}\le e^{-2D}$ on $\mathcal E_S$. Here $C_1=C_0+C_\kappa+1$, using $D\ge1$ for sufficiently large $S$;
thus $C_1$ is independent of $n,S$ for fixed $\kappa$.

For the required distance, on $\mathcal E_S$
for sufficiently large $S$, so that
$\max_f\widehat\lambda_f\le2\bar\lambda_0$ and
$\min_f\bar N_f\ge\mu_{\min}/2$.
Set $\Delta_f=\sqrt{\lambda_f}-\sqrt{\widehat\lambda_f}$ and
$H_\lambda=\sum_f\Delta_f^2$.
Since $\psi(u)\ge(\sqrt u-1)^2$,
$H_\lambda\le
(\max_f\widehat\lambda_f/\min_f\bar N_f)W_S(\lambda)
\le(4\bar\lambda_0/\mu_{\min})W_S(\lambda)$.
The identity
$\lambda_f-\widehat\lambda_f
=2\sqrt{\widehat\lambda_f}\Delta_f+\Delta_f^2$,
together with $\sum_f\Delta_f^4\le H_\lambda^2$, gives
$\|\lambda-\widehat\lambda\|_2^2
\le8(\max_f\widehat\lambda_f)H_\lambda+2H_\lambda^2
\le16\bar\lambda_0H_\lambda+2H_\lambda^2$.
Thus, for any fixed $K>0$, the condition $W_S(\lambda)\le KD/S$
implies $H_\lambda\le4K\bar\lambda_0\gamma_{n,S}$ and hence
\[
 \|\lambda-\widehat\lambda\|_2^2
 \le64K\bar\lambda_0^2\gamma_{n,S}
      +32K^2\bar\lambda_0^2\gamma_{n,S}^2.
\]
Take $K=\max\{C_0,B_1\}$.
Since $W_S(\lambda_0)\le C_0D/S$, this bound applies both to
$\lambda_0$ and uniformly to every $\lambda$ satisfying
$W_S(\lambda)\le B_1D/S$.
The squared triangle inequality then gives
\[
 \begin{aligned}
 \|\textup{LTri}(\Lambda-\Lambda_0)\|_2^2
 &\le2\|\lambda-\widehat\lambda\|_2^2
      +2\|\widehat\lambda-\lambda_0\|_2^2\\
 &\le C\bar\lambda_0^2(\gamma_{n,S}+\gamma_{n,S}^2)
 \le Mr_{n,S}^2
 \end{aligned}
\]
eventually, since $\gamma_{n,S}\to0$ and
$r_{n,S}^2=\bar\lambda_0^2\gamma_{n,S}$.

Consequently, on $\mathcal E_S$, the posterior probability of
$\|\textup{LTri}(\Lambda-\Lambda_0)\|_2^2>Mr_{n,S}^2$
is bounded by the posterior probability of
$W_S(\lambda)>B_1D/S$, which is at most $e^{-2D}$.
Bounding the posterior probability by $1$ on $\mathcal E_S^c$ gives
\[
 \mathbb E_0\Pi_S\!\left(
 \|\textup{LTri}(\Lambda-\Lambda_0)\|_2^2>Mr_{n,S}^2
 \,\middle|\,Y_S\right)
 \le P_0(\mathcal E_S^c)+e^{-2D}
 \le\frac{q+v_n}{D}+e^{-2D}
 \longrightarrow0,
\]
since $(q+v_n)/D\le1/\log(e+nS)\to0$ and
$D\ge p\log(e+nS)\to\infty$.
If $r_{n,S}^2\to0$, this proves the stated consistency.
\end{proof}

 \subsection{Proof of Theorem 3}
\begin{proof}
  Let $\bar\eta=\eta/\sum_{e\in E}\eta_e$.
  By (iii) and the restriction map $R$, the true law of the restricted ordering is
  $p^{\mathcal C}_{\bar\eta}:=R[\mathrm{PL}(\bar\eta)]$ on the finite set $\Omega_{\mathcal C}$.
  By (iv), the model induces $p^{\mathcal C}_{\bar\theta(\Lambda)}=R[\mathrm{PL}(\bar\theta(\Lambda))]$.

  Because $\Omega_{\mathcal C}$ is finite, the map $\bar\theta\mapsto p^{\mathcal C}_{\bar\theta}$ is continuous,
  and by (iv) the parameterization is identifiable on the simplex. By (v) there exists $\Lambda^\dagger$ with positive prior support and  $\bar\theta(\Lambda^\dagger)=\bar\eta$.
  Since the posterior for $\Lambda$ concentrates in probability on the set of KL minimizers,
  for every neighborhood $U$ of $\bar\eta$,
  $
  \Pi\!\big(\bar\theta(\Lambda)\in U \,\big|\, R_{1:S}\big)\;\xrightarrow[S\to\infty]{P}\;1.
  $
  Consequently, for any posterior draw $\widehat\Lambda_S$,
$
  \left\|\,p^{\mathcal C}_{\bar\theta(\widehat\Lambda_S)}-p^{\mathcal C}_{\bar\eta}\,\right\|_{\mathrm{TV}}
  \;\xrightarrow[S\to\infty]{P}\;0.
 $
  Since $\iota_v=\mathbf 1\{R\in\mathcal A_v\}$ almost surely,
  $
  \mathbb{P}_{Q_{\widehat\Lambda_S}}(\iota_v=1)
  =\sum_{r\in\mathcal A_v} p^{\mathcal C}_{\bar\theta(\widehat\Lambda_S)}(r),
  \qquad
  p_{F^0}(v)=\sum_{r\in\mathcal A_v} p^{\mathcal C}_{\bar\eta}(r).
  $
  Thus
  \(
  \Big|\mathbb{P}_{Q_{\widehat\Lambda_S}}(\iota_v=1)-p_{F^0}(v)\Big|
  & =\Big|\sum_{r\in\mathcal A_v}\!\big(p^{\mathcal C}_{\bar\theta(\widehat\Lambda_S)}(r)-p^{\mathcal C}_{\bar\eta}(r)\big)\Big|\\
  & \le \sum_{r\in\Omega_{\mathcal C}}\!\big|p^{\mathcal C}_{\bar\theta(\widehat\Lambda_S)}(r)-p^{\mathcal C}_{\bar\eta}(r)\big| \\
  & = 2\,\left\|p^{\mathcal C}_{\bar\theta(\widehat\Lambda_S)}-p^{\mathcal C}_{\bar\eta}\right\|_{\mathrm{TV}},
  \)
  which converges to $0$ in probability.
  \end{proof}

\section{Alternative edge-finding algorithms specific to the proposed model}
In our implementation, we can take shortcut that bypasses the need for Gaussian elimination. Since we start with the persistence barcode generated by the high-performance persistent homology software GHUDI \citep{gudhi:urm}, the additional cost of finding the edges is low. For the zero-dimensional bars, the spanning tree edges have been sorted according to the lengths, hence
 we further find the admissible edges via simple thresholding with an $\mathcal  O(n_0)$ cost using vectorized matrix operation. For the one-dimensional bars, we find the edge set by taking the set differences of two connected components. Since the connected components are given in the homology software output, the additional edge-finding cost for each one-dimensional bar is close to constant, leading to a total cost $\mathcal O(n_1)$.

\vspace{-0.5cm}
\section{Additional details for the data application}

\begin{figure}[H]
  \centering
  \begin{subfigure}[t]{0.3\textwidth}
      \centering
      \includegraphics[width=\textwidth]{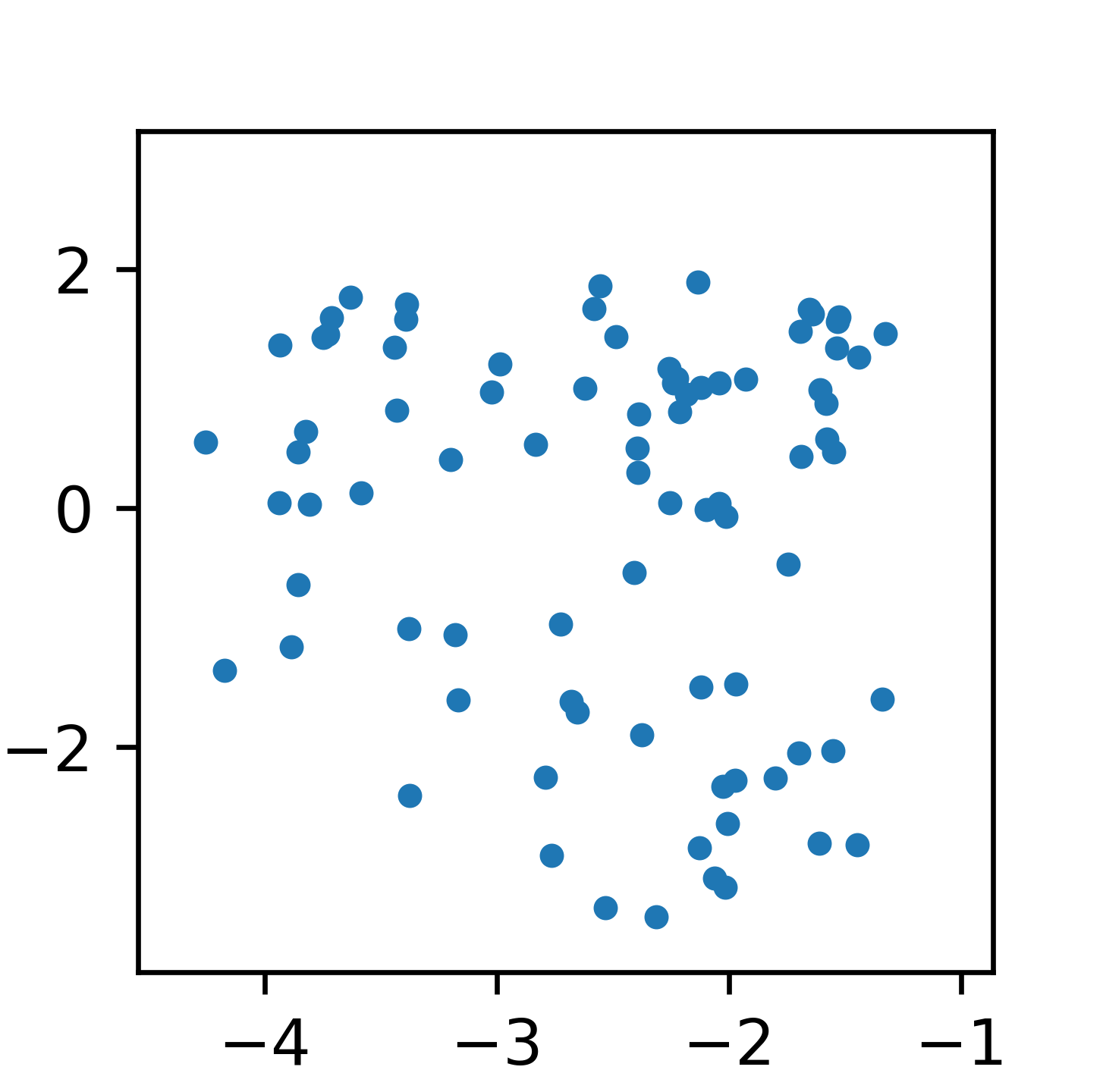}
      \caption{Latent positions for the cognitively normal group}
  \end{subfigure}
  \begin{subfigure}[t]{0.3\textwidth}
    \centering
    \includegraphics[width=\textwidth]{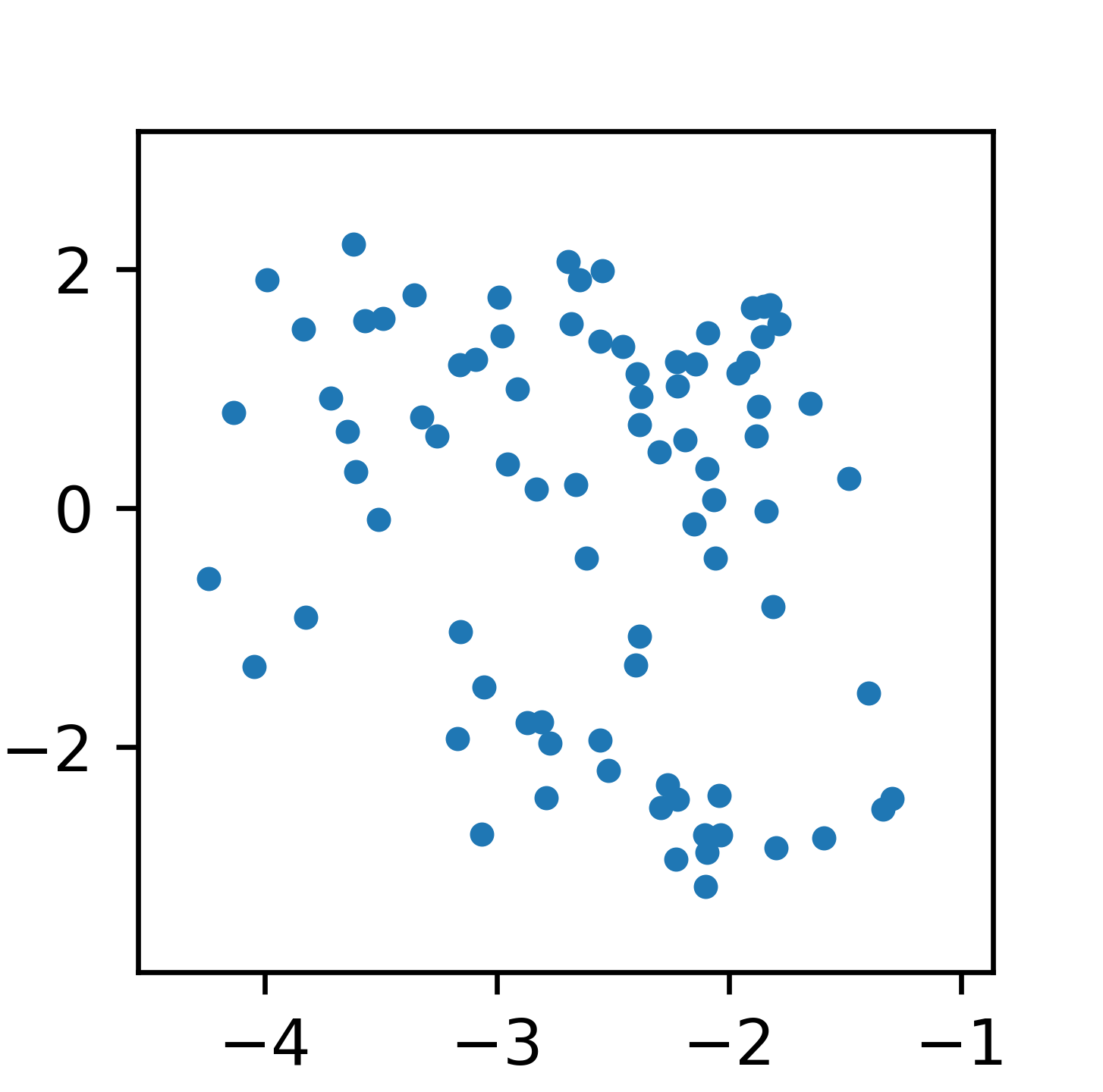}
    \caption{Latent positions for the mildly cognitively impaired group}
\end{subfigure}
\begin{subfigure}[t]{0.3\textwidth}
  \centering
\includegraphics[width=\textwidth]{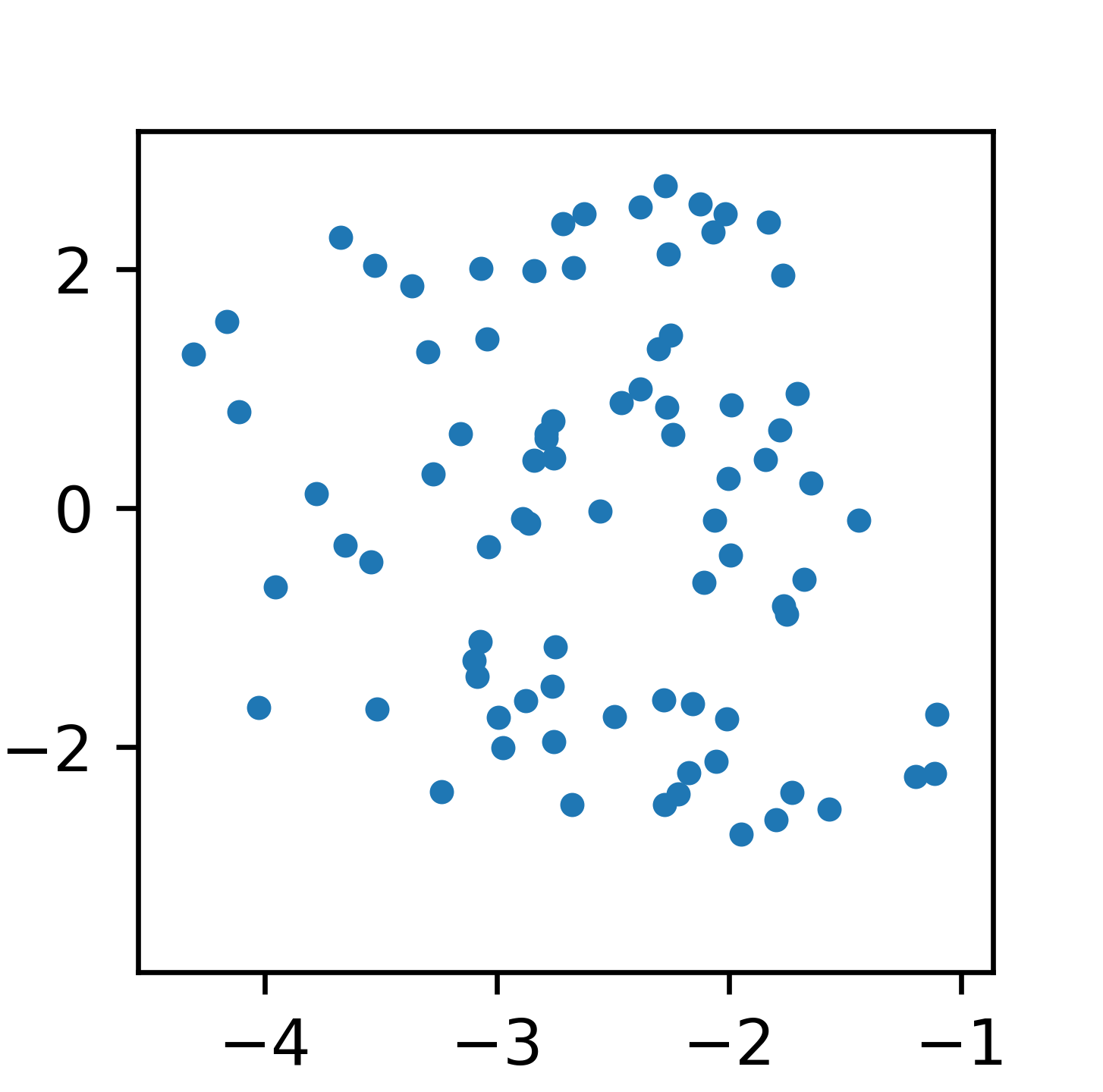}
  \caption{Latent positions for the Alzheimer's disease group}
\end{subfigure}
\caption{Visualization of the latent positions for the three groups in two dimensions. The posterior mean of $\Lambda$ is first calculated, then a low rank approximation is obtained using truncated SVD at rank 2.
}
  \label{fig:adni_latent_position}
\end{figure}

\begin{figure}[H]
  \centering
  \begin{subfigure}[b]{0.32\textwidth}
      \centering
      \includegraphics[width=\textwidth]{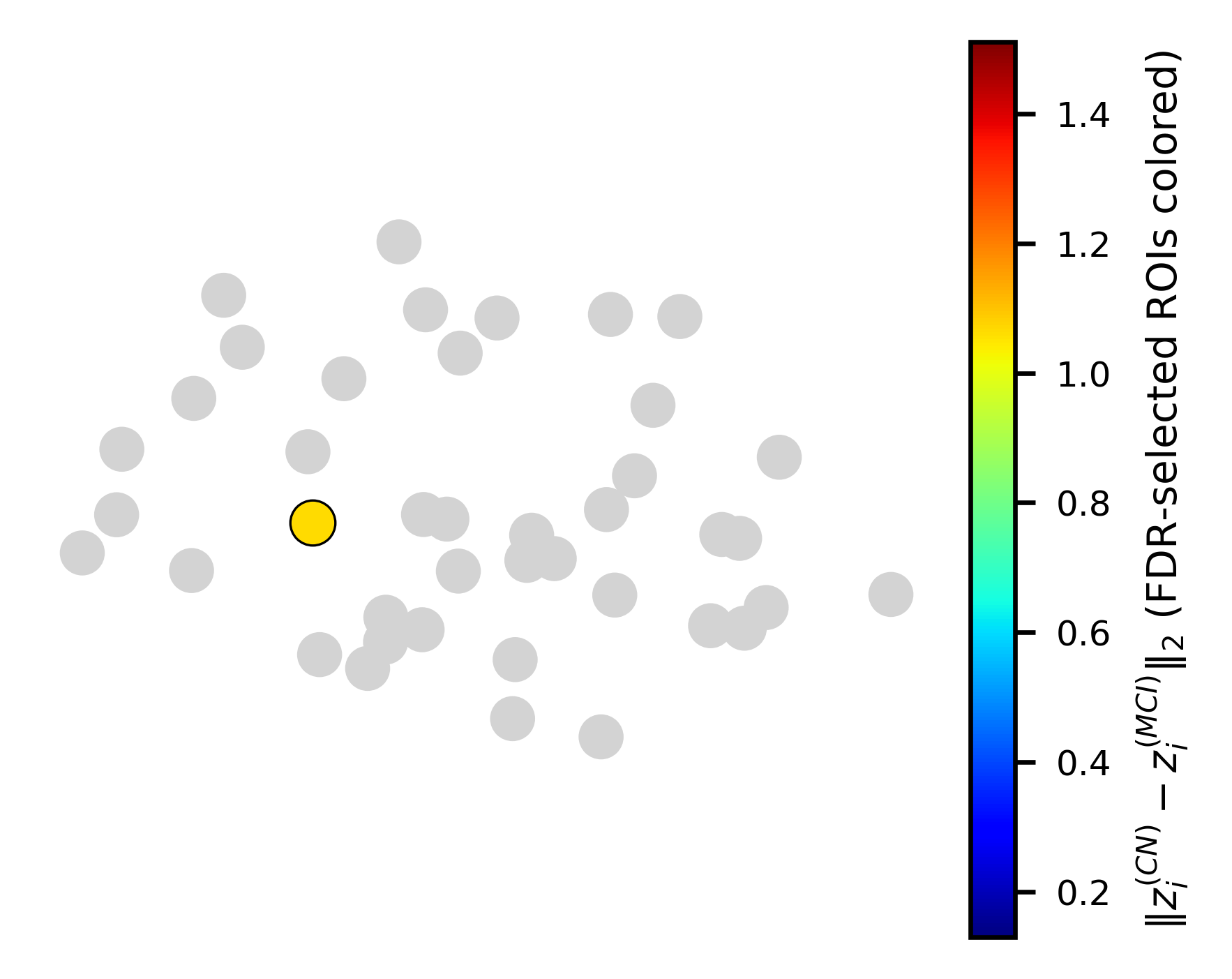}
      \caption{Left lateral, CN vs MCI}
    \end{subfigure}
  \hfill
  \begin{subfigure}[b]{0.32\textwidth}
      \centering
      \includegraphics[width=\textwidth]{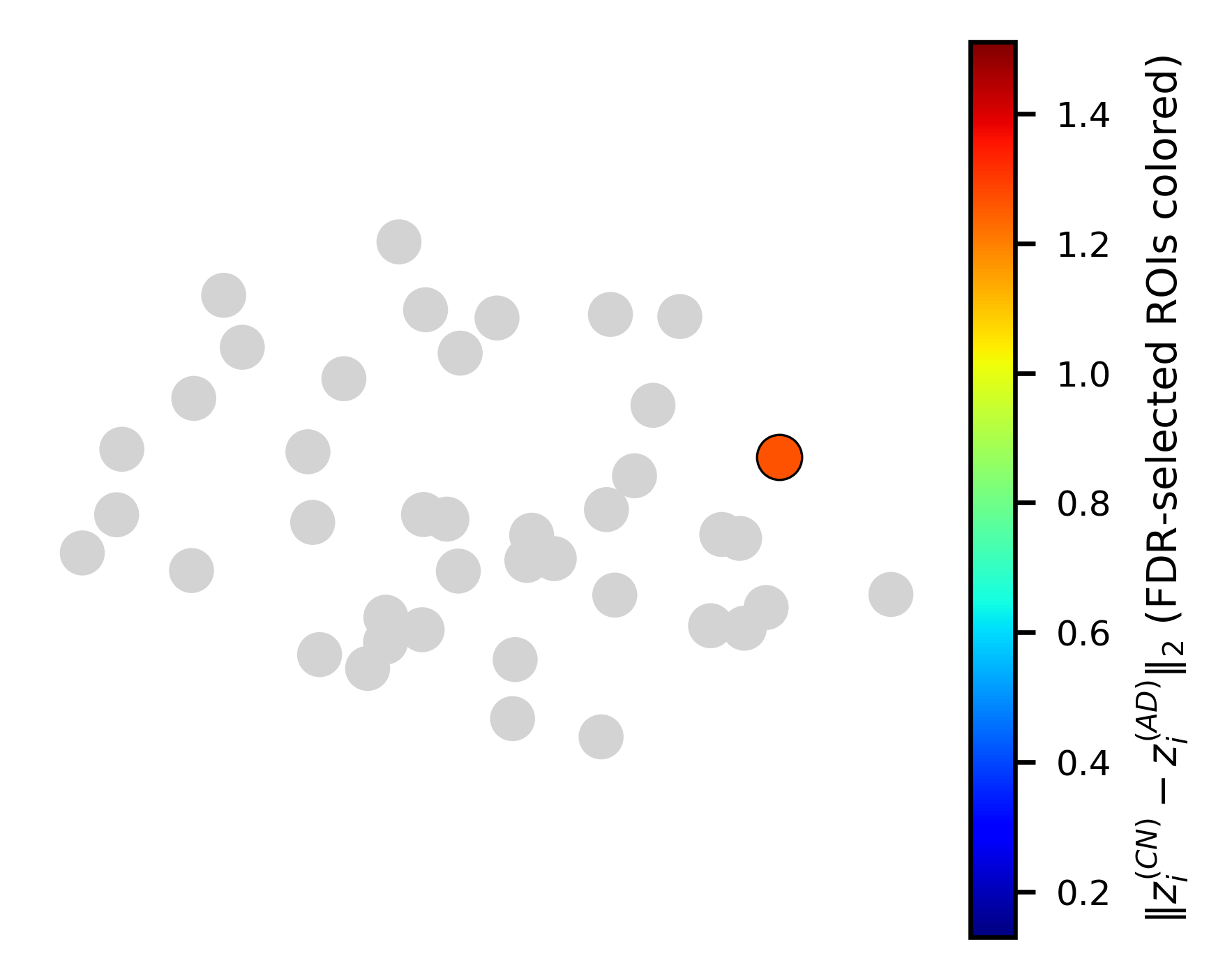}
      \caption{Left lateral, CN vs AD}
  \end{subfigure}
  \begin{subfigure}[b]{0.32\textwidth}
    \centering
    \includegraphics[width=\textwidth]{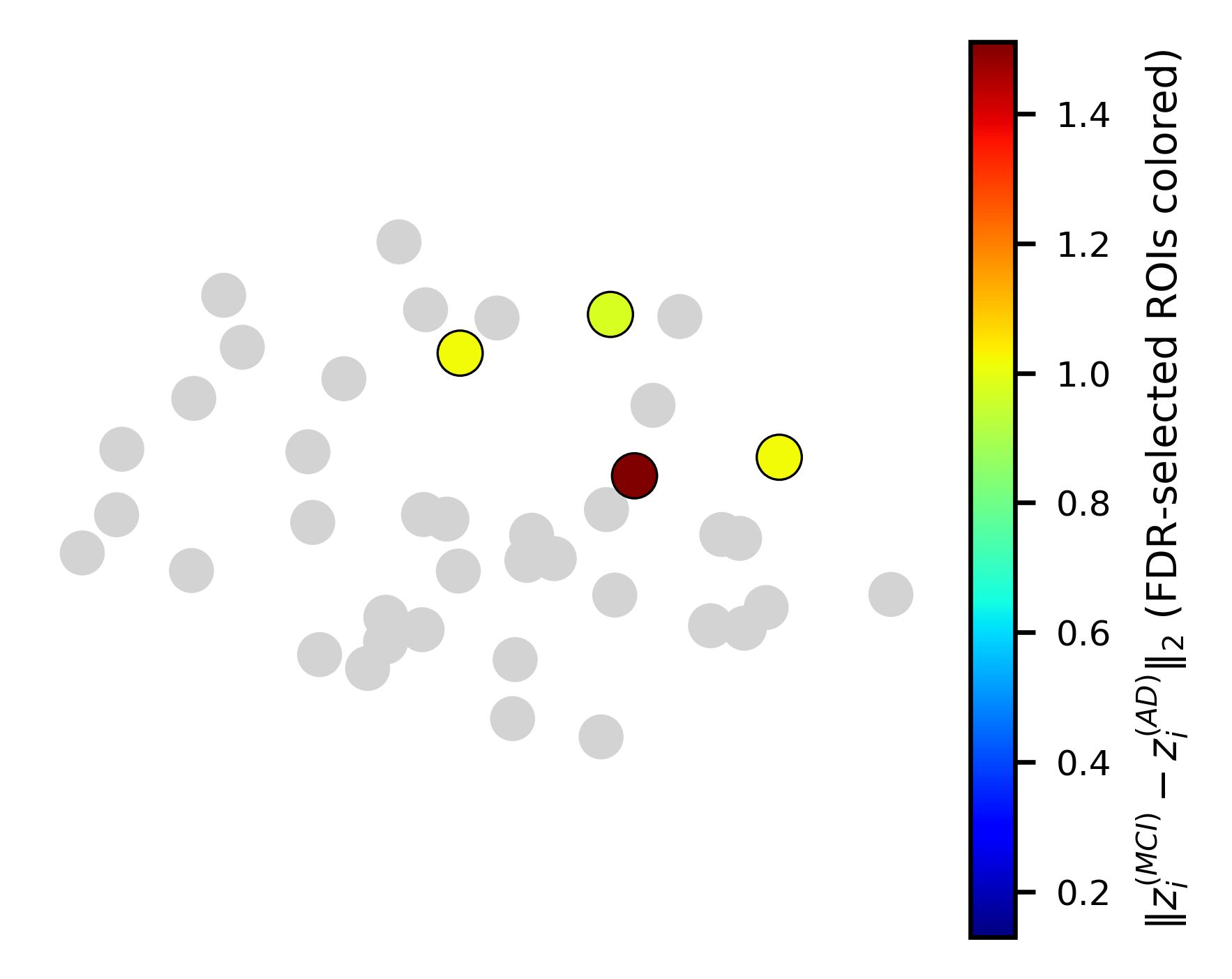}
    \caption{Left lateral, MCI vs. AD}
\end{subfigure}
\begin{subfigure}[b]{0.32\textwidth}
  \centering
  \includegraphics[width=\textwidth]{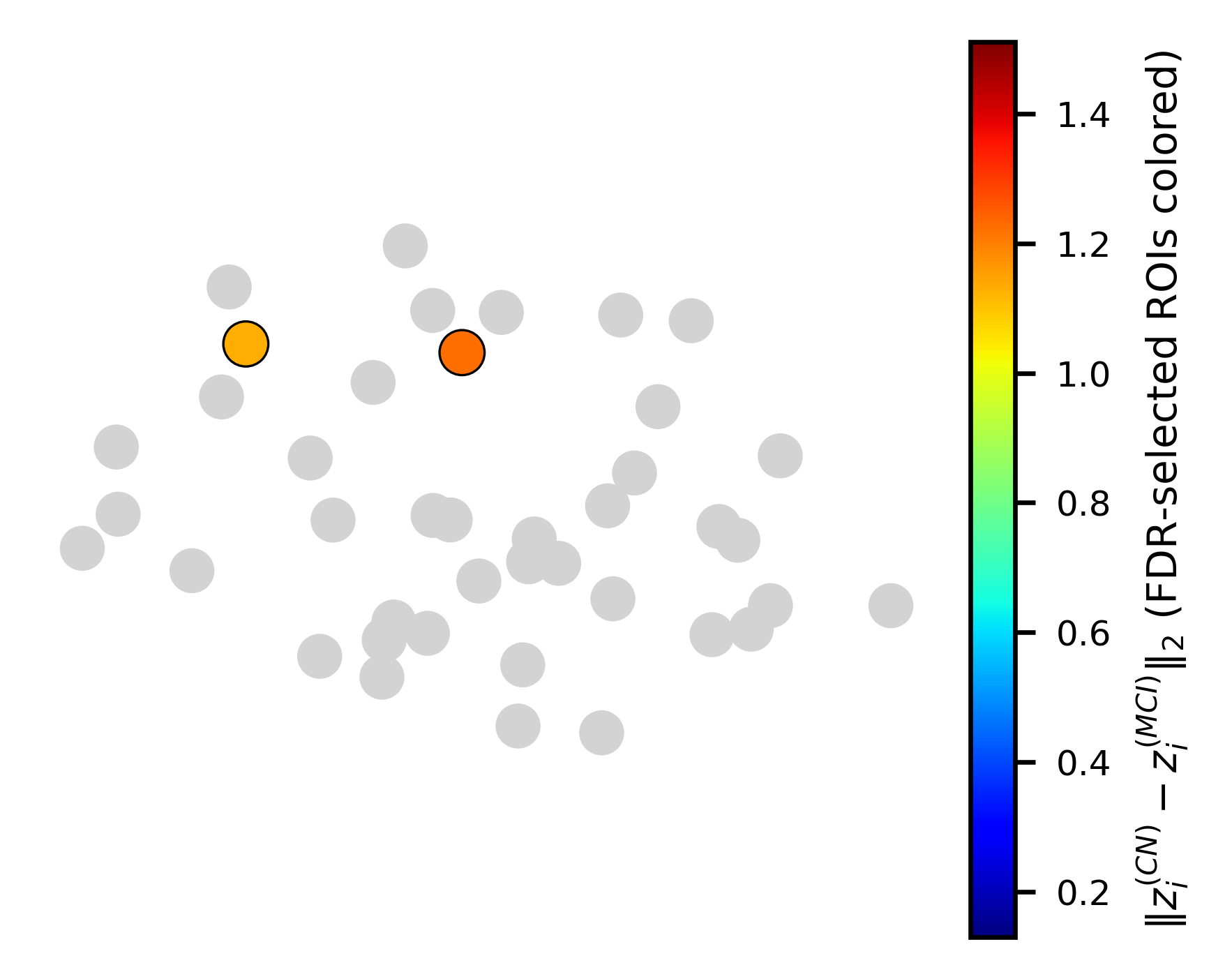}
  \caption{Right lateral, CN vs MCI}
\end{subfigure}
\begin{subfigure}[b]{0.32\textwidth}
  \centering
  \includegraphics[width=\textwidth]{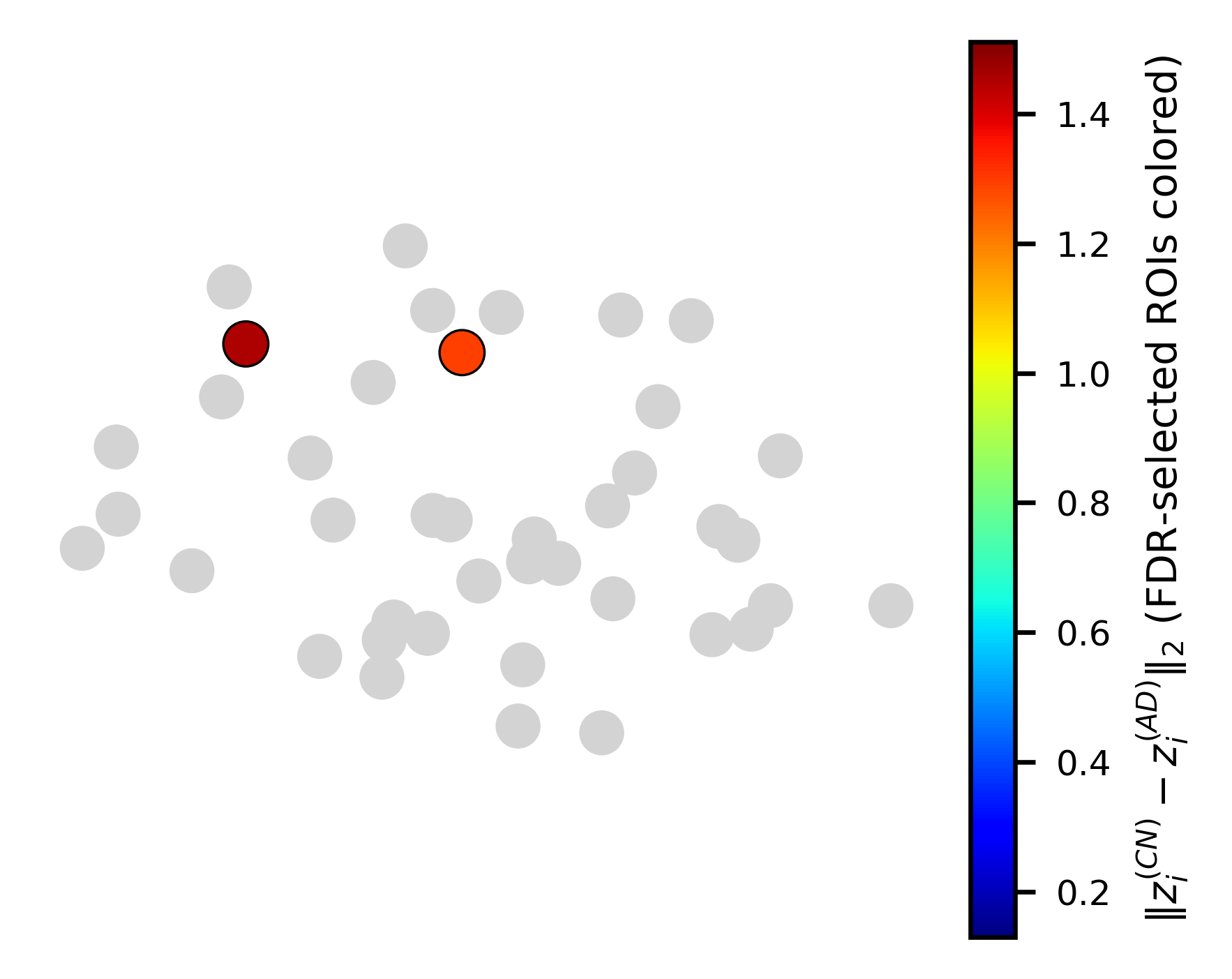}
  \caption{Right lateral, CN vs AD}
\end{subfigure}
\begin{subfigure}[b]{0.32\textwidth}
  \centering
  \includegraphics[width=\textwidth]{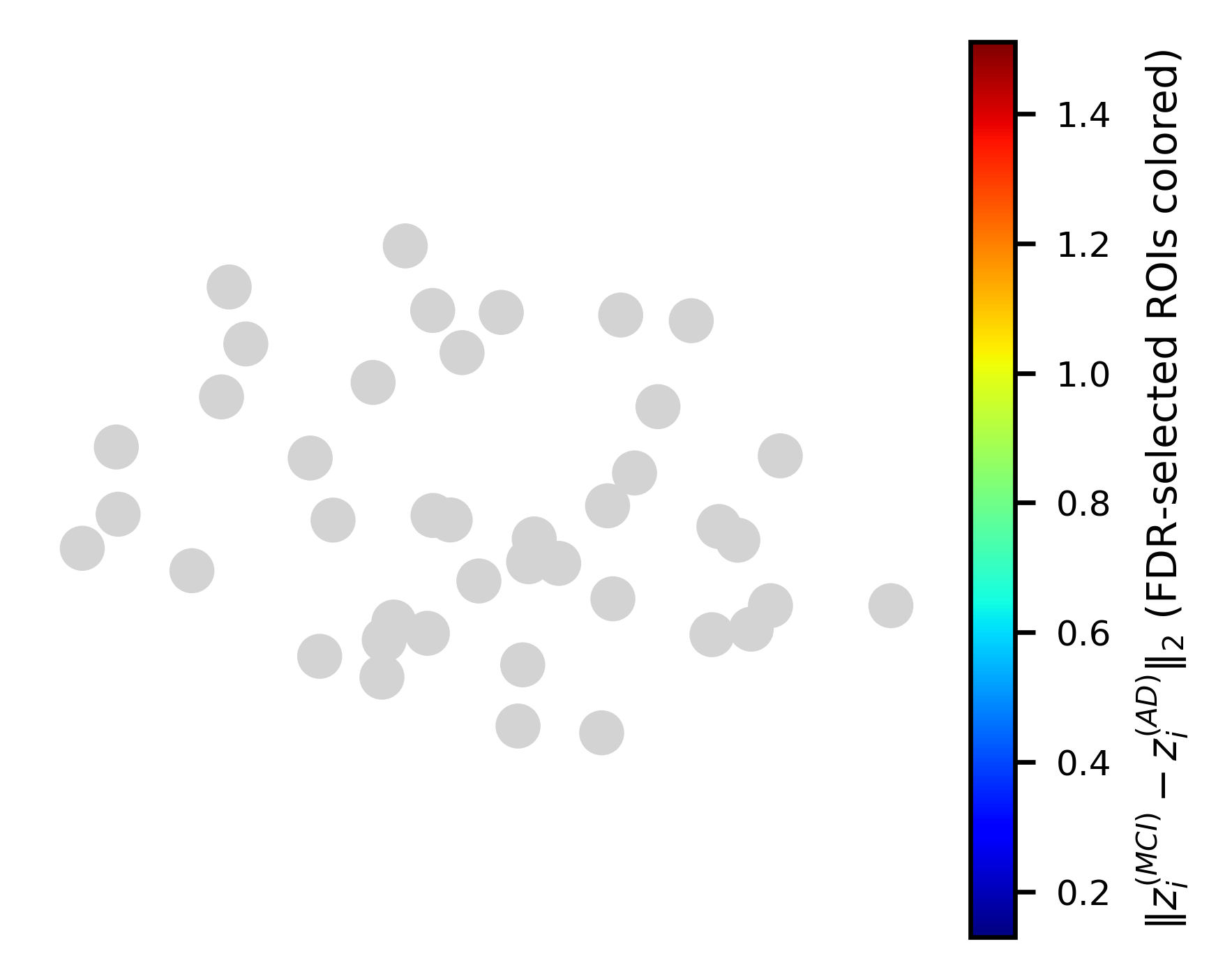}
  \caption{Right lateral, MCI vs AD}
\end{subfigure}
\caption{Lateral views of ROI-wise differences in the latent positions.
Left-lateral and right-lateral brain maps color the magnitude of the latent displacement $\|z_i^{(p)} - z_i^{(q)}\|_2$ between groups. The same color scale is used across panels.}
  \label{fig:adni_lat_diffs}
\end{figure}

 \section{Computational performance for simulation with both zero-dimensional and one-dimensional features}

\begin{figure}[H]
  \centering
  \includegraphics[width=0.5\textwidth]{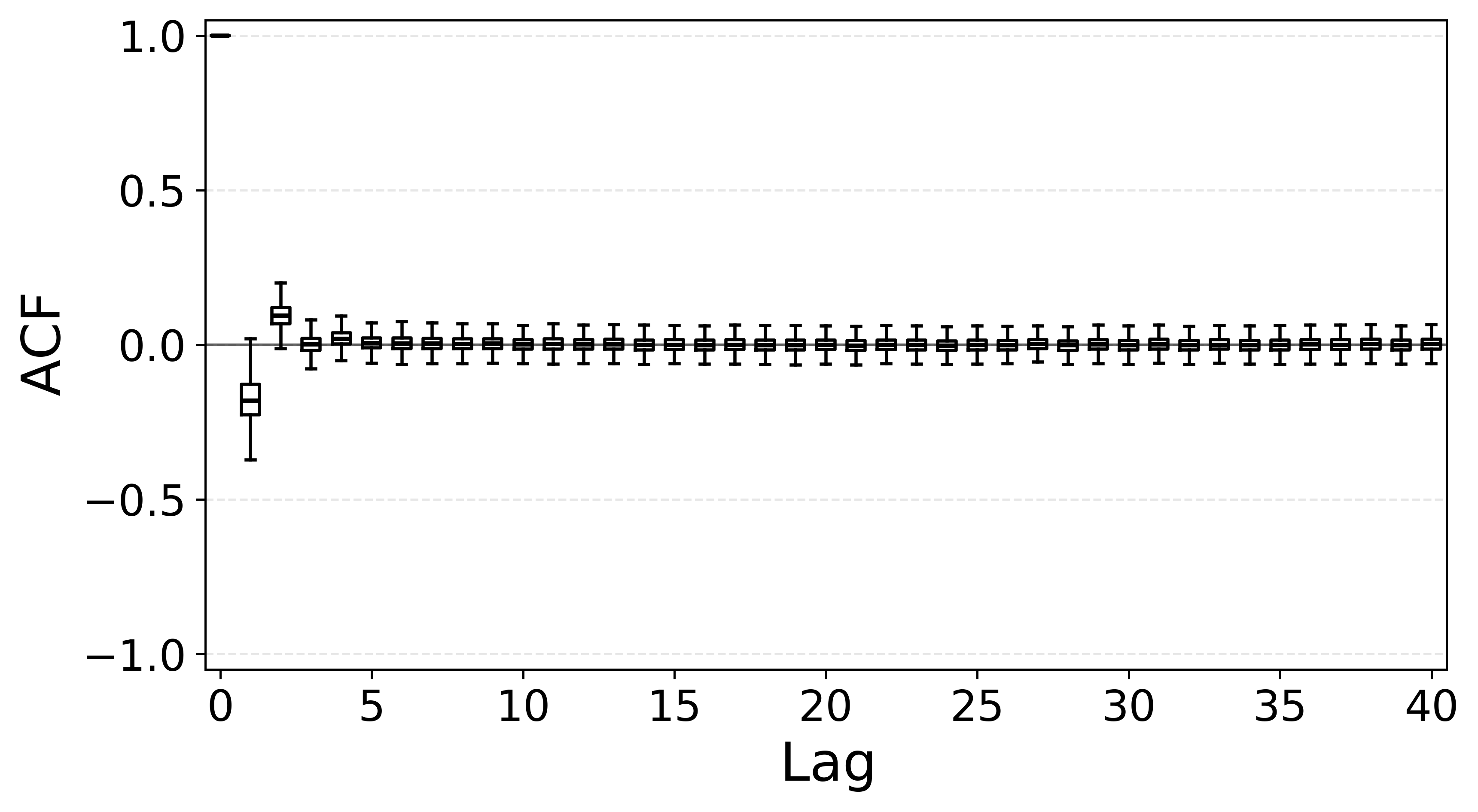}
        \caption{The boxplots of the autocorrelation function for all elements of $\Lambda$.}
  \label{fig:Lambda_acf_boxplot_lag40_h0h1}
\end{figure}

 We conduct simulations involving both zero-dimensional and one-dimensional features. We first generate oracle points $n=150$ on two circles, both centered at origin in $\mathbb{R}^2$ but with radii 1 and 2, respectively. We then generate 10 subjects of data by adding independent noise from $\text{N}(0,0.05^2 I)$ to each point. We obtain the persistence barcode, and then fit our model to the data.  Figure~\ref{fig:Lambda_acf_boxplot_lag40_h0h1} shows the autocorrelation function for all elements of $\Lambda$. The computational performance is comparable to the one we demonstrate in the main text.

\section{Determining the embedding dimension for Laplacian Eigenmaps}
\label{sec:supp-embedding}

\begin{figure}[H]
  \centering
  \begin{subfigure}[t]{0.45\textwidth}
    \centering
    \includegraphics[width=\textwidth]{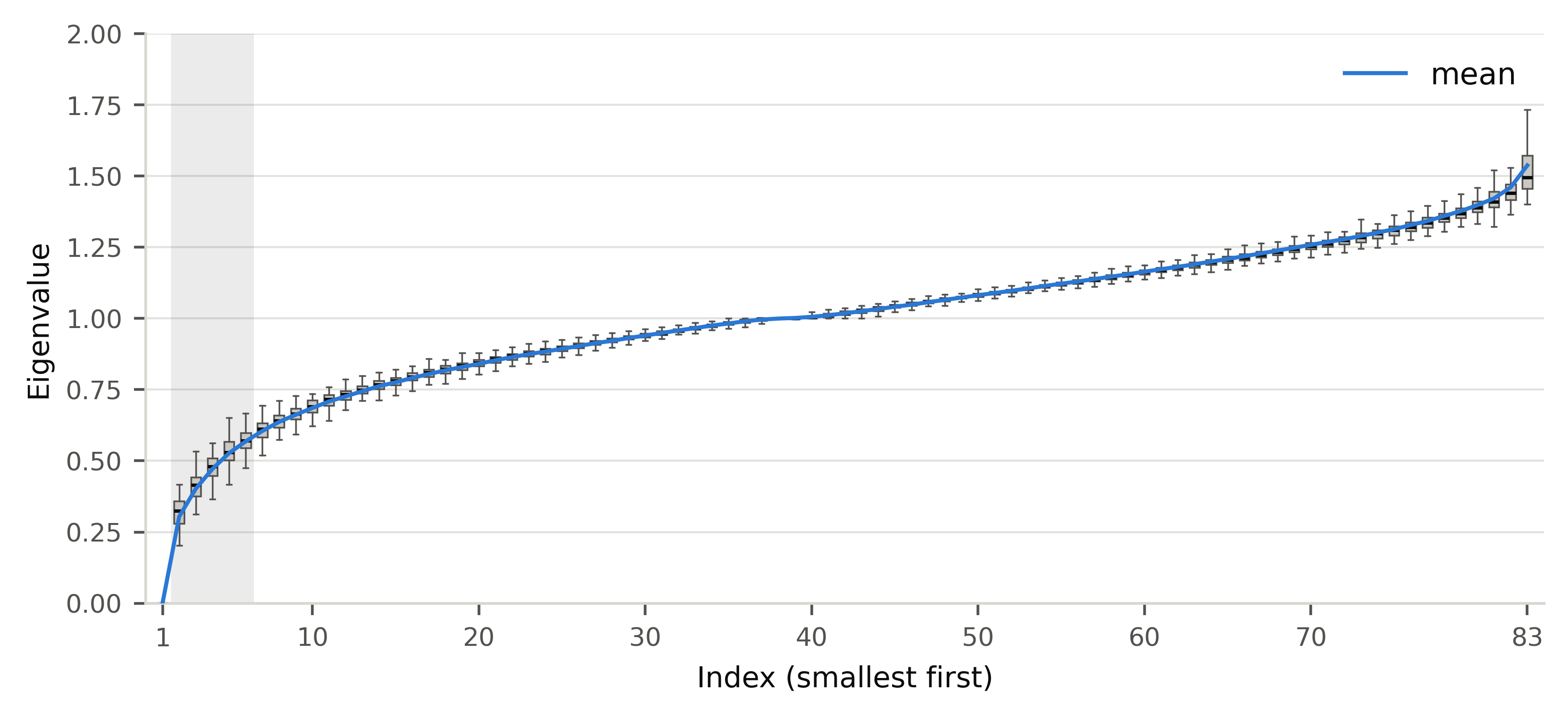}
    \caption{Boxplot of all $83$ eigenvalues for 104 subjects. Index 1 is the trivial zero eigenvalue. The shaded band marks the five eigenvalues used for the $k = 5$ embedding.}
    \label{fig:scree_box}
  \end{subfigure}
  \vspace{1em}
  \begin{subfigure}[t]{0.45\textwidth}
    \centering
    \includegraphics[width=\textwidth]{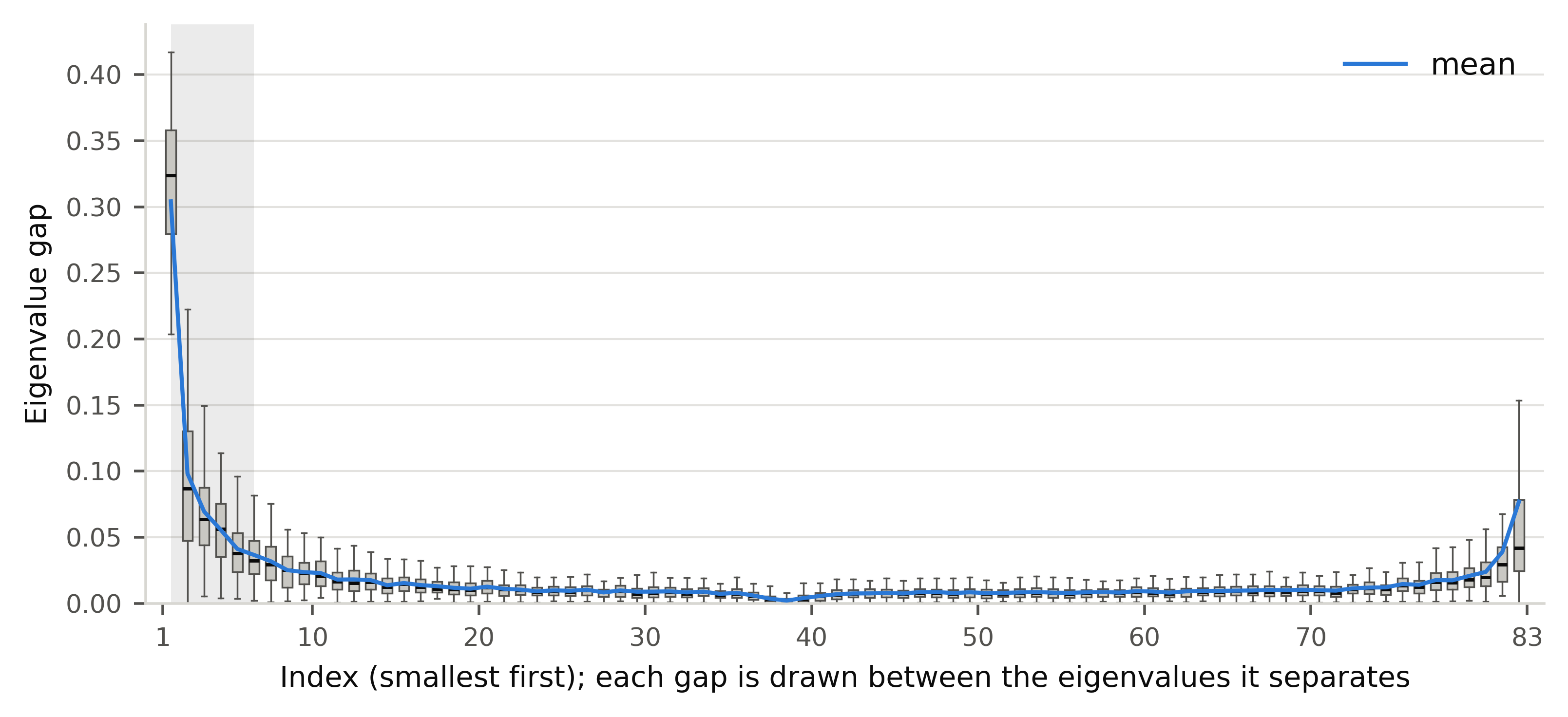}
    \caption{Boxplot of the $82$ gaps between consecutive eigenvalues for 104 subjects.}
    \label{fig:scree_gap}
  \end{subfigure}
  \caption{(a) Spectrum of normalized Laplacian eigenvalues across all subjects. (b) Corresponding consecutive eigenvalue gaps. In both panels, the shaded band indicates the region of the five eigenvalues used for $k=5$ embedding.}
  \label{fig:scree_combined}
\end{figure}

The Laplacian eigenmaps embedding of symmetric matrix $W$ uses the eigenvectors of the normalized Laplacian $\mathcal N(W) = I - D^{-1/2}WD^{-1/2}$ belonging to its $k$ smallest eigenvalues, where $D$ is the diagonal matrix of degrees of $W$. The close-to-zero eigenvalues are associated with the meaningful low-dimensional structure, such as connected components, of the data. Most of the noisy eigenvalues are concentrated near and above one. Figure~\ref{fig:scree_box} shows the complete spectrum of $\mathcal N(W)$ for all subjects as boxplots across subjects. Figure~\ref{fig:scree_gap} shows the gaps between consecutive eigenvalues, which are used to determine the embedding dimension $k$. The gaps are relatively large until the 5th eigenvalue, which lead us to choose $k = 5$ for the embedding. We repeat the experiments described in the main text using $k=8$ as well, such as the localization performance and the AUC scores, to be almost indistinguishable from the results using $k=5$.

\section{Comparison with alternative models}
\subsection{Comparison with competing-Weibull likelihood}
\label{sec:weibull}
The exponential form in \eqref{eq:competing_exponentials}
and~\eqref{eq:triangulation_likelihood} is a working choice, and this section
asks whether it drives the conclusions. We refit the same data under a strictly
larger family: a competing-Weibull model in which each edge length carries a
shared shape parameter $\xi > 0$, with $\xi = 1$ recovering the
competing-exponential model.

The two models share an identical pipeline: the same precomputed admissible sets,
minimum-spanning-tree winners, loop vertices, and triangulating edge sets. The
observed times $d^0_i$, $b^1_i$, and $d^1_i$ enter the exponential likelihood
directly and enter the Weibull likelihood through their $\xi$-th powers. This is
the only difference between the two models. Both are fitted by maximum likelihood,
with $\xi$ estimated by profile likelihood.

We evaluate both models in three settings. In Setting~A, points are
drawn near one or two Gaussian centers in $\mathbb{R}^2$, and the groups differ
in how many points are assigned to each center. Its filtration produces
connected components but no loops, so Setting~A only involves the zero-dimensional
likelihood~\eqref{eq:competing_exponentials}. In Setting~B, points are drawn
around a single circle and one contiguous arc of them is displaced in the
second group. In Setting~C, points are drawn near two concentric circles and an
arc of the outer circle is displaced. Both B and C produce loops alongside
components, so they involve the one-dimensional
likelihood~\eqref{eq:triangulation_likelihood}. Each setting uses ten subjects
per group and is replicated over 50 Monte Carlo datasets at separation levels
$\Delta = 1$ and $\Delta = 2$.

Changing the distribution could move either of the two things the paper reports,
and we track both. The first is the \emph{grouping}: each subject is assigned to
the group under whose fitted $\Lambda^p$ its barcode is most likely, and we
record whether the two models assign every subject identically. The second is
the \emph{localization}, scored as in Table~\ref{tab:localization} of the main
text: for a contrast between groups $p$ and $p'$ we rank the $n$ vertices by the
per-vertex latent difference $\hat\delta_i = \lVert z^{(p)}_i - z^{(p')}_i
\rVert_2$ and take the area under the curve against the set $G$ of vertices
whose oracle position differs between the two groups. The estimator of
$\hat\delta_i$ is the one difference: there it is the posterior median, whereas
both fits here are maximum likelihood, so $\hat\delta_i$ is evaluated at the
fitted coordinates. The absolute level of the AUC is therefore not comparable
across the two tables, but that is not what we need: both models here are ranked
by the same estimator, and what matters is whether they agree with each other,
so we report the gap
$|\Delta\text{AUC}|$ between them on the same dataset. Both quantities are
computed for every pair of groups within a dataset, giving 300 contrasts in
Setting~A, which has three groups, and 100 in each of Settings~B and~C.

The comparison is informative only if the two models genuinely differ, and they
do. The likelihood ratio test rejects $H_0 \colon \xi = 1$ in every replicate of
every setting: the simulated bar lengths are not exponential, and the extra
shape parameter is doing real work. The two are substantively different fits
rather than two routes to the same one. We read $\hat\xi$ only as bounded away
from one, since its magnitude is sensitive to how $\Lambda$ is regularized.

\begin{table}[!ht]
\centering
\caption{Agreement between the competing-exponential and competing-Weibull
likelihoods. Entries are medians over group-pair contrasts, with interquartile
ranges in brackets; the quantities are defined in the text.}
\label{tab:weibull-battery}
\begin{tabular}{lccc}
\hline
 & localization gap & & identical \\
Setting & $|\Delta$AUC$|$ & $\hat\xi$ & grouping \\
\hline
A: Gaussian centers ($H_0$)
  & 0.036 [0.016, 0.065] & 2.94 & 100\% \\
B: Circle ($H_0{+}H_1$)
  & 0.171 [0.065, 0.302] & 3.29 & 100\% \\
C: Two circles ($H_0{+}H_1$)
  & 0.087 [0.038, 0.151] & 5.10 & 100\% \\
\hline
\end{tabular}
\end{table}

What the change of distribution does not move is the inference, as
Table~\ref{tab:weibull-battery} shows. The grouping is identical everywhere: in
all three settings, in every one of the 500 contrasts, both models place every
subject in the same group.

The localization agrees closely in the setting the paper relies on. In
Setting~A, the design behind Section~\ref{sec:sim}, the median gap is $0.036$
with interquartile range $[0.016, 0.065]$, which on the scale of an AUC leaves
the ordering of the vertices effectively unchanged. The gaps widen where loops
contribute, to $0.171$ in Setting~B and $0.087$ in Setting~C. In both, the
signed difference has an interquartile range spanning zero, so neither model
localizes better than the other; the wider gaps reflect variability from
replicate to replicate rather than an advantage for either.

The claim we draw is therefore not that the simulated barcodes follow the
competing-exponential likelihood, which the likelihood ratio test rules out, but
that the quantities the paper reports are insensitive to replacing it with a
strictly larger family that fits the same data better.

\subsection{Comparison with full generative models}
\label{sec:generative}

Equation~\eqref{eq:generative} factorizes the likelihood of the observed point cloud into the conditional distribution $G$ of the cloud given its persistent homology features, and the persistent homology likelihood $L$. Our inference is based on $L$ only, a partial likelihood. We now assess this reduction: the invariance it gains and the information it discards.

We retain the three-group design of Section~\ref{sec:sim} and apply to each subject's point cloud a random orthonormal transformation, which we call the nuisance transformation.
For subject $s$, let $\bar y^s = n^{-1}\sum_{i=1}^n \tilde y^s_i$ be the centroid of the cloud, and let $U_s$ be a $2\times 2$ orthonormal matrix, $U_s^\top U_s = I$, drawn independently of the group label and of the cloud.
The transformed points are
\(
    \tilde y^s_i(U_s) = \bar y^s + U_s(\tilde y^s_i - \bar y^s), \qquad i = 1,\dots,n.
\)
The transformation keeps the centroid of the cloud fixed at $\bar y^s$, so that it carries no information about the group label.

We consider three distribution for $U_s$.
In the sign-flip regime, $U_s = \mathrm{diag}(\pm 1, \pm 1)$ with independent, equiprobable signs, so each coordinate is multiplied by $-1$ with probability one half.
In the rotation regime, $U_s = \bigl(\begin{smallmatrix} \cos\omega_s & -\sin\omega_s \\ \sin\omega_s & \cos\omega_s \end{smallmatrix}\bigr)$ with $\omega_s$ uniform on $(-\pi, \pi]$, which rotates the cloud by the angle $\omega_s$.
In the rotation--reflection regime, $U_s$ is this rotation matrix, multiplied on the right by $\mathrm{diag}(-1, 1)$ with probability one half independently of $\omega_s$. That is, half of the clouds are reflected across the vertical axis before they are rotated, and $U_s$ is uniformly distributed over all $2\times 2$ orthonormal matrices.
In each regime the transformation preserves all pairwise distances, so the persistent homology features $y^s$ are unchanged and the transformation alters only $G$ in \eqref{eq:generative}.

We compare our model with two generative models fitted directly to the point cloud.
The first is a per-vertex Gaussian model, $\tilde y^s_i \sim \mathrm{N}(\mu^{(p)}_i, \sigma_p^2 I)$ independently over vertices $i$ and subject $s$ in group $p$.
Its maximum likelihood estimates are the vertex-wise samplem mean $\mu^{(p)}_i$ over the subjects of group $p$ and the pooled residual variance $\sigma_p^2$.
In the absence of the nuisance, this is the model that generated the data and is therefore correctly specified.
The second is a four-component Gaussian mixture on the pooled points of each group with unrestricted covariances, which uses no vertex labels.
We report the three-class accuracy on $10$ held-out subjects per group, and the AUC for recovering the $75$ differing vertices. We rank the vertices by $\lVert\hat\mu^{(p)}_i-\hat\mu^{(p')}_i\rVert$ for the per-vertex model and by $\lVert z_i^{(p)}-z_i^{(p')}\rVert$ for ours, and report the AUC against the differing set.

Table~\ref{tab:frame} reports mean and standard deviations over $32$ replicates for the two generative models and $16$ for ours.
Without the nuisance, the per-vertex model recovers the differing vertices exactly, while our model attains an AUC of $0.782$, as expected when a correctly specified model is compared with a misspecified one.
Under the three regimes, our results are identical to those without the nuisance. Our likelihood is defined on the persistent homology features $y^s = T(\tilde y^s)$, and the Vietoris--Rips filtration $T$ depends on the point cloud only through $D^{(s)}$, which every regime leaves unchanged. This is the invariance the reduction in \eqref{eq:generative} gains.

The per-vertex model is not invariant to the nuisance transformation. Its fitted mean for vertex $i$ in group $p$ is $\hat\mu^{(p)}_i = S_p^{-1}\sum_{s}\tilde y^s_i(U_s)$, where the sum runs over the $S_p$ subjects of group $p$.
Because $\mathbb{E}\{\tilde y^s_i(U_s) \mid \tilde y^s\} = \bar y^s$, $\hat\mu^{(p)}_i$ converges in probability to the expected centroid of group $p$ as $S_p$ grows, the same point for every vertex.
The estimated difference $\hat\mu^{(p)}_i - \hat\mu^{(p')}_i$ then converges to a vector that does not depend on $i$, so it no longer identifies the differing vertices.
With $S_p = 10$, recovery is accordingly unstable: the mean AUC lies between $0.38$ and $0.57$ across the three regimes, with a standard deviation of about $0.36$ across replicates.
The model absorbs the nuisance into its noise estimate, with $\hat\sigma_p$ rising from $0.066$ to about $0.42$, and the ratio of the root-mean-square spread of $\hat\mu^{(p)}_i$ to $\hat\sigma_p$ falls from $8.66$ to below $0.5$. Accuracy falls less, to between $0.73$ and $0.75$ against a chance rate of $0.333$, so the nuisance mainly costs localization.

The Gaussian mixture has no AUC entry. It models the distribution of a single point rather than the position of vertex $i$, so it has no vertex-specific parameter whose difference between groups could be ranked. It is nonetheless the more accurate of the two generative models under the nuisance transformation, with accuracy between $0.83$ and $0.87$.

We repeat the study with the vertex positions and the noise drawn from a variance-matched mixture of elliptical Laplace and $t_4$ distributions, for which neither competitor is correctly specified. Both generative models' entries are reproduced within one standard deviation, so what breaks the per-vertex model is the nuisance transformation, not the distributional assumption.

The invariance is not specific to the Vietoris--Rips filtration; it is inherited from whatever the filtration takes as input, so each filtration in Section~\ref{sec:background} contributes an invariance of its own. The generalized Vietoris--Rips and the alpha filtrations are again functions of the distance matrix, or of the power distance in the weighted case, and are therefore invariant to the same rigid motions and to any isometric deformation of the cloud. The sublevel filtration is a function of the vertex values $\phi(w_i)$ on a fixed graph $G$, so the invariance it confers is to any transformation of the data that leaves those values unchanged: for the radial filtration, rotations and reflections about the reference point; for the kernel density filtration, any isometry of the cloud, since a density estimate built from distances is itself isometry-equivariant. In each case the reduction in \eqref{eq:generative} discards exactly the part of the mechanism that the chosen filtration cannot see, and the partial likelihood is invariant to it without any change to the model.

\begin{table}[t]
\centering
\begin{tabular}{lccc}
\toprule
Transformation & Ours & Per-vertex Gaussian & Gaussian mixture \\
\midrule
\multicolumn{4}{l}{\textit{Recovery of the differing vertices (AUC)}} \\
\quad none & $0.782\,(0.038)$ & $1.000\,(0.000)$ & --- \\
\quad sign flips & $0.782\,(0.038)$ & $0.542\,(0.356)$ & --- \\
\quad rotation & $0.782\,(0.038)$ & $0.378\,(0.370)$ & --- \\
\quad rotation and reflection & $0.782\,(0.038)$ & $0.567\,(0.359)$ & --- \\
\addlinespace
\multicolumn{4}{l}{\textit{Three-class accuracy}} \\
\quad none & $1.000\,(0.000)$ & $1.000\,(0.000)$ & $1.000\,(0.000)$ \\
\quad sign flips & $1.000\,(0.000)$ & $0.733\,(0.107)$ & $0.872\,(0.092)$ \\
\quad rotation & $1.000\,(0.000)$ & $0.744\,(0.088)$ & $0.831\,(0.081)$ \\
\quad rotation and reflection & $1.000\,(0.000)$ & $0.754\,(0.073)$ & $0.837\,(0.081)$ \\
\bottomrule
\end{tabular}
\caption{Effect of nuisance transformations, each drawn independently of the subject's group label. AUC is defined Section~\ref{sec:sim}. Entries are means over $32$ replicates ($16$ for ours), with standard deviations across replicates in parentheses. The exchangeable Gaussian mixture has no vertex-indexed parameter, hence no AUC entry.}
\label{tab:frame}
\end{table}

\section{Modeling higher-order features}\label{sec:sim-h2}

The regimes above lie in $\mathbb{R}^2$, where the second homology is empty by
construction.  We close with one in $\mathbb{R}^3$ that produces
two-dimensional features, and ask whether extending~\eqref{eq:triangulation_likelihood}
to $k=2$ recovers a group difference that the zero- and one-dimensional bars
alone do not.  The
oracle is four disjoint spherical shells, $40$ points area-uniform on each and
$n=160$ in total, placed far enough apart to remain in separate connected
components until after every void has died.  Group~$B$ differs from group~$A$
by reshaping one shell into a spheroid of the same surface area, with $q$ the
ratio of its axes, so that $q=1$ is the identity and smaller $q$ flattens the
shell more.

Holding the area fixed is what makes the comparison informative.  The same
$40$ points then sit at the same local density, so the minimum spanning tree
barely moves, which is the only feature \eqref{eq:competing_exponentials} depends on; the void enclosed by the
shell, meanwhile, changes shape. At $q=0.70$ the shell is flattened by $22\%$
along one axis while its surface area stays at $4\pi$, and the two groups'
spanning-tree edge lengths remain almost indistinguishable, with total weights
$63.71$ and $64.00$ and a two-sample Kolmogorov--Smirnov statistic of $0.013$.
Only the reshaped shell's two-dimensional bar shifts; the other three are
unchanged.

Subjects are $\mu^{(g)} + N(0,\sigma^2 I_3)$ with $\sigma=0.05$, ten per
group, and each group is fitted separately under the prior~\eqref{eq:prior_z} with $m=5$,
$\kappa=6$ and $\alpha=0.2$, using two chains of $300$ warm-up and $300$
sampling iterations.  Every replicate redraws the oracle, and within a
replicate all arms are fitted to identical data, so the comparisons are paired
across $16$ draws.  Since $\Lambda=ZZ^\top$ is identified only up to rotation,
we align the two groups in each posterior draw by orthogonal Procrustes and
record $\delta_i = \lVert z_i^{A}R - z_i^{B}\rVert$, the quantity used to
localize in Section~6.  Ranking the $160$ vertices by the posterior mean of
$\delta_i$, we report the probability that a vertex of the reshaped shell
outranks an untouched one, for which $0.5$ is chance.

\begin{table}
\centering
\caption{Recovery of the reshaped shell, mean (standard deviation) over
16 draws of the underlying point configuration, both fits using
identical data. Chance is 0.5. The first column uses the likelihood of
equations~\eqref{eq:competing_exponentials}
and~\eqref{eq:triangulation_likelihood}; the second adds the
two-dimensional term. The first two rows are calibration: identical
geometry, without and with unequal subject noise across groups
($\sigma = 0.05$ against $0.08$). $q$ is the axis ratio of the reshaped
shell; smaller $q$ is a larger departure from the sphere.}
\label{tab:h2-arms}
\small
\begin{tabular}{l cc}
\toprule
 & $H_0{+}H_1$ & $H_0{+}H_1{+}H_2$ \\
\midrule
$q=1.00$ (identity) & $0.502\,(0.226)$ & $0.490\,(0.091)$ \\
$\sigma$ mismatch   & $0.504\,(0.189)$ & $0.500\,(0.123)$ \\
\addlinespace
$q=0.85$ & $0.558\,(0.202)$ & $0.639\,(0.102)$ \\
$q=0.70$ & $0.651\,(0.208)$ & $0.830\,(0.090)$ \\
$q=0.55$ & $0.777\,(0.146)$ & $0.931\,(0.070)$ \\
\bottomrule
\end{tabular}
\end{table}

The first two rows of Table~\ref{tab:h2-arms} are calibration.  At $q=1$ the
groups differ only through independent noise draws, and every arm sits at
chance.  In the second row the geometry is again identical but the groups have
different subject noise, $\sigma=0.05$ against $\sigma=0.08$; every arm stays
at its $q=1$ value, the largest paired shift being $+0.010$ ($p=0.74$), so
recovery responds to the change in shape and not to heterogeneity in noise.
Neither row is used to compare arms.

Across the three reshaped cells, adding the two-dimensional term raises
recovery at every dose, by $+0.082$, $+0.179$ and $+0.154$, the last improving
$15$ of the $16$ draws ($p = 8\times10^{-4}$).  The gain belongs to the second
homology rather than to higher-order features in general: the $H_2$ arm
exceeds the $H_0$ arm by $+0.403$ at $q=0.55$ in every draw, and it is the
only arm whose recovery rises with the deformation in all $16$, by $+0.088$,
$+0.272$ and $+0.412$ relative to the identity cell, while $H_0$ moves by at
most $0.021$ over the same sweep and $H_1$ not until the strongest
deformation.

What the regime establishes is thus specific.  The recovery target is the
reshaped shell rather than a vertex within it, since $B^k_i$ spans the cycle's
connected component and carries no marker of which vertices bound the void.
Subject to that resolution, when the data contain persistent two-dimensional
features, modeling them with the two-dimensional form
of~\eqref{eq:triangulation_likelihood} locates the source of a group
difference that the zero- and one-dimensional terms largely miss.

\begin{figure}
\centering
\includegraphics[width=0.9\textwidth]{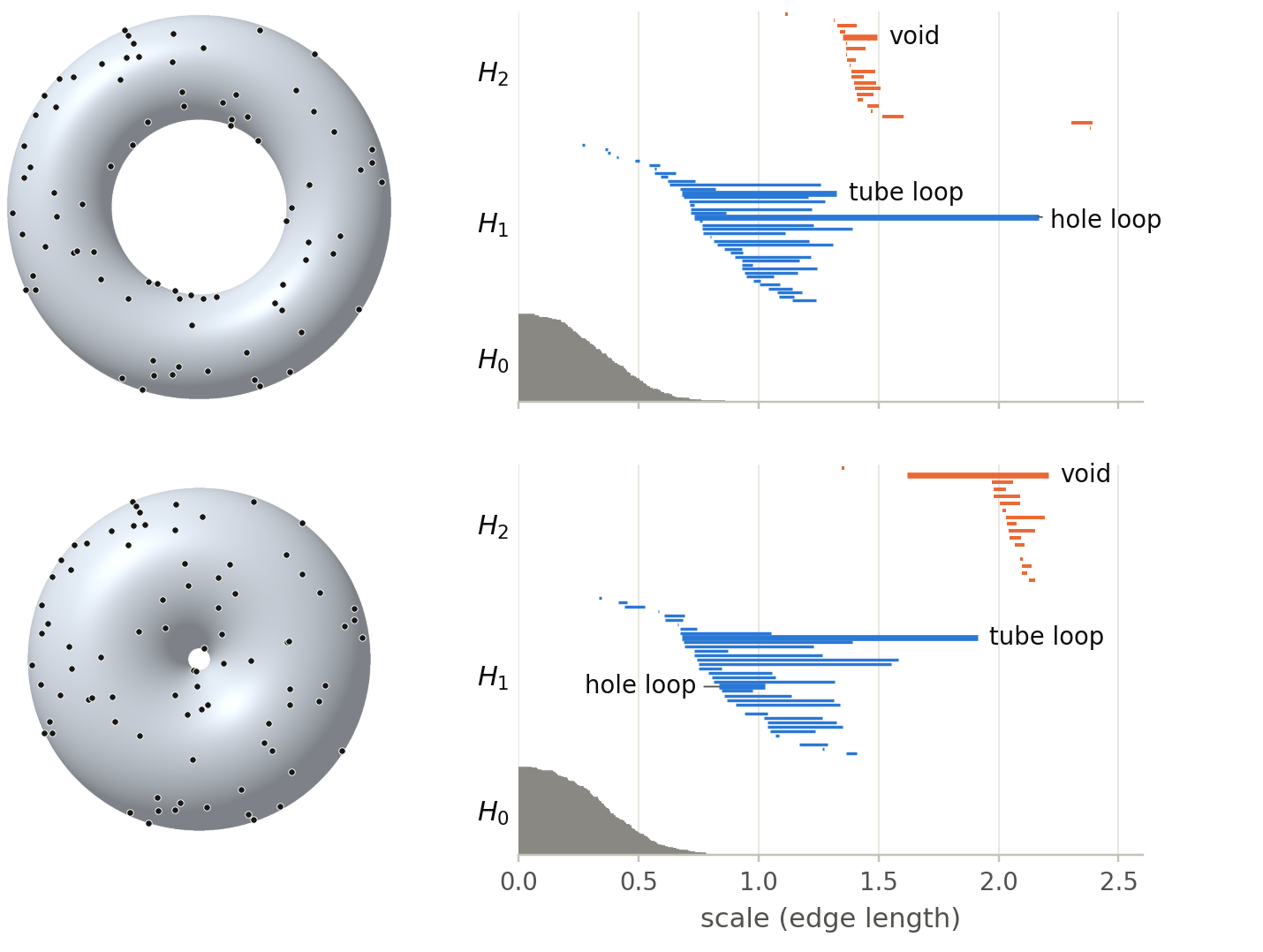}
\caption{The tori at $t=1$.  Left: the reshaped torus seen from above,
thin in group~$A$ (top; tube radius $0.75$, ring radius $2$) and fat in
group~$B$ (bottom; tube radius $1.15$, ring radius $1.30$), at equal
surface area, with the $160$ vertices of one draw before noise.  Right:
the persistence barcode of the same draw after noise, with the most
persistent void, loop around the tube (tube loop) and loop around the
central hole (hole loop) labelled.  The hole loop dominates the thin
torus, whose voids are short; the fat torus has a longer void and tube
loop, and a short hole loop.}
\label{fig:h2-torus}
\end{figure}

A sphere has no loops, so we also consider a surface on which loops and a
void coexist (Figure~\ref{fig:h2-torus}).  Each subject is a cloud of
$320$ points, $160$ spread uniformly over each of two disjoint tori,
perturbed by $N(0,0.05^2 I_3)$ noise, with ten subjects per group.  One
torus is the same in both groups, with ring radius $1.5$ and tube radius
$1$.  The other has tube radius $1-0.25t$ in group~$A$ and $1+0.15t$ in
group~$B$, and ring radius $1.5$ divided by its tube radius, which keeps
the surface area, and hence the density of points, unchanged.  At $t=1$ it
is a thin torus with a large central hole in group~$A$ and a fat torus
with a small one in group~$B$: on the thin torus the loop around the hole
is the most persistent feature and the void is short-lived, while on the
fat torus the void lasts three times longer and the loop around the hole a
fifth as long.  Before any model is fitted, the two groups then differ in
the death of the void by $15$ standard deviations, in that of the loops
around the tube and around the hole by $20$ and $15$, and in the minimum
spanning tree by less than one.  We take $t=0$, where the groups do not
differ, and $t=0.10,0.25,0.50,1$, and repeat every setting over $32$
independent draws of the tori.

We fit our model with the hierarchical prior in
Section~\ref{sec:multigrp} to the simulated data, using the likelihood of
equations~\eqref{eq:competing_exponentials} and~\eqref{eq:triangulation_likelihood},
the latter also applied at $k=2$, under $m=5$, $\kappa=\kappa_0=6$ and $\alpha=0.2$.  In
each setting, we use the optimization-based warm start to find the
conditional mode of the posterior distribution, initialized at the mode for
all subjects pooled, and estimate the latent coordinates by this mode.  For
localization, we align the latent coordinates of the two groups by an
orthonormal transformation, rank the vertices by
$\hat\delta_i=\|\tilde z^A_i-\tilde z^B_i\|_2$, the distance used in
Section~\ref{sec:data}, and report the AUC of $\hat\delta_i$ against the
vertices of the reshaped torus.  For classification, we simulate ten
further subjects per group that are not used in the fit, and assign each
to the group under which its likelihood at the estimated $\hat\Lambda^p$
is larger, the classifier of Section~\ref{sec:data}.

\begin{table}
\centering
\caption{Localization and classification on the tori, mean (standard
deviation) over $32$ draws.  Localization is the AUC of $\hat\delta_i$
against the vertices of the reshaped torus; classification is the accuracy
on ten held-out subjects per group.  Chance is $0.5$ for both.}
\label{tab:h2-torus}
\small
\begin{tabular}{l cc}
\toprule
 & Localization & Classification \\
\midrule
$t=0$ (no difference) & $0.49\,(0.13)$ & $0.50\,(0.09)$ \\
\addlinespace
$t=0.10$ & $0.62\,(0.14)$ & $0.69\,(0.12)$ \\
$t=0.25$ & $0.81\,(0.11)$ & $0.95\,(0.06)$ \\
$t=0.50$ & $0.92\,(0.07)$ & $1.00\,(0.02)$ \\
$t=1$ & $0.91\,(0.07)$ & $1.00\,(0.00)$ \\
\bottomrule
\end{tabular}
\end{table}

When the groups do not differ, both localization and classification are
at chance.  Both improve with $t$: the model classifies $69\%$ of the
held-out subjects at $t=0.10$, $95\%$ at $t=0.25$ and over $99\%$ from
$t=0.50$, while the AUC rises from $0.62$ at $t=0.10$ to above $0.9$ from
$t=0.50$.  Hence, on data where loops and a void coexist, the model fitted
with all three terms both classifies new subjects and localizes the source
of the group difference.

\section{Comparison with additional classifiers}\label{sec:supp-competitors}

The $K$-nearest neighbor classifier in Section~\ref{sec:data} depends on the data only through the bottleneck distance. To assess whether other representations of the same data separate the groups, we consider two additional classifiers. The first approach uses regression with the connectivity matrix as a matrix-valued predictor \citep{zhou2014regularized}. We specify a three-class multinomial logistic regression model in which the usual covariate-contribution term $X^{T}{\beta}_c$ is replaced by $\text{trace}(B_c^T W^s)$, where $W^s$ is the connectivity matrix for subject $s$ and $B_c$ is a low-rank symmetric coefficient matrix for class $c$.
The second is fitted on the persistence summaries. We convert each subject's $H_0$ and $H_1$ diagrams into persistence landscapes \citep{bubenik2015statistical} and fit a kernel ridge classifier to the resulting feature vectors. All tuning parameters, including the rank of $B_c$, are selected by cross-validation inside each training fold. Both models are evaluated by 5-fold cross-validation, as are the lasso logistic and ordinal regression models in Section~\ref{sec:data}.

Neither classifier improves materially on the majority-class rule (Table~\ref{tab:supp-competitors}). Matrix regression attains 50.0\% on the 3-class task and 77.9\% on the binary task of separating cognitively normal from diseased subjects.
Kernel ridge regression on persistence landscapes attains 44.2\% and 77.9\%, respectively. These values are best judged against the majority-class baseline. On these $S = 104$ subjects, always predicting the largest class gives 44.2\% on the 3-class task, and declaring every subject diseased gives 79.8\% on the binary task. Across the five comparison methods in Table~\ref{tab:supp-competitors}, at most 6 of the 21 cognitively normal subjects are classified correctly, so their accuracy on the normal-versus-diseased task comes almost entirely from the diseased majority.

Beyond predictive accuracy, the methods differ in the region-level inference they support.
The lasso logistic and ordinal regression models give point estimates of a coefficient for each region pair, and matrix regression gives a score for each region through the row norms of $\hat{B}_c$.
The bottleneck distance and the persistence landscape are global summaries of the the persistence diagram and give no region-level quantity.
None of these comparison methods quantifies uncertainty about region-level quantities. In our model, the latent coordinates $z_i^{(p)}$ are indexed by region, so the fit gives the region-level posterior inference in Section~\ref{sec:data}.

\begin{table}[ht]
\centering
\small
\resizebox{\textwidth}{!}{%
\begin{tabular}{llccl}
\hline
Method & Input representation & 3-class (\%) & CN vs.\ diseased (\%) & Region-level output \\
\hline
$K$-nearest neighbors        & bottleneck distance   & 48.1 & 73.1 & none \\
Lasso logistic regression    & raw connectivity      & 43.3 & 78.8 & edge-level point estimate \\
Ordinal regression           & raw connectivity      & 48.1 & 78.8 & edge-level point estimate \\
Matrix regression            & raw connectivity      & 50.0 & 77.9 & region-level point estimate \\
Kernel ridge                 & persistence landscape & 44.2 & 77.9 & none \\
Proposed model               & barcode               & 76.9 & 86.5 & region-level posterior \\
\hline
\end{tabular}
}
\caption{Classification accuracy on the ADNI structural connectivity data ($S = 104$ subjects, $n = 83$ regions; 21 cognitively normal, 46 mildly cognitively impaired, 37 Alzheimer's disease). The 3-class column is accuracy over the three diagnostic groups. The CN vs.\ diseased column is accuracy after merging the mildly cognitively impaired and Alzheimer's disease groups. The last column records what each method reports about an individual region.}
\label{tab:supp-competitors}
\end{table}

\end{document}